\documentclass[a4paper,12pt]{article}
\pdfoutput=1
\usepackage{amssymb,amsmath,bm}
\usepackage{a4wide}
\usepackage{color}
\usepackage{slashed}
\usepackage{cite}
\usepackage[mathscr]{euscript}
\usepackage[normalem]{ulem}
\usepackage{hyperref}
\usepackage{units}
\usepackage{relsize}
\usepackage{bbold}
\usepackage[latin1]{inputenc}
\usepackage{amsfonts}
\usepackage{lscape}
\usepackage{amsthm}
\usepackage{booktabs}
\usepackage{array}
\usepackage{rotating}
\usepackage{float}
\usepackage{multirow}
\usepackage{xfrac}
\usepackage{wasysym}
\usepackage{graphicx,rotating,amsmath,multirow,xcolor,colortbl,xcolor}
\usepackage{hyperref,pdflscape,slashed}

\usepackage{tikz}
\usetikzlibrary{matrix,decorations.pathreplacing}

\usepackage{cite}

\definecolor{rosso}{cmyk}{0,1,1,0.4}
\definecolor{rossos}{cmyk}{0,1,1,0.55}
\definecolor{rossoc}{cmyk}{0,1,1,0.2}
\definecolor{blu}{cmyk}{1,1,0,0.3}
\definecolor{blus}{cmyk}{1,1,0,0.6}
\definecolor{bluc}{cmyk}{1,1,0,0.1}
\definecolor{verde}{cmyk}{0.92,0,0.59,0.25}
\definecolor{verdec}{cmyk}{0.92,0,0.59,0.15}
\definecolor{verdes}{cmyk}{0.92,0,0.59,0.4}
\definecolor{Gray}{gray}{0.95}

\font\tenrsfs=rsfs10 at 12pt
\font\sevenrsfs=rsfs7
\font\fiversfs=rsfs5
\newfam\rsfsfam
\textfont\rsfsfam=\tenrsfs
\scriptfont\rsfsfam=\sevenrsfs
\scriptscriptfont\rsfsfam=\fiversfs
\def\mathscr#1{{\fam\rsfsfam\relax#1}}

\usepackage[small,bf]{caption}
\hypersetup{colorlinks,bookmarksopen,bookmarksnumbered,
linkcolor=blus,pdfstartview=FitH,urlcolor=blue,citecolor=verde}

\newcommand{\tev}{\textrm{TeV}}
\newcommand{\gev}{\textrm{GeV}}

\newcommand{\lsim}{\stackrel{<}{_\sim}}
\newcommand{\gsim}{\stackrel{>}{_\sim}}

\newcommand{\ie}{{\em i.e.}}
\newcommand{\hc}{{\rm h.c.}}

\newcommand{\be}{\begin{equation}}
\newcommand{\ee}{\end{equation}}
\newcommand{\bea}{\begin{eqnarray}}
\newcommand{\eea}{\end{eqnarray}}
\newcommand{\beq}{\begin{equation}}
\newcommand{\eeq}{\end{equation}}
\newcommand{\beqa}{\begin{eqnarray}}
\newcommand{\eeqa}{\end{eqnarray}}

\def\id{\mathbb{1}}

\def\eq#1{eq.~(\ref{#1})}
\def\fig#1{fig.~\ref{#1}}

\def\tab#1{table~\ref{#1}}

\counterwithin*{equation}{section}

\def\eq#1{eq.~(\ref{#1})}
\def\fig#1{fig.~\ref{#1}}
\def\sec#1{sect.~\ref{#1}}

\def\O{\mathcal{O}}
\def\R{\mathcal{R}}
\def\LUV{\Lambda_{\rm UV}} 
\def\mst{m_*}
\def\mS{m_S}

\def\Fs{F_ {{\rm susy}\! \! \! \! \! \! \! \!\diagup \,\,\,}}
\def\Ks{\kappa_ {{\rm susy}\! \! \! \! \! \! \! \!\diagup \,\,\,}}
\def\msoft{m_{\rm soft}}
\def\SSMnH{{\rm SSM}_{H \!\!\!\!\!\not\,\,\,\,\,}} 
\def\CFTs{{\rm CFT}_ {{\rm susy}\! \! \! \! \! \! \! \!\diagup \,\,\,}} 
\def\mgra{m_{3/2}}
\def\TR{T_{\textrm{RH}}}
\def\gSM{g_{\textrm{SM}*}}

\def\hlambda{\hat\lambda}
\def\muRG{\mu_{\scriptscriptstyle{\textrm RG}}}
\def\scalar#1{{\tilde \varphi}_{\raisebox{-2pt}{$\scriptstyle {#1}$}}}
\DeclareMathOperator*{\Max}{max}

\begin{document}

{\hfill CERN-TH-2026-188}

\vspace{2cm}

\begin{center}
\boldmath

{\textbf{\LARGE The Super-Composite Higgs}}

\unboldmath

\bigskip

\vspace{1 truecm}

{\bf Kaustubh Agashe}$^a$, {\bf Gian F. Giudice}$^{b,c}$, {\bf Riccardo Rattazzi}$^d$, {\bf Raman Sundrum}$^a$
 \\[5mm]

$^a$ {\it Maryland Center for Fundamental Physics, University of Maryland, \\ College Park, MD 20742, USA}\\[2mm]
$^b$ {\it New York University Abu Dhabi, UAE}\\[2mm]
$^c$ {\it CERN, Theoretical Physics Department, Geneva, Switzerland}\\[2mm]
$^c$ {\it Theoretical Particle Physics Laboratory (LPTP), Institute of Physics, \\ EPFL, Lausanne, Switzerland}\\[2mm]

\vspace{2cm}

{\bf Abstract}
\end{center}

\begin{quote}
We construct the Super-Composite Higgs, a scenario designed to stabilize the Higgs mass and generate the flavor structure within a single dynamical framework. The basic idea is to combine supersymmetry and compositeness, where the Higgs mass is protected by supersymmetry and Yukawa couplings are generated through partial compositeness. This combination offers mutual advantages and ameliorates several drawbacks of each individual theory. The Super-Composite Higgs makes experimentally testable predictions across three different areas of fundamental physics. At the precision frontier, the electron EDM is predicted to be within reach of future experiments. At the observational cosmology frontier, measurable effects in CMB and large-scale structures are expected from an ultra-light gravitino. At the high-energy frontier, an upper bound on supersymmetric particle masses, derived from cosmological considerations, indicates that squarks should be lighter than about 10 TeV, which is within reach of the FCC-$hh$.

\end{quote}

\thispagestyle{empty}
\vfill

\newpage
{\small
\tableofcontents
}
\newpage
%%%%%%%%%%%%%%%%%%%%%%%%%%%%%%%%%%%%%%%%%

\section{Introduction}
\label{sec:intro}

BSM model building is the exploration of the space of possible short-distance completions of the Standard Model (SM), with the goal of explaining its structural properties. Provided this endeavor is undertaken with clarity of assumptions and intents, the outcome has unavoidably value for refining the very nature of the questions being asked. In many instances, the knowledge gained from this exploration can offer insights into how experimental searches probe the space of hypotheses and, in particular, into which searches are the most effective. It is with this perspective in mind that we approach the subject of this paper.

One of the grand ambitions of fundamental physics is to understand the remarkably hierarchical structure of Nature's laws. On the cosmological front, there is the sheer size and age of the Universe in the face of the cosmological constant problem, and the high degree of uniformity on large distances. On the particle physics frontier, Nature exhibits vast mass hierarchies from its heights at the inferred Planck scale to the ghost-like neutrinos. In this paper, we focus on two challenges in BSM physics: {\it (i)} the electroweak (EW) naturalness problem of understanding the mechanism triggering spontaneous symmetry breaking and its radiative stability over many decades in energy, and {\it (ii)} the flavor puzzle of explaining the distinctive hierarchical features observed in the quark and lepton masses and mixings. 

Our main goal is to construct a scenario that addresses the origin of the weak scale and the origin of flavor within the same dynamics, and with a chance of being  directly tested by experiments. It will lead us to explore regions of theory space that have not yet received the attention they deserve. In particular, we argue that generating electroweak and flavor hierarchies in concert is most plausibly 
 accomplished by supersymmetric strongly-coupled BSM physics despite its formidable theoretical challenges. 

Let us consider the experimental prospects for discovering the BSM physics at the root of the hierarchy problem and flavor puzzle. The gambling principle of naturalness within effective field theory (EFT) suggests that the hierarchy problem should be solved by new physics at the TeV scale and, while it is unsettling that we have not seen such new physics yet, whether virtually in EW or Higgs precision tests or on-shell in new-particle searches at the LHC, this is not yet a definitive crisis. For example, consider supersymmetry, a celebrated BSM paradigm for solving the hierarchy problem. 
From a bottom-up perspective, the key measure of EW fine-tuning in this paradigm is captured by the contributions to the physical Higgs boson mass $m_h$ from higgsinos and stop quarks at tree-level and one-loop, respectively:
\begin{equation}
    \delta m_h^2 = - 2 \mu^2 + \frac{3 y_t^2}{2 \pi^2} \, {\tilde m}_t^2 \, \ln \frac{M}{{\tilde m}_t} + ... \, .
    \label{massHbu}
\end{equation}
Here $y_t$ is the top Yukawa coupling, $\mu$ sets the scale for the higgsino mass, and $M$ is the scale at which the mediation of supersymmetry breaking occurs.

The amount of tuning corresponding to these two contributions to the Higgs mass is
\bea
&&\left( \frac{m_h^2}{\delta m_h^2}\right)_{\textrm{higgsino}} =  \left( \frac{300~\gev}{\mu}\right)^2 \times 9\%
\label{higgsinoc}
\\
&&\left( \frac{m_h^2}{\delta m_h^2}\right)_{\textrm{stop}} =  \left( \frac{1.2~\tev}{{\tilde m}_t}\right)^2 \times \left\{
\begin{array}{l}
3\permil ~~\textrm{for}~M=10^{16}~\gev \\
2\% ~~\textrm{for}~M=10^{5}~\gev
\end{array}
\right.
\label{stopc}
\eea
The experimental bounds on higgsinos are model-dependent, but \eq{higgsinoc} demonstrates that higgsinos contribute to fine-tuning at most at the percent level. The usual source of grievance over supersymmetry comes from the stop contribution. Even for ${\tilde m}_t$ as low as its experimental limit of 1.2 TeV, the tuning is at a disturbing per-mil level, once high-scale mediation is taken for granted.  However, when we consider a low mediation scale of about 100 TeV, the tuning becomes milder, at the level of a few percent.

Admittedly, from a top-down perspective, known supersymmetric models suffer significantly greater fine-tuning, especially when one considers the origin of the Higgs quartic coupling which often requires much larger stop masses. However, it is unclear if these additional sources of tuning are a shortcoming of the supersymmetric paradigm as a whole or just of the current state of the art in theoretical model-building. The message of
 the bottom-up perspective is that, independently of specific model-dependent choices, lowering the mediation scale comes with a reduction of tuning in supersymmetry.  Indeed, our goal is to re-examine the problem within the supersymmetric paradigm and 
present a plausible top-down framework which approaches the bottom-up hopes. 

By contrast, there are strong reasons to think that the physics underlying the flavor puzzle will be a greater reach in energy. In the SM, all flavor-violating physics arises through the Yukawa matrices. This yields a distinctive experimental structure in which flavor-changing neutral currents and CP violation are highly suppressed by the GIM mechanism, while flavor-changing charged currents are comparatively sizable and parametrized by the CKM matrix. These striking features are beautifully borne out by a host of flavor/CP experiments. The flavor puzzle is thereby substantially embodied by the extreme hierarchical structure of the Yukawa matrices, which the SM accommodates but does not explain. By definition then, a full answer to this puzzle would entail that the hierarchical Yukawa matrices emerge from some more violently flavor-violating BSM physics. 

In models addressing the origin of the weak scale, flavor often surfaces only as a nuisance. This is because BSM models tackling EW naturalness often lack the structural minimality of the SM, which governs its phenomenological adequacy via the GIM mechanism. However, the minimality of flavor in the SM reflects its inability to explain the observed flavor pattern. Considering all that, why then not view the need to modify flavor in natural models as an opportunity to explain its structure? Rather than retreat defensively into scenarios that mimic the SM and its GIM protection at high scales, why not play offensively and try to explain flavor from relatively low scale dynamics? This is the challenge we set for ourselves. 

We choose to pursue our stated goal by marrying supersymmetry and Higgs compositeness in a theory that we call Super-Composite Higgs. It is quickly explained how we were carried to these grounds. Models of composite Higgs perfectly illustrate the dangerous opportunities for flavor that are offered by natural models. In a composite Higgs theory, the Yukawa interactions are strongly irrelevant deformations,  hence insufficient sources of  fermion masses, unless one relies on interesting, but elaborate, multi-scale models. Remarkably, the unique alternative is partial fermion compositeness, which offers the possibility of a full-fledged theory of flavor. 

Partial compositeness~\cite{Kaplan:1991dc} is the hypothesis that quarks and leptons are quantum superpositions of elementary and composite particles, with these superpositions and associated flavor hierarchies unfolding from strong-coupling renormalization-group (RG) evolution between the far UV and the compositeness scale. Unsurprisingly, partial compositeness comes together with new sources of flavor and CP violation which, unfortunately, can be tamed only at the price of extreme unnaturalness in the EW breaking \cite{Glioti:2024hye}. This occurs because, in composite Higgs models, a single scale $m_*$ controls both flavor and naturalness. 

Combining supersymmetry and Higgs compositeness can mitigate the EW naturalness problem and allow for a dynamical theory of flavor. In this way, supersymmetry and compositeness share a symbiotic relationship. Strong-coupling physics generates flavor hierarchies via partial compositeness and then proceeds (via a process of sequential ``tumbling'') to break supersymmetry. Soft masses for the supersymmetric partners are generated through a dynamical mechanism analogous to gauge mediation, which ensures a high degree of flavor universality. 

However, the Achilles heel of almost any calculable mechanism of supersymmetry-breaking mediation is the origin of the higgsino mass parameter $\mu$ which, for phenomenological reasons, must be comparable to other soft masses, even though it is protected by an independent symmetry, namely the Peccei-Quinn chiral symmetry $U(1)_{\textrm{PQ}}$. The puzzle is why there is a near-coincidence of scales among the soft masses, $\mu$, and the PQ-violating Higgs mass-mixing $B_{\mu}$. This poses the so-called $\mu$--$B_\mu$ problem. 

Surprisingly, compositeness comes to the rescue of supersymmetry by offering a viable solution to the $\mu$--$B_\mu$ problem. This is yet another advantage of the mutual assistance between supersymmetry and compositeness.

We do not explicitly know the specific form of strong dynamics needed for a fully realistic model along the lines hypothesized in this paper. This is unsurprising, given the limited theoretical control over non-perturbative phenomena. Much of our intuition relies on AdS/CFT holographic duality to recast the 4D strong dynamics as  (modestly) weakly-coupled and higher-dimensional warped compactifications of Randall-Sundrum 1 (RS1) type~\cite{Randall:1999ee}, generalizing non-supersymmetric RS1 constructions. However, our study will largely follow a 4D description of strong supersymmetric dynamics. 

The need for a formulation of the strongly-coupled dynamics in Super-Composite Higgs introduces another aspect of novelty in our work, beyond the original goal of addressing naturalness and flavor simultaneously. The novel aspect resides in the extension to supersymmetry of the SILH approach~\cite{Giudice:2007fh} to describe or, better, to depict composite Higgs models. It consists of two basic steps. The first is to hypothesize the UV completion of the Higgs sector into an abstract  gapped CFT weakly coupled to the rest of the SM. The second is the description of the low-energy phenomenology through an effective Lagrangian with well-defined power counting rules, but with many $O(1)$ undetermined parameters. 

With this approach, we renounce strict quantitative control and also
 take the risk that the hypothesized UV scenario may not correspond to any truly existing 4D CFT.\footnote{One can speculate, based for instance on indications from the bootstrap \cite{Kos:2015mba} and naturalness arguments, that non-supersymmetric CFTs constitute a discrete set. The set of such theories, with a finite number of degrees of freedom quantified for instance by the $a$ and $c$ anomaly coefficients, is therefore expected to be finite.  With supersymmetry one can naturally have manifolds of CFTs (see {\it e.g.} ref.~\cite{Leigh:1995ep}), but  associated with exactly marginal deformations. Lacking the latter one expects, again,  a discrete set of CFTs.}
However, we gain the ability  to explore model building away from under the lamppost of weakly-coupled Lagrangians. Moreover, 
the risk of wishful hypotheses and the relinquishing of full quantitative control are tempered by the existence of concrete constructions  based on warped compactifications. As a matter of fact, warped compactifications, through their holographic interpretation, have been the indispensable inspiration of this whole approach. They offer a remarkable setup where  
one can describe the physics of a strongly coupled CFT, within the EFT canon, virtually reaching energies as high as the Planck scale. However, this advantage comes at the price of surrendering control over states associated with operators of large enough dimensions. 

The prejudice that CFTs  form a discrete set implies that generic warped models, which like all EFTs feature continuous parameters, are probably not UV completable. Somewhat relatedly, the calculability of warped models within EFT implies a specific sparse operator spectrum, a property which holographically corresponds to a large value of the 't Hooft coupling. This situation may be difficult to realize in 4D, besides very specific cases like $N$=4 gauge theories. 
But not enough is known about these issues. Concerning the first issue, one could speculate that perhaps  complete CFTs densely populate the set of holographic constructions. Indeed, recent studies classifying Wilson-Fisher fixed points \cite{Osborn:2020cnf,Pannell:2023tzc} indicate, encouragingly, that their number  grows rather fast with the number of fields. Concerning the second issue, we must  stress that the 4D description that we will employ in this paper encompasses any theory that may exist, including those that do not possess a calculable weakly coupled 5D dual.
It is thus with some of the thrill that accompanies the most daring and uncertain explorations that we undertake our study.

As this paper is rather long and detailed, we feel it is important to provide some guidance for readers. The paper is divided into two main parts. Sections~\ref{sec:intro} through \ref{sec:allthat} serve as a conceptual and technical introduction. In \sec{sec:why} we present our motivations for combining supersymmetry and compositeness into a single framework. Sections~\ref{sec:frameCFT} and \ref{sec:frame5D} give an overview of our scenario, first as a purely 4D theory and then as a 5D theory with
warped compactification. In \sec{sec:tumbling}, we use a calculable toy example to illustrate the plausibility of our tumbling hypothesis. Section~\ref{sec:allthat} is technical in nature, laying out the power counting rules used throughout the paper and reviewing the RG flow induced by partial compositeness at large $N$. 

Sections ~\ref{sec:SILSH} through \ref{sec:cosmo} constitute the essential core of the paper, also addressing the phenomenological consequences of our findings. These sections are largely self-sufficient and can be read independently.
Section~\ref{sec:SILSH} describes our hypotheses and outlines the structure of the low-energy theory. Section~\ref{sec:Minimal} discusses the parameters of the theory and its physical mass spectrum. The implications for flavor and CP-violating processes are described in \sec{sec:flavor}. Section~\ref{sec:extrasym} explores some modifications of the minimal setup. Section~\ref{sec:cosmo} focuses on signatures that are relevant for experiments in observational cosmology and high-energy colliders. Finally, \sec{sec:conc} provides an executive summary.

\section{Why combining supersymmetry and compositeness?}
\label{sec:why}

Despite its striking successes, the SM suffers from several structural limitations that indicate the need for an extension of the theory. As discussed in \sec{sec:intro}, naturalness and flavor are the two key unresolved issues, which constitute the main motivations for further exploring nature at short distances using high-energy colliders. We can clarify these issues further by emphasizing three specific aspects.

\medskip
\textbf{\textit{(i)} The EW Naturalness Problem}

This is the well-known problem that quantum corrections destabilize the Higgs mass. The SM does not provide a solution to this issue, as EW breaking is merely parametrized by the Higgs potential and no dynamical explanation is offered for its short-distance origin. Addressing the naturalness problem requires physics beyond the SM. LHC results have added pressure to this issue by revealing that all known solutions involving new weak-scale particles suffer from tunings that are generally worse than one percent. This raises concerns about a `little hierarchy' between the weak scale and the masses of new particles. We will not suggest here any remedy to this little hierarchy problem, which may result from an unfortunate accident, anthropic reasoning, or poorly understood features of current models. For now, we simply acknowledge the existence of a little Higgs hierarchy problem. Nevertheless, it is remarkable that supersymmetry or compositeness can explain the big Higgs hierarchy---spanning 15 orders of magnitude, from the Planck mass to the multi-TeV region---even though they fail to address the issue at the last decade of the energy scale.

\medskip
\textbf{\textit{(ii)} The Flavor Puzzle}

We refer to the {\it flavor puzzle} as the question of why the flavor parameters---specifically, the masses and mixings of quarks and leptons---have the values they do. This is a good scientific question, not just because every physicist has the ambition to calculate all the free parameters of a theory, but especially because the flavor structure displays a peculiar hierarchical pattern that hints at hidden features of a fundamental underlying theory. While the SM can parametrize the observed flavor structure, it does not provide any dynamical explanation for its peculiarity. The flavor puzzle can only be addressed by physics beyond the SM.

\medskip
\textbf{\textit{(iii)} The Flavor Problem}

The SM flavor structure, which originates from Yukawa couplings, results in a GIM-suppression of flavor-changing neutral currents. This property is crucial for an agreement with experimental data. However, it is also quite fragile, and its occurrence in the SM makes it quite unique. Indeed,
in scenarios addressing EW naturalness this property is generally lost. This issue, which will refer to  as the {\it flavor problem}, indicates that Higgs naturalness is basically a paradox. On the one hand, the SM is automatically adequate to describe flavor, but violently unnatural. On the other hand, BSM scenarios that cure naturalness invariably come at the price of renouncing this automatic adequacy.
\medskip

Low-energy supersymmetry and Higgs compositeness are likely the two most successful dynamical approaches to addressing the naturalness problem. However, neither theory is entirely satisfactory, even when we set aside the little hierarchy problem.

Regarding supersymmetry, several key issues arise in relation to the three points mentioned above.

\begin{itemize}

\item[\textbf{(I)}] Supersymmetry does not address the flavor puzzle because the flavor pattern is simply parametrized by Yukawa couplings, just as in the SM. 

\item[\textbf{(II)}] The generic structure of soft terms results in a severe flavor problem.

\item[\textbf{(III)}] Models that seek to resolve the flavor problem by using non-generic and calculable structures of soft terms (such as gauge~\cite{Dine:1993yw,Dine:1994vc,Dine:1995ag,Giudice:1998bp} or gaugino~\cite{Kaplan:1999ac,Chacko:1999mi} mediation, or variants of anomaly~\cite{Randall:1998uk,Giudice:1998xp} mediation) inevitably lead to a $\mu$--$B_\mu$ problem~\cite{Dvali:1996cu} unless specific artificial adjustments are incorporated into the theory. 
\end{itemize}

Issues \textbf{(II)} and \textbf{(III)} are two faces of a single {\it enten-eller} problem: {\it either} soft terms are generic, which results in $\mu$ and $B_\mu$ being correctly predicted by the mechanism of ref.~\cite{Giudice:1988yz}, but also in unacceptable levels of flavor violations; {\it or} soft terms are computable and meet the criteria for minimal flavor violation~\cite{DAmbrosio:2002vsn}, but the ratio $B_\mu/\mu$ ends up to be equal to the supersymmetry-breaking scale $\mS$, which is much larger than the desired soft mass.

\medskip

Let us now examine Higgs compositeness, which not only provides a solution to naturalness but also, unlike supersymmetry, addresses the flavor puzzle through the mechanism of partial compositeness~\cite{Kaplan:1991dc}. However, the theory is not without its own shortcomings.

\begin{itemize}

\item[\textbf{(IV)}] An appealing aspect of compositeness is that addressing the flavor puzzle is mandatory, and not merely an option. Consequently, the flavor problem poses an inescapable challenge, and it pushes the scale of compositeness to such high values that it undermines the solution to naturalness unless additional hypotheses or specific patterns of flavor symmetries are incorporated into the theory.

\item[\textbf{(V)}] The hypothesis of Higgs compositeness requires a dynamical justification for the separation between its mass and the confinement scale. Usually this separation is achieved through an approximate non-linearly realized symmetry, in which the Higgs acts as a pseudo-Goldstone boson. This framework makes the Higgs potential calculable in terms of SM parameters. However, in realistic constructions, it is difficult to produce a tunable potential, leading to complications in the model-building process.

\end{itemize}

The aim of this paper is to demonstrate how the combination of supersymmetry and compositeness can mutually reinforce one another. While preserving the benefits of each individual theory, their merging has the potential to overcome the limitations associated with each. The resulting theory, here referred to as {\it Super-Composite Higgs}, possesses the remarkable ability to simultaneously address items \textit{(i--iii)} without encountering issues \textbf{(I--V)}. We use the term \textit{address} instead of \textit{solve} because, as we will discuss, the parameter freedom intrinsic in the modeling of the strong sector
impedes our ability to make precise predictions, allowing only for power-counting estimates. In spite of these limitations, it is noteworthy that the combined dynamics of supersymmetry and compositeness may offer a unified theory of naturalness and flavor. In this respect, the Super-Composite Higgs is an economical and realistic framework underlying particle physics hierarchies.

Our scenario can be described using the two different but related viewpoints of warped 5D constructions or quasi-CFT 4D theories. The most explicit realization of the Super-Composite Higgs is based on a variant of the Randall-Sundrum model and its holographic interpretation. However, in practice, for most of the relevant phenomenological applications, the explicitness of a 5D perspective does not provide significant quantitative control. Indeed, throughout most of our paper, we adopt a 4D viewpoint, using the 5D perspective primarily to clarify the physical interpretation of our results. Furthermore, our 4D description is quite general and also includes (hypothetical) theories that do not admit a 5D dual.

In the following, we present and compare the broad features of the 4D and 5D perspectives.

\boldmath
\section{Super-Composite Higgs: 4D viewpoint}
\unboldmath
\label{sec:frameCFT}

\begin{figure}[t]
\begin{center}
\includegraphics[width=0.65\columnwidth]{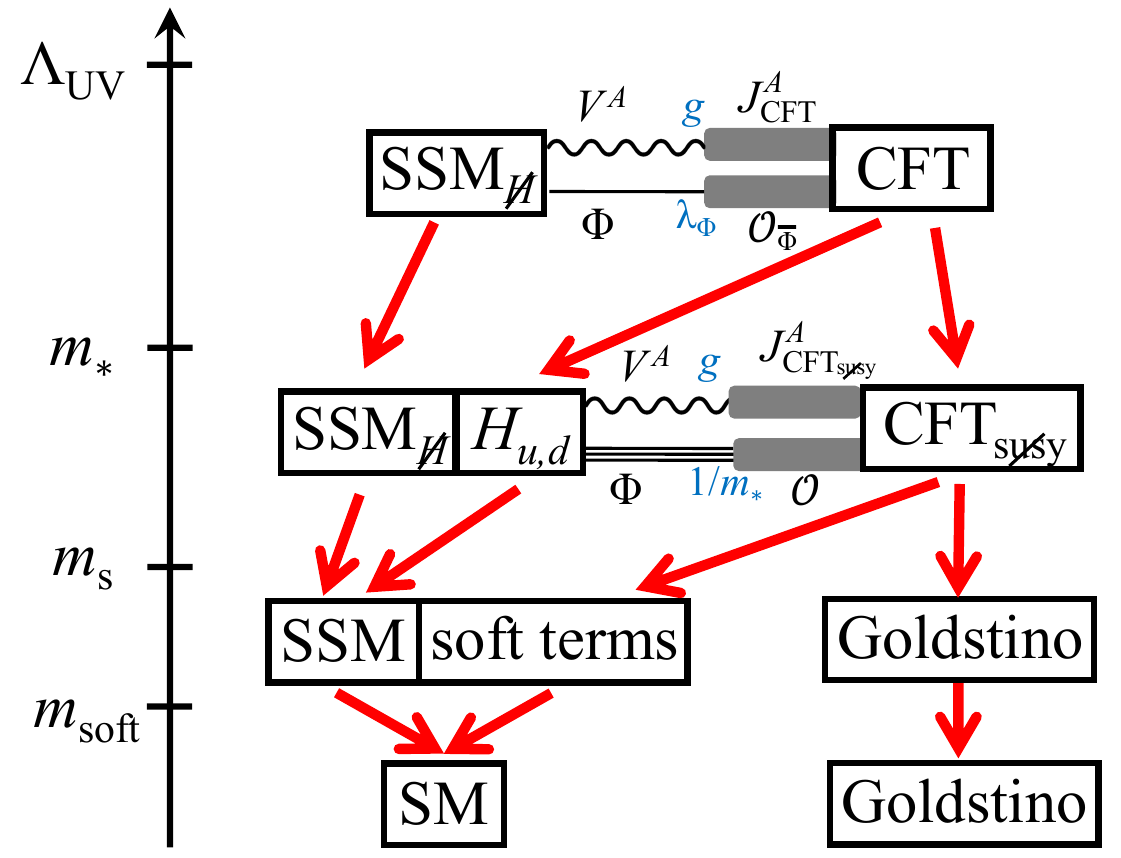}
\end{center}
\caption{A schematic representation of the Super-Composite Higgs framework, with different effective theories valid between successive energy thresholds. The links between sectors represent the way sectors interact with each other. The CFT sector communicates with $\SSMnH$ through SM gauge interactions, with coupling constant $g$, and a linear mixing between composite operators $\O_{\bar \Phi}$ and matter superfields $\Phi$, with coupling constant $\lambda_\Phi$. The $\CFTs$ sector communicates with SSM ($=\SSMnH+ H_{u,d}$) through SM gauge interactions and irrelevant interactions suppressed by powers of $1/\mst$.}
\label{fig:Overall}
\end{figure}

In this section we will provide a brief overview of the Super-Composite Higgs framework, focusing on the perspective of 4D strongly-coupled field theories. Our aim here is to summarize the main concepts, while a more detailed and systematic exploration will be given in \sec{sec:SILSH}.

The structure of the Super-Composite Higgs is illustrated schematically in \fig{fig:Overall}. The theory is defined at an energy cutoff $\LUV$, where it emerges from an unspecified fundamental dynamics. Below $\LUV$, the Super-Composite Higgs consists of two sectors. The first sector, $\SSMnH$, is a supersymmetric extension of the SM (SSM) that does not include any Higgs field. The second sector, referred to as ``CFT'', is a quasi-conformal supersymmetric theory that operates near a strongly interacting fixed point in its UV and is characterized by a set of low-lying (in scaling dimension) operators. This sector is weakly coupled to $\SSMnH$ through a weak gauging of a $SU(3)\times SU(2)\times U(1)$ (sub)-group  of its global symmetry and  through superpotential mixings between $\SSMnH$ chiral matter superfields and composite chiral primaries.

We further assume that relevant deformations, which we won't specify here, {\it naturally} lead to two IR thresholds in the renormalization group (RG) flow of the CFT sector. 
At the first threshold, which occurs at a scale $\mst$, some of the CFT degrees of freedom become gapped, and the CFT flows into two new sectors. The first sector contains weakly coupled massless Higgs superfields, which combine with $\SSMnH$ to yield the familiar supersymmetric SM. The second sector is again a quasi-conformal theory $\CFTs$, charged under the SM gauge group. This sector will eventually lead to spontaneous supersymmetry breaking at a further lower scale $\mS$, thereby providing soft terms to the supersymmetric SM, typically of size $\msoft$. The two sectors interact with each other through SM gauge forces and a general set of irrelevant interactions suppressed by the scale $\mst$. The basic features of the resulting dynamics will be largely controlled by hypotheses on  operator dimensions and possible symmetries. In addition we will assume for definiteness, and similarly to ref.~\cite{Giudice:2007fh}, an underlying large-$N$ structure, which  offers a well-defined power counting rule through an effective coupling $g_*\sim 4\pi/\sqrt N$. We believe this corresponds to the minimal structure one can assume for the strong sector.

The mixings between the  $\SSMnH$ matter  fields and the chiral composites provide the seeds for the Yukawa couplings, conforming to the mechanism of partial compositeness. This is how the Super-Composite Higgs addresses the flavor puzzle, resolving item \textbf{(I)} in the list described in \sec{sec:why}. Meanwhile, supersymmetry protects the Higgs mass down to the scale $\msoft$, addressing the naturalness problem without the need of a weak-scale compositeness scale, thus dealing with item \textbf{(IV)}. 

There are various sources for soft terms. Gauginos acquire masses through an induced mixing with the $\CFTs$ sector, in a manner fully analogous to partial compositeness. Moreover, SM gauge interactions provide masses for squarks and sleptons at one loop. Notably, this pattern of soft terms has the same form as conventional gauge mediation and retains minimal flavor violation.  On the other hand, partial matter compositeness introduces new sources of flavor and CP violation, partly associated, but not fully, with soft supersymmetry breaking.  Interestingly,  the dynamical alignment ensured by partial compositeness and the  suppression by inverse powers of $\mst$ guarantee that
 the resulting effects remain within current experimental bounds. This addresses item \textbf{(II)}. 

The $\mu$ and $B_\mu$ parameters are generated after supersymmetry breaking from effective operators that are suppressed by inverse powers of the compositeness scale $\mst$. The requirement that $\mu$ have typical size $\msoft$ necessitates a modest scale separation between $\mS$ and $\mst$. This separation is crucial for addressing the flavor problem that plagues ordinary compositeness. Moreover, the dynamical characteristics of the $\CFTs$ sector automatically predict that the ratio $B_\mu/\mu^2$ is order unity, thus providing a novel solution to item \textbf{(III)}.

Supersymmetry introduces an interesting twist to the issue of how to separate the Higgs mass scale from that of the other composite states, {\it i.e.}~item \textbf{(V)}.  In the absence of supersymmetry,  this separation is achieved by assuming that the Higgs is a pseudo-Goldstone boson from some approximate non-linearly realized symmetry, the simplest case being the breaking pattern $SO(5)\to SO(4)$.
Supersymmetry offers alternative solutions to the issue of scale separation.
As long as supersymmetry is preserved, the Higgs mass can arise only from the $\mu$-term, which is holomorphic and protected by the non-renormalization theorem. In a perhaps  simpler fashion than in the non-supersymmetric case, the challenges associated with a light Higgs are then tackled by only requiring that $\mu=0$ in the supersymmetric limit. This could be either the result of an approximate PQ-like symmetry or just an accidental coincidence allowed by the non-genericity of the superpotential. Once $\mu$ is set to zero, it remains zero as long as supersymmetry is exact, even in the absence of other symmetries to protect it. The origin of $\mu$ can then be tied to that of the soft terms once supersymmetry is spontaneously broken.

Below the scale $\mst$, the only remnant of the $\CFTs$ sector is a massless Goldstino. The Goldstino acquires mass by merging with the gravitino but remains part of the physical spectrum at low energies. The presence of the Goldstino leads to distinctive signatures at high-energy colliders and has cosmological implications, which we will discuss in \sec{sec:cosmo}.

The structure of the Super-Composite Higgs sketched in \fig{fig:Overall} may seem quite complex at first glance. However, the key point is that the entire framework is governed by the same dynamics. Dividing the theory into individual building blocks is intended to bring some clarity and enable a certain degree of calculability. In reality, these different building blocks are part of a unified tumbling structure from which naturalness and flavor hierarchies emerge.

Before delving into a more detailed study of the theory, in the next section we will present the AdS/CFT dual description of our framework.

\section{\boldmath Super-Composite Higgs: AdS$_5$ dual viewpoint\unboldmath}
\label{sec:frame5D}

\subsection{Non-supersymmetric holographic composite Higgs}

The paradigm of the Super-Composite Higgs sketched out in the previous section 
admits a holographic realization, which we schematically recall here.
We first review the non-supersymmetric 
structure and then augment it with supersymmetry.

\begin{figure}[t]
\begin{center}
\includegraphics[width=0.65\columnwidth]{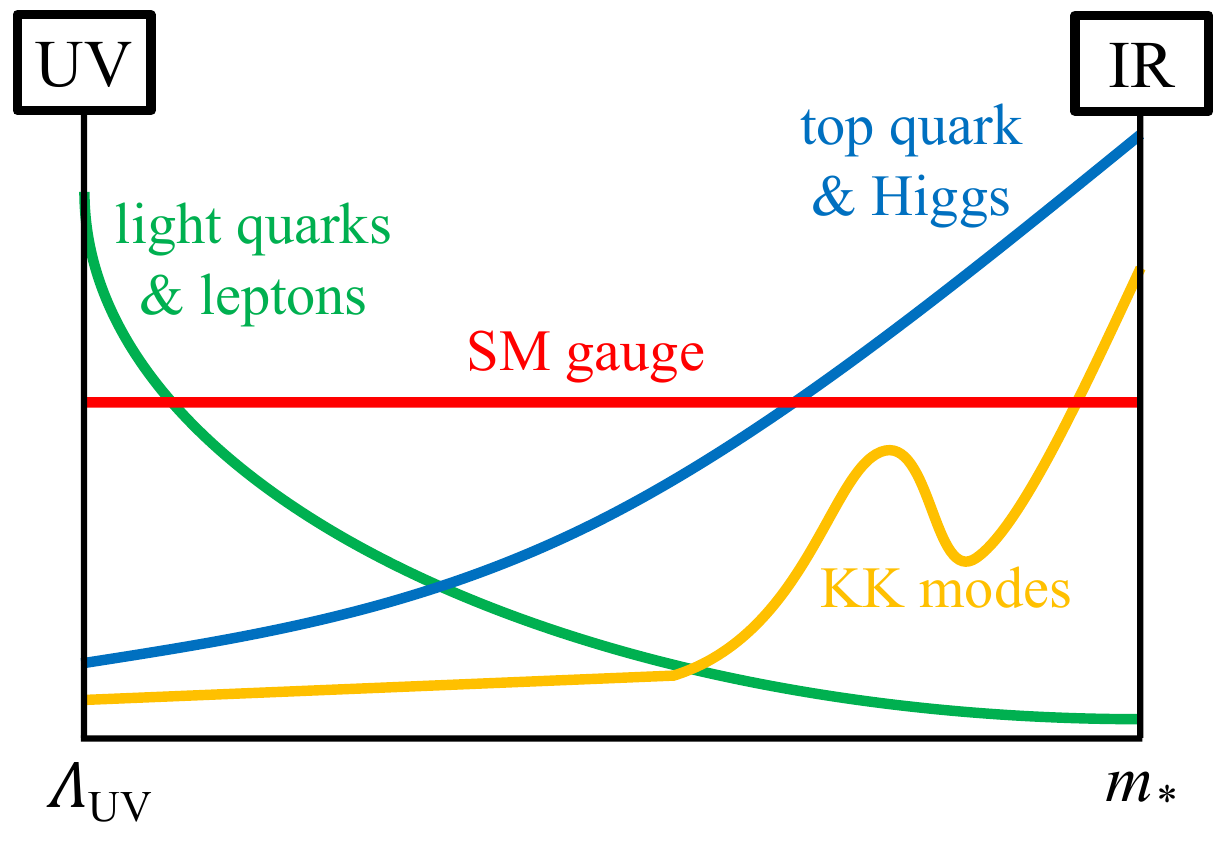}
\end{center}
\caption{Sketch of the non-supersymmetric warped 5D model showing the field profiles along the fifth dimension, which stretches between the IR brane, located at $1/\mst$, and the UV brane, located at $1/\LUV$.}
\label{fig:nonsusy}
\end{figure}

The model is framed in RS1 spacetime, where the geometry is approximately a slice of AdS$_5$ (see \fig{fig:nonsusy}) \cite{Randall:1999ee} (for reviews, see, for example \cite{Sundrum:2005jf,Gherghetta:2010cj}).
With suitable boundary conditions, the 5D graviton field, upon a Kaluza-Klein (KK) decomposition, results in a 
4D graviton zero-mode, localized near one end of the extra dimension, thereby called the UV brane \cite{Randall:1999vf}. We then choose the 5D Planck mass $M_5$ and AdS curvature scales $k$ to be both smaller than $M_P$ so that to reproduce the right scale of 4D gravity,
$M_P^2 = M_5^3 / k$. 
The SM fermions arise as chiral zero-modes of 5D fermions, with profiles in the extra dimension which are exponentially sensitive to their 5D mass terms \cite{Grossman:1999ra,Gherghetta:2000qt}. 
Similarly, the SM gauge fields are zero-modes of 5D gauge fields, but their profiles are flat in the extra dimension.

The SM Higgs field can be naturally light 
if it is localized near the other end of the extra dimension (IR brane) due to the gravitational red-shift factor
$e^{ - k L }$, where
$L$ is the proper length of the extra dimension \cite{Randall:1999ee}.
The Goldberger-Wise mechanism \cite{Goldberger:1999uk} can be used to naturally stabilize the proper size of the extra dimension at $k L \sim 30$, thereby yielding the Planck--Fermi hierarchy.  

Overall, the working hypothesis is that the 5D couplings and mass parameters are roughly comparable 
to the AdS$_5$ curvature scale. Any hierarchies in couplings or masses of the 4D particles (either zero or excited KK modes of SM fields)
arises only from the extra-dimensional profiles. In particular, the hierarchical structure of Yukawa couplings
must arise from the profile overlap of SM left and right-handed zero-modes with the IR brane-localized Higgs field. 
 Thus, the 
 SM fermions peaked near the IR (UV) brane have larger (smaller) Yukawa couplings to the SM Higgs.
 Consequently, the third-generation quarks are primarily localized near the IR brane, while lighter quarks and charged leptons live near the UV brane, as schematically illustrated in \fig{fig:nonsusy} \cite{Gherghetta:2000qt}. 
In this manner, the hierarchy of SM fermion masses and mixing angles can be easily accommodated without any large hierarchies in the fundamental 5D theory, thus addressing the SM flavor puzzle.

Massive KK modes of the graviton, gauge and fermion 5D fields represent the BSM physics of the higher-dimensional mechanism for generating the hierarchical structure of the SM.
The large red-shift favors the lightest of these KK modes to be localized near the IR brane such that in principle their masses (denoted by $M_{ \rm KK }$) could be within collider reach.
However, there are strong (indirect) bounds on $M_{ \rm KK }$ stemming from contributions of virtual exchanges of these KK particles to various precision tests of the SM: EW data require $M_{ \rm KK } \gsim 5$--10 TeV (see, for example, \cite{Davoudiasl:2009cd}), hadronic flavor/CP violation $M_{ \rm KK }\gsim O(10)$ TeV and leptonic flavor/CP violation $M_{ \rm KK } \gsim O(100$--$1000)$ TeV (see, for example, \cite{Glioti:2024hye}, for updated
bounds).

In the most predictive realization of such models, where the SM Higgs field is identified with the extra-dimensional component of a 5D gauge field ($A_5$) \cite{Contino:2003ve,Agashe:2004rs}, the full radiative stability of the weak scale would require $M_{ \rm KK } \sim O( \hbox{TeV} )$.
Hence, satisfying the constraints from precision tests requires significant fine-tuning, which 
reflects the little hierarchy problem in this framework.

\subsubsection*{4D dual of non-supersymmetric 5D model}

We can translate the above sketch of the warped extra-dimensional model into the language of 4D strong dynamics and compositeness 
using the AdS/CFT duality \cite{Sundrum:2011ic} and its deformations \cite{Arkani-Hamed:2000ijo, Rattazzi:2000hs}. 
The approximately AdS$_5$ bulk is 
dual to a 4D sector (here referred to as CFT), which is approximately conformal over a large energy range. The IR brane corresponds, in the 4D picture, to the IR gapping and confinement at 
$\mst$.
The gauge fields in the bulk are dual to global/flavor symmetries of the CFT, weakly-coupled to external elementary SM gauge bosons. 
The Higgs field localized near the IR brane is dual to a light composite; in particular, in its incarnation as $A_5$, it is dual to a pseudo-Goldstone boson (see, {\it e.g.}~ref.~\cite{Contino:2010rs,Panico:2015jxa} for reviews).
The 
KK modes are the massive composites of the (gapped) CFT 
sector, and thus we identify $M_{ \rm KK }$ of the 5D model with the compositeness scale 
$\mst$. The coupling among KK modes is furthermore identified with the effective coupling among reasonances of the purely 4D description : $g_{KK}\sim g_*=4\pi/\sqrt N$. This establishes the relation ``large $N$'' $\leftrightarrow$ ``weakly-coupled 5D EFT''. 
The 
profiles of SM zero-mode fermions in the extra dimension translate into their varying degrees of partial compositeness \cite{Kaplan:1991dc}. Therefore, the SM fermions are quantum superpositions of elementary and composite fermions, with the top/bottom quark (localized near the IR brane) being mostly composite, while light fermions (residing near the UV brane) are mostly elementary. 
The degree of compositeness of the SM fermions is determined by the scaling dimensions of the fermionic CFT operators which couple to the different elementary fermions.
These operators are dual to 5D fermionic fields, with their scaling dimensions being dual to the 5D fermion masses \cite{Contino:2004vy}.

\subsection{Supersymmetric warped 5D model}

\begin{figure}[t]
\begin{center}
\includegraphics[width=0.7\columnwidth]{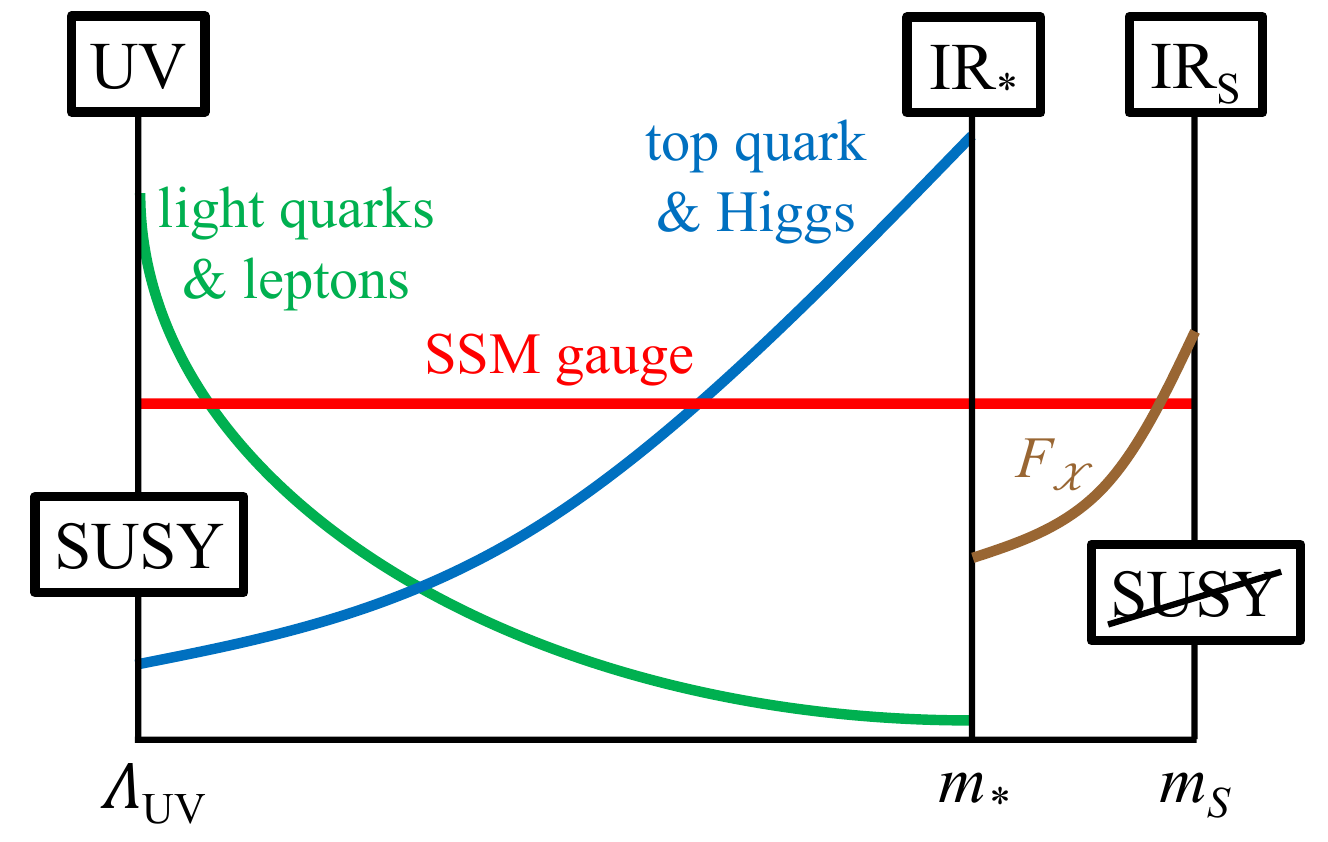}
\end{center}
\caption{Sketch of the supersymmetric warped 5D model showing the field profiles along the fifth dimension, which stretches between the IR$_S$ brane, located at $1/\mS$, and the UV brane, located at $1/\LUV$, with an intermediate IR$_*$ brane at $1/\mst$.}
\label{fig:susy}
\end{figure}

Note that, in the non-supersymmetric warped composite Higgs model discussed above, $M_{ \rm KK }$ (or equivalently $\mst$) plays simultaneously two roles: it is both 
the scale for the origin of flavor hierarchies and the scale controlling the radiative stability of the SM Higgs, realized as $A_5$.
However, as mentioned above, strong flavor/CP bounds arising from the former role of $\mst$ result in a severe little hierarchy problem for the latter purpose.
Incorporating supersymmetry into the model allows us to disentangle
these two roles. While we retain $\mst$ as the flavor scale, a significantly lower scale for supersymmetry breaking can resolve most of the little hierarchy problem of the non-supersymmetric warped model.
Then,
the Higgs realization as $A_5$ and the complication associated to its model-building are no longer required. Instead, the Higgs could be just the zero-mode of a bulk scalar with a profile chosen to peak near the IR brane and with its mass protected by supersymmetry.
A caveat is that we need the supersymmetric $\mu$ term to be justified by the non-genericity of the superpotential or forbidden by PQ symmetry. As discussed later on, the $\mu$ term will be generated from supersymmetry breaking at the appropriate scale.

Let us consider the supersymmetric version of the model sketched in \fig{fig:susy}, with supersymmetry broken at a scale $\mS < \mst$ on the IR brane. 
An attractive feature of this scenario is that, since the Higgs fields are localized on the same brane as the supersymmetry-breaking, the Higgs 
$\mu$-term and various soft mass terms ($B \mu$, $m^2_{ H_{ u, d } }$) are all comparable to each other (with stops also acquiring a similar mass).
In contrast, gauge fields are not localized near the supersymmetry-breaking brane, hence they only feel supersymmetry-breaking diluted 
by the volume (length) of the extra dimension in units of the curvature scale, \textit{i.e.}~$k L \approx 
\ln \left( \LUV / \hbox{TeV} \right) \sim 30$, see
\fig{fig:susy}.
Having gaugino masses suppressed with respect to Higgs mass parameters is unfavorable for
EW symmetry breaking.
The authors of ref.~\cite{Okada:2011ed} proposed a similar 
model, although with a much higher IR brane scale. However, they did not address the issue of a hierarchy between the Higgs and gaugino mass terms.
Rectifying this problem is a central concern of this paper.
Note that most of the other squarks and sleptons masses arise from (warped) gaugino mediation \cite{Gherghetta:2000kr}. This ensures that these soft masses are predominantly flavor-universal, thus addressing the supersymmetric flavor problem.

Here, we reduce the Higgs-related mass terms so as to make them comparable to gaugino masses by extending the warped throat and introduce two separate IR branes, referred to as IR$_*$ and IR$_S$, see \fig{fig:susy}. SM gauge fields propagate across the entire extra-dimensional space, while Higgs and matter fields are restricted to the region between the UV and IR$_*$, whose location corresponds to
$\mst$.\footnote{See ref.~\cite{Agashe:2016rle} for a non-supersymmetric version of such a model.}
The IR$_S$ brane corresponds to $\mS \sim {\cal O}(100)$ TeV , with supersymmetry broken around this scale by a ``hidden-sector'' localized there. This allows for a gaugino mediation mechanism, proceeding as before.
The 
key new idea is that the order parameter for supersymmetry breaking is taken to be the effective auxiliary component $F_{\cal X}$  of a bulk hypermultiplet superfield ${\cal X}$ (see {\it e.g.} ref.~\cite{Arkani-Hamed:2001vvu}) whose  profile 
 is localized towards the IR$_S$ brane, 
but which extends to the IR$_*$ brane. This results in a small overlap of $F_{\cal X}$ 
with the Higgs and top quark fields.
We will show how this setup allows all Higgs mass terms to be uniformly reduced by the tail of the $F_{\cal X}$ profile, 
making them 
comparable to the soft mass terms from gaugino mediation. 
This is a simple extension of the general principle of using extra-dimensional overlaps in order to 
generate hierarchies.
This mechanism only requires a mild single coincidence for this suppression factor to be 
$\sim 1 / \ln \left( \LUV / \hbox{TeV} \right)$ so as to be comparable to the volume dilution for 
gauge fields discussed above.
Of course, 
gaugino mediation is not suppressed by the $F_{\cal X}$ profile tail since gauge fields extend all
 the way to the IR$_S$ brane where the $F_{\cal X}$ profile is maximized.

It turns out that the $F_{\cal X}$ direct coupling to the Higgs sector on the IR$_*$ brane also results in significant flavor-aligned $A$-terms of $O( \hbox{TeV} )$ (after factoring out the SM Yukawa couplings), which are attractive since they can be used to enhance the stop contribution to the physical Higgs boson through stop mixing. 
Moreover, the tail of the $F_{\cal X}$ profile coupling
to matter fields on the IR$_*$ brane results in flavor non-universal 
soft masses proportional
to the size of the corresponding Yukawa couplings, but not aligned with them. These
flavor-violating effects act as perturbations to the flavor-blind soft mass terms generated
by gaugino mediation. 
A similar framework in the context of a {\em flat} extra dimension (with a GUT-scale 
compactification/KK scale instead) was dubbed ``flavorful supersymmetry'' 
in ref.~\cite{Nomura:2008pt}.\footnote{The model in ref.~\cite{Nomura:2008pt} does not lead to 
large $A$-terms, unlike our setup,   
and employs a different mechanism from ours for obtaining Higgs mass terms comparable 
to the other soft terms.}

There are however important differences with earlier work. In standard gaugino-mediation, there is a hierarchy between the KK scale and the lower fundamental scale of supersymmetry breaking, whereas here we are identifying these two scales as $m_S$.  Furthermore, $m_S$  is not far above the scale of superpartner masses or far below $m_*$. As we shall see, this relatively compact set of scales implies that the phenomenology is not dominated by just soft-breaking terms, as is usually the case, but includes significant {\it hard} supersymmetry-breaking effects which determine some of the most stringent constraints and interesting predictions of our scenario.

The 
supersymmetric-CP problem is simply solved by assuming that CP is preserved in the entire bulk (including the IR$_*$ and IR$_S$ branes), but broken only on the UV brane.
Thus, a sufficient 
CKM phase can be generated via UV brane-localized kinetic terms for SM fermions (see ref.~\cite{Frigerio:2018uwx} for the corresponding CFT dual picture), while keeping the soft mass terms generated in the IR predominantly real/CP-preserving.
Finally, each interval of this extra dimension can be stablized by a supersymmetric version of the Goldberger-Wise mechanism \cite{Son:2008mk}\footnote{For stabilization of such a non-supersymmetric setup, see ref.~\cite{Lee:2021wau}.}, without the need for any additional tuning, given the mild hierarchy between the IR$_*$ and IR$_S$ branes.
Thus, such a mild extension of the warped throat combines solutions to the supersymmetric flavor/CP problems, the origin of flavor hierarchies and the Planck--Fermi hierarchy problem with superpartner masses only constrained by current LHC searches.

\subsubsection*{Dual of supersymmetric warped model}
The extension of the warped throat outlined above is, from a 4D perspective, dual to the tumbling scenario 
between the scales $\mst$ and $\mS$, which will be discussed in \sec{sec:tumbling}. In the 5D picture, $\mS$ is identified with $M^{ \rm gauge }_{ \rm KK }$  \cite{Agashe:2016rle}.
While the Higgs and top quark are composites of the CFT strong dynamics, 
all SM matter, including Higgs and top, is weakly-coupled with respect to the $\CFTs$ sector below $\mst$. This situation is dual to having SM matter and Higgs residing only in the region between the UV and IR$_*$ branes. 
 The $\CFTs$ sector below $\mst$
is still charged under the SM gauge group, which is dual to the SM gauge fields living in the extended throat.
The 
supersymmetry-breaking on the IR$_S$ brane, located at $1/ M^{ \rm gauge }_{ \rm KK }$,
in the 5D model is dual to $\CFTs$, which spontaneously breaks supersymmetry at the scale $\mS$ \cite{Gherghetta:2000kr},
resulting in gauge-mediated supersymmetry breaking from the 4D perspective.
Indeed, we will argue that the 5D warped gaugino mediation discussed earlier is dual 
to 4D direct gauge mediation, in which messengers are strongly-coupled and are an intrinsic part of the dynamical supersymmetry-breaking sector \cite{Nomura:2004zs,Abel:2010vb,Benini:2009ff}.
This joint sector represents the dual to the warped throat bounded between the IR$_*$ and IR$_S$ branes.

\renewcommand{\arraystretch}{1.2}
\begin{table}[t!]
\centering
\begin{tabular}{|c|}
\hline
{\bf Dual descriptions of the Super-Composite Higgs}\\
\hline
\end{tabular}
\begin{tabular}{|c|c|}
\hline
{\bf 4D strongly-coupled theory} & {\bf 5D warped geometry} \\
\hline
\hline
RG energy scale & Extra-D coordinate \\ \hline
UV cutoff scale $\LUV$ & UV brane \\ \hline
Compositeness scale $\mst$ & IR$_*$ brane \\ \hline
SUSY-breaking scale $\mS$ & IR$_S$ brane \\ \hline
CFT sector & Bulk space between UV and IR$_*$ branes \\ \hline
$\CFTs$ sector & Bulk space between IR$_*$ and IR$_S$ branes \\ \hline
4D coupling $g_*={4\pi}/{\sqrt N}$ & KK coupling $g_{KK}$ \\ \hline
Mostly elementary SM matter & 5D profile peaked at UV brane \\ \hline
Mostly composite Higgs \& top & 5D profile peaked at IR$_*$ brane \\ \hline
Yukawa hierarchical pattern generated by & Yukawa hierarchical pattern generated by \\[-1.5mm] 
SM matter mixing with CFT operators 
& overlap between SM matter  \\[-1.5mm] 
with large scaling dimensions & and Higgs 5D profiles \\ \hline
Only $\mst$-suppressed irrelevant & 
SM matter \& Higgs fields, living in\\[-1.5mm]  
interactions between $\CFTs$ and & 
the bulk between IR$_*$ and IR$_S$ branes \\[-1.5mm]
SM matter \& Higgs & 
have masses $\gsim 1/R_{\text AdS} \equiv k$
\\ \hline
$\CFTs$ charged under SM gauge & Gauge fields propagate everywhere in 5D \\ \hline
Composite SUSY-breaking messengers & Gauge/gaugino KK modes \\ \hline
$\CFTs$ confinement scale $\mS$ & Gauge/gaugino KK mode masses $M^{ \rm gauge }_{ \rm KK }$ \\ \hline
4D gauge mediation & Warped 5D gaugino mediation \\ \hline
$\CFTs$ SUSY-breaking operator ${\cal O}_{\cal X}$ & Bulk ${\cal X}$ between IR$_*$ and IR$_S$ branes \\ \hline
Suppression of Higgs soft masses due to & Suppression of Higgs soft masses due to \\[-1.5mm]
${\cal O}_{\cal X}$ scaling dimensions & overlap between Higgs and $F_{\cal X}$ 5D profiles \\ \hline
\end{tabular}
\renewcommand{\arraystretch}{1}
\caption{A dictionary to translate the Super-Composite Higgs between the languages of 4D strongly-coupled theories and 5D warped geometries.}
\label{tab:dual}
\end{table}

Finally, we come to the dual of the $F_{\cal X}$ profile mechanism, which is responsible for the suppression of Higgs mass terms.
The bulk field ${\cal X}$   is dual to a chiral operator ${\cal O}_{\cal X}$ of $\CFTs$, which acts as an order parameter for supersymmetry-breaking.
Since the Higgs field is not part of the $\CFTs$ and has only irrelevant couplings to 
${\cal O}_{\cal X}$, once supersymmetry is broken at $\mS$, the
Higgs-related soft mass terms are roughly equal to $ \mS 
\left( \mS / \mst \right)^{ \Delta_{ {\cal O}_{\cal X}} }$, where $\Delta_{ {\cal O}_{\cal X} }$ denotes the scaling dimension of ${\cal O}_{\cal X}$.
These Higgs mass terms can readily be of order of gaugino and sfermion masses for $\Delta_{ {\cal O}_{\cal X} } \gsim 1$, as we will detail later on.
Couplings to ${\cal O}_{\cal X}$ also generate flavor-violating contributions to 
supersymmetry breaking terms.
However, these are further suppressed by
partial compositeness 
and, therefore,
the soft masses for 1st/2nd generations
are dominated by flavor-universal gauge mediation.
This is dual to the suppression of matter soft terms 
due to the small overlap on the IR$_*$ brane between
1st/2nd generation SM profiles
and the $F_{\cal X}$ profile. 
Nevertheless, the residual flavor-violating supersymmetry breaking, both hard and soft, play important roles phenomenologically. 

In \tab{tab:dual}, we summarize the dictionary to translate between the dual 4D and 5D languages of the Super-Composite Higgs.

\section{Toy model of tumbling}
\label{sec:tumbling}

In this paper we are exploring whether a partnership between (partial) compositeness and supersymmetry can plausibly and realistically explain the striking hierarchical structure of particle physics, including a solution to the EW hierarchy problem. In particular, we identify the new strong dynamics sector responsible for (partial) compositeness of the supersymmetric SM (SSM) particles with the sector that dynamically breaks supersymmetry. When we integrate out the massive states of the strong sector we expect to be left with just the SSM, with hierarchical Yukawa matrices and with some pattern of soft supersymmetry-breaking mass terms. The question is, what is this pattern?

The couplings between composite operators of the strong sector and SSM particles, required for partial compositeness, transmit supersymmetry-breaking to the SSM. We can therefore think of the composite states interpolated by the composite operators, or equivalently their strongly-coupled constituent ``preons'', as the ``messengers'' of supersymmetry-breaking. In particular, the SSM gauge superfields couple to conserved currents of the strong sector, leading to gauge mediation as an important contribution to the transmitted pattern of supersymmetry-breaking. Because the messengers are an integral part of the supersymmetry-breaking sector, this is a form of {\it direct} gauge mediation. Beyond gauge interactions, the SSM fermions and sfermions mostly have irrelevant couplings to the strong sector, which are quite  weak at the scale of dynamical supersymmetry-breaking. Therefore, the resulting sfermion  soft terms will be dominated by gauge mediation, which then mitigates the supersymmetric flavor problem. But the Higgs and (EW singlet) top superfields are entirely (light) composites of the strong sector and therefore one might expect them to feel supersymmetry-breaking even more strongly than gauge-mediation effects. However, EW naturalness, consistency with sparticle search limits, and the degree to which the supersymmetric flavor and CP problems are solved all depend on a balance between the various supersymmetry-breaking effects.

The size of the different supersymmetry-breaking effects is sensitive to the mass scales characterizing SM  partial compositeness, dynamical supersymmetry-breaking, and its transmission via the messengers.
Since the same strong sector is responsible for all these functions, we might expect all these scales to be roughly the same,
\begin{equation}
   \mst \sim m_M \sim \mS\, ,
\end{equation}
where $\mst$ denotes the scale at which the composite Higgs and the composites that mix with SM fermions are formed,
$m_M$ represents the rough mass of messenger composites (made of messenger preons) responsible for gauge mediation, and $\mS$ represents the rough scale of (spontaneous) supersymmetry-breaking splittings in the strong sector. This is somewhat analogous to the rough coincidence of scales we see in real-world QCD, $\Lambda_{c} \sim \Lambda_{\chi} \sim \Lambda_{\rm QCD}$, where $\Lambda_{c}$ is the scale characterizing quark confinement, $\Lambda_{\chi}$ the scale characterizing spontaneous chiral symmetry breaking, and $\Lambda_{\rm QCD}$ the rough scale at which quarks and gluons become strongly coupled.

With this coincidence of scales, gauge mediation is typified by gaugino masses of size
\begin{equation}
M_{\lambda} \sim \frac{N_M\, \alpha_{\textrm{SM}}\, \mS^2}{4 \pi\, m_M} \sim \frac{N\, \alpha_{\textrm{SM}}\, \mst}{4 \pi} \, .
\end{equation}
Here, the middle expression is the usual parametric form for gauge mediation, where $N_M$ represents the number of messengers,  with an $O(1)$ factor dressing the standard perturbative expression which is determined by the details of the strong dynamics. We will consider strong sectors with a large-$N$ expansion, where $N$ represents the number of preon ``colors''. Preon messengers therefore have this minimal multiplicity from the SM gauge viewpoint:
\begin{equation}
   N_M \sim N\, .
\end{equation}
On the other hand, the Higgs superfields, as composites of the strong sector, inherit its going rate of supersymmetric particle splittings. Assuming that Peccei-Quinn symmetry is also broken by the strong dynamics, we then estimate Higgs-related soft terms in the IR effective SSM as
\begin{equation}
\label{higgssoft}
   \mu \sim \mst, ~~~~ A_{ij} \sim y_{ij} \mst, ~~~~B_\mu \sim m_{H_u}^2 \sim m_{H_d}^2 \sim \mst^2\, ,
\end{equation}
where the $A_{ij}$ are flavor-aligned $A$-terms proportional to the SSM Yukawa matrices.

Clearly, a typical choice of such soft terms leading to EW symmetry breaking would then predict $v_{\rm EW} \sim \mst$, in sharp contrast to the need to significantly separate the origin of flavor structure from EW symmetry breaking, if we are to be safe from flavor/CP-violation bounds. Furthermore, the light fermions and gauge fields would have superpartners from gauge mediation which are significantly lighter by a factor $\sim N \alpha_{\textrm{SM}}/(4 \pi)$. The largest this factor can be is determined by the need to avoid Landau poles in SM gauge couplings up to the far-UV cutoff $\LUV$, due to messenger contributions to their running,
\begin{equation}
\frac{N\, \alpha_{\textrm{SM}}}{4 \pi} \sim \frac{N_M\, \alpha_{\textrm{SM}}}{4 \pi}
\lsim \frac{1}{ \ln(\LUV^2/m_M^2)} \sim \frac{1}{ \ln(\LUV^2/\mst^2)} \sim \frac{1}{40{\textrm {--}}60}\, .
\end{equation}
It is true that fine cancellations can raise the new physics scales linked to $\mst$ beyond current reach for direct and precision searches, but then supersymmetry is in no way an improvement over the non-supersymmetric partial compositeness scenario from the viewpoint of the EW hierarchy problem.

These dismal conclusions are evaded if we allow for the possibility of ``tumbling'' within the strong dynamics sector, whereby there are modest hierarchies between the scales characterizing Higgs and SM fermion partial compositeness, gauge messengers, and supersymmetry-breaking,
\begin{equation}
   \mst \gsim m_M \gsim \mS\, .
\end{equation}
Now, at the scale $\mst$, the Higgs superfields are formed  by the strong dynamics, but a residual IR strong dynamics continues below this scale, remains coupled to SM gauge fields and goes on to break supersymmetry dynamically.
The composites of the IR strong dynamics therefore act as messengers of gauge mediation, with mass scale $\sim m_M$. The supersymmetry-breaking splittings within the IR strong sector are of order $\mS$. That is, below $\mst$ the theory reduces to a standard direct gauge mediation scenario consisting of the SSM and a separate supersymmetry-breaking sector.
The crucial point is that the Higgs superfields have broken away from the IR strong sector below $\mst$, with only irrelevant couplings to it. These mediate Higgs soft terms
suppressed by these irrelevantly weak couplings, allowing them to be comparable in size to gauge mediation effects, rather than of order $\mst$ as they were without tumbling.

In the rest of the paper we will invoke such a possibility and show how the suppressed Higgs soft terms (now comparable to gauge mediation) can robustly retain their {\it relative} similarity as in \eq{higgssoft}, which is attractive  in reconciling realism with improved EW naturalness.
 Strong dynamics tumbling between multiple phases at different scales was first posited within Technicolor theories of dynamical EW symmetry breaking~\cite{Raby:1979my}. Later, a form of tumbling was exhibited under greater theoretical control in the Klebanov-Strassler duality cascade~\cite{Klebanov:2000hb}. In this section, we provide an explicit theoretically-controlled toy model of tumbling strong dynamics closer in spirit to the form we will need to invoke for our realistic construction, so as to demonstrate its broad plausibility.

 Our central theoretical module will be supersymmetric QCD with a large number of colors $N$ and massive flavors $F$. Above the scale of these masses, the running gauge coupling is governed by the NSVZ $\beta$-function~\cite{Novikov:1983uc},
\begin{equation}
\frac{d \alpha^{-1}}{d \ln \muRG} = \frac{3N - \left[1 +  \gamma(\alpha)\right] F}{2 \pi - N \alpha}\, .
\end{equation}  
We choose  $F$ to admit a Banks-Zaks \cite{Belavin:1974gu,Banks:1981nn} 
RG fixed point, $F = (3 - \epsilon)N$, with $\epsilon \ll 1$.
Given the perturbative expansion of the anomalous dimension from quark wavefunction renormalization,
$\gamma = N \alpha/(2 \pi) + {\cal O}(N^2 \alpha^2)$, we see that there is a controlled fixed point at
\begin{equation}
   \alpha_* \approx \frac{2 \pi \epsilon}{3 N}, ~~~~ \gamma_* \approx \frac{\epsilon}{3}\, .
\end{equation}

We assume that the quark flavors are roughly degenerate, with mass $\approx m$. Below this threshold, they are integrated out, leaving pure supersymmetric Yang-Mills (SYM) as the IR EFT. Since this EFT is asymptotically free, it runs to a strong coupling regime at an RG-invariant scale approximated by
\begin{equation}
\Lambda_{\textrm{SYM}} \sim \muRG \, e^{- \frac{2 \pi}{N \alpha(\muRG)}} \approx m\, e^{- \frac{2 \pi}{N \alpha_*}} \approx m\, e^{- {3}/{ \epsilon}}\, ,
\end{equation}
where it confines and strongly breaks its (already anomalous) $U(1)_R$ symmetry via gaugino condensation,
\begin{equation}
   \langle {\cal W}_{\alpha}^2 \rangle = \frac{N}{16 \pi^2}\, \Lambda_{\textrm{SYM}}^3  = \frac{N}{16 \pi^2} \,m^3 \,e^{- 9/{\epsilon}}\, .
\end{equation}
We thereby arrive at a simple tumbling structure with two parametrically-separated characteristic scales distinguishing conformal fixed-point behavior from confinement,
\begin{equation}
   m \gg \Lambda_{\textrm{SYM}} \approx m \, e^{- 3/ \epsilon}\, .
\end{equation}

In assessing this first attempt at a model of tumbling and the plausibility of some version of tumbling playing a role in Nature, there are a few concerns. {\it (i)} The hierarchy $e^{- 3/ \epsilon}$ was only achieved by tuning $F$ close to $3N$, $\epsilon \ll 1$. {\it (ii)} The SQCD theory is not a {\it strongly}-coupled theory tumbling between two phases, because at $m$ the theory is still weakly coupled, $\alpha(m) \approx \alpha_* \approx 2 \pi \epsilon/(3N)$.
{\it (iii)} Even the ability to tune $\epsilon \ll 1$ requires large $N$, another special requirement.
However, {\it (i)} only implies that it is less plausible to expect a very large separation of scales, and that for a less-tuned and more plausible $\epsilon \lsim 1$ we would get a modest hierarchy. Indeed in this paper we will only need to invoke tumbling between scales separated by relatively small factors. In turn, the fixed-point and dynamics around $m$ is no longer very weakly coupled for $\epsilon \lsim 1$, as measured by the large-$N$ 't Hooft coupling,
$N \alpha_* \approx 2 \pi \epsilon/3$, thereby resolving {\it (ii)}.
And now that we do not need to tune $\epsilon$ to be very small, we also do not need very large $N$, mitigating {\it (iii)}. In summary, there is a robust and plausible range of parameters $F, N, m$  of SQCD in which it is strongly coupled, approximately conformally invariant in the UV and confining in the IR, with these two regimes separated by a modest hierarchy.

With this basic tumbling structure in place as a toy model of the strong BSM dynamics, we can incorporate ``composite'' Higgs superfields. We do this by introducing elementary Higgs superfields ${H}_{u,d}$ and gauge superfields with their usual SSM quantum numbers. We will refer to the SQCD sector above as the ``hidden'' sector so as not to confuse it with the SQCD within the SSM. We take some of the $F$ hidden-sector flavors to form vectorlike representations under the SM gauge symmetry, with the quantum numbers of EW doublet and up-type singlet quarks and their conjugates, ${\cal Q} + \bar{{\cal Q}}$, ${\cal U} + \bar{{\cal U}}$ (with no multiplicity beyond the $N$ hidden-colors). In this way, the hidden sector is now gauged under the SM gauge symmetry.

The Higgs superfields can now Yukawa-couple to these hidden-quarks via a superpotential,
\begin{equation}
    W = y_u \,{H}_u \,\bar{{\cal U}} \,{\cal Q} + y_d \,{ H}_d \,\bar{{\cal Q}}\, {\cal U}\, .
    \label{yukcc}
\end{equation}
In the $N \gg 1, \epsilon \ll 1$ approximations above, these couplings are governed in the UV by the RG equations,
\begin{equation}
\frac{dy}{d \ln \muRG} \approx y \left( \frac{N y^2}{16 \pi^2} - 2 \gamma_*\right)\, ,
\end{equation}
reflecting that, due to the anomalous dimension $\gamma_*$ of the hidden quarks, the Yukawa couplings are relevant in the UV of hidden SQCD. Thus, if $y_u$, $y_d$ start at weak values in the UV they will flow in the fixed-point regime of hidden SQCD to their own fixed-point values,
\begin{equation}
   \frac{y^2_{u,d}}{16 \pi^2} \rightarrow \frac{y^2_*}{16 \pi^2} \approx \frac{2 \epsilon}{3 N}\, .
\end{equation}

In this way, for moderate $\epsilon \lsim 1, N \gsim 1$, the Higgs fields strongly couple to the hidden SQCD and form a single strongly-coupled sector above the hidden quark mass scale $m$. The elementary Higgs thereby strongly mixes with the hidden quark bilinears due to the couplings in \eq{yukcc}, so that the Higgs is significantly {\it partially composite}. But below $m$, the hidden quark superfields are integrated out and the Higgs superfields decouple from the IR hidden gauge dynamics. We therefore identify $m$ as the Higgs compositeness scale $\mst$,
but the hidden sector continues below this scale to confine at $\Lambda_{\textrm{SYM}} \approx m e^{- 3/ \epsilon}$. The residual non-renormalizable coupling of the Higgs fields to the hidden SYM in the IR is given by one-loop matching at $m$,
\begin{equation}
\sim \frac{y_*^2 \alpha_*}{4 \pi m^2} \int d^2 \theta {H}_u { H}_d {\cal W}^2_{hid} \approx
\frac{16 \pi^2  \epsilon^2}{9 N^2 m^2} \int d^2 \theta { H}_u { H}_d {\cal W}^2_{hid}\, .
\end{equation}

At this stage, we have a toy model of a (partial) composite Higgs strong dynamics, which tumbles over a hierarchy $\sim e^{3/\epsilon}$ between a Higgs compositeness scale and the scale of confinement and gaugino condensation. But this strong dynamics does not break supersymmetry. Nevertheless, we see that the SYM $U(1)_R$ symmetry  protects against a supersymmetric $\mu \int d^2 \theta {H}_u {H}_d$ term for the Higgs fields, until this symmetry is strongly broken by gaugino condensation,
\begin{equation}
\mu \sim   \frac{ \epsilon^2}{9 N m^2} \, \Lambda_{\textrm{SYM}}^3 \sim
\frac{ \epsilon^2}{9 N} \, m\, e^{- 9/\epsilon}\, .
\end{equation}
This illustrates how the modest tumbling hierarchy gives a suppression by a power of that hierarchy in the generation of a $\mu$ term, relative to the no-tumbling expectation $\mu \sim \mst = m$. The
degree of suppression depends on the degree of non-renormalizability of the residual couplings of the Higgs fields to the IR strong sector, in this example given by dimension-$6$ couplings.

The final improvement of our toy model will be to introduce dynamical supersymmetry breaking and direct gauge mediation. We remain within SQCD by using the metastable ISS supersymmetry-breaking mechanism~\cite{Intriligator:2006dd}. This requires that the IR hidden dynamics consists of several light flavors $N < F_{IR} < \nicefrac{3}{2}\, N$. To do this, we assume that the original $F = (3 - \epsilon) N$ hidden quark flavors separate between
$F - F_{IR}$ heavy hidden quark flavors with mass $\sim m$ as before, including those that Yukawa-couple to the Higgs fields, and $F_{IR}$ light quarks with small masses $\sim m_{\rm IR} < \Lambda_{\rm IR}$, where $\Lambda_{\rm IR}$ is the strong interaction scale in the IR SQCD below $m$. In this way, the UV hidden sector starts with SQCD strongly coupled to the Higgs fields at a conformal fixed-point, the Higgs fields decoupling below $m$, and the surviving IR SQCD going on to dynamically break supersymmetry (with metastable vacuum) below $m_{\rm IR} < \Lambda_{\rm IR}$. We take some of the light hidden flavors to also carry SM gauge quantum numbers, so they are effectively messengers for gauge mediation to the SSM, $m_M = m_{\rm IR}$. It should be noted however that the resulting gauge mediation spectrum is not realistic in this toy model, in particular because the SM gaugino masses are suppressed due to a residual discrete $R$-symmetry in the ISS metastable vacuum.

We have thereby illustrated how a single strong sector can generate  composite Higgs fields at a compositeness scale $\mst = m$, below which the Higgs fields decouple the IR strong dynamics modulo non-renormalizable weak couplings suppressed by $\mst$, the IR strong dynamics continuing to lower scales and breaking supersymmetry. The IR dynamics gauge-mediation messengers and supersymmetry-breaking scales are smaller than the Higgs compositeness scale, with $\mS \lsim m_M < \mst$. The remainder of this paper explores this broad possibility of unifying Higgs compositeness and supersymmetry-breaking within a single strong dynamics sector, with a relatively modest tumbling between the associated scales.

\section{\boldmath $4\pi$'s, $g_*$'s, $N$'s and all that\unboldmath}
\boldmath
\label{sec:allthat}
\subsection{Field normalizations}
\unboldmath
Throughout the paper, we will make frequent use of a version of na\"ive dimensional analysis to estimate coefficients in effective actions. In order to avoid confusion and ambiguity, we must explain our normalization choice and relate it to other choices.
Such different choices are conveniently illustrated by considering a toy renormalizable scalar field theory described by an overall mass scale $\mst$ and an overall coupling strength $g_*$ (we will later indicate by $\mst$ and $g_*$ also the corresponding parameters of our effective theory), whose classical action has the form 
\beq
{\cal L} =\frac{1}{2} (\partial \phi)^2-\frac{\mst^2}{2}\phi^2-\frac{a \mst g_*}{3!}\phi^3+\frac{g_*^2}{4!}\phi^4\,,
\label{canonical}
\eeq
with $a=O(1)$. We can think of the above as the renormalized Lagrangian at $\mu=\mst$. The connected $n$-point correlators in momentum space have then the general form
\beq
\langle \hat\phi(p_1)\dots\hat \phi(p_n)\rangle_c= g_*^{n-2}\left [F_n^{(0)}(\{p\},\mst)+\frac{g_*^2}{16\pi^2}F_n^{(1)}(\{p\},\mst)+\dots\right ] \, ,
\eeq
where $F_n^{(0)}\,,F_n^{(1)}\,,\dots$ are $O(1)$ functions encapsulating the contributions at tree level, one-loop, etc.
The field in \eq{canonical} is normalized canonically, corresponding to the two-point function scaling like $g_*^0$ and the one point function (the vev)  scaling like $\mst/g_*\equiv v$. Depending on the goals, there are two other normalizations worth considering.

The first is the {\it{mass normalization}} $\bar \phi\equiv g_* \phi $ in which the Lagrangian can be written as
\beq
{\cal L}=\frac{\mst^4}{g_*^2} L(\bar \phi/\mst,\partial/\mst)\,.
\eeq
This normalization makes evident that $g_*^2$ is the loop counting  parameter. Indeed, by working with redundant units and assigning to $S=\int d^4x {\cal L}$ units of [action] and to $\mst$ units of 1/[length], $g_*^2$ has units of 1/[action], which is consistent with $ \hbar g_*^2$ being an adimensional parameter controlling the loop expansion. The advantage of the mass normalization is that the field $\bar \phi$ has unit of mass, so that $\langle \bar \phi\rangle$  scales purely like the mass parameter $\mst$ without powers of the coupling. This is to be contrasted with the canonical normalization, where $\langle \phi\rangle\propto \mst/g_*$.  The scaling  of field vevs (including for composite operators) purely as powers of the mass scale(s) renders the mass normalization best suited for power counting. Notice  more generally that connected $n$-point  correlators of mass normalized fields scale like $g_*^{2n-2}$.

The other option, related to the canonical and mass normalized fields by $\phi_g\equiv \phi/g_*=\bar \phi/g_*^2$, we shall call the {\it{coupling normalization}}. Its main property is that, for any $n$, correlators scale with the same power of $g_*$
\beq
\langle \hat\phi_g(p_1)\dots\hat \phi_g(p_n)\rangle_c= \frac{1}{g_*^2}\left [F_n^{(0)}(\{p\},\mst)+\frac{g_*^2}{16\pi^2}F_n^{(1)}(\{p\},\mst)+\dots\right ]\,.
\label{couplingNorm}
\eeq

\boldmath
\subsection{Large $N$}
\unboldmath
\label{sec:Large N}
In the previous section we considered an elementary field with a weak coupling $g_*$, but a parallel discussion holds for composites in large-$N$ theories. Consider for definiteness large-$N$ QCD. Here composites can be classified according to the number of color traces they entail. So mesons $M_{ij}\equiv {\bar q}_i^\alpha q_j^\alpha$ (with $\alpha$ and $i,j$  color and flavor indices, respectively) are single trace operators, while glueballs $S=G_{\mu\nu}^AG^{A\mu\nu}$ and dimesons $D_{ijkl}=M_{ij}M_{kl}$ are double-trace operators, and so on.

Consider now the correlators of the meson fields $M_{ij}\equiv {\bar q}_i q_j$ defined with $q$ being canonically normalized in the
action. Applying standard large-$N$ reasoning one concludes 
\beq
\langle \hat M_{i_1 j_i}(p_1)\dots \hat M_{i_n j_n}(p_n) \rangle = \frac{N}{16\pi^2}\left [ {\tau^I}_{i_1...\dots j_n} F_I^{(0)}(\{p\},\mst)+O(1/N)\right ] \, ,
\eeq
where ${\tau^I}$ are the set of flavor invariant tensors and where we have identified the hadron mass scale with $\mst$. Comparing to \eq{couplingNorm}, it is immediate to interpret $M$ as a field normalized according to the coupling normalization with  
\beq
g_*\sim \frac{4\pi}{\sqrt N}\,,
\eeq 
an identification  we will assume in what follows. The meson field in the mass normalization is then given by $\bar{M}_{ij}\equiv g_*^2 M_{ij}=
(16\pi^2/N) M_{ij}$. Considering a source $J$ coupled to the meson through $\Delta {\cal L}=J\cdot M$, the generating functional for $M$ correlators  has then the form $W[J]=(1/g_*^2)\left (\hat W[J,\mst]+O(g_*^2/16\pi^2)\right)$ so that its Legendre transform, the effective action for meson operators, reads
\beq
\Gamma[M]=\frac{1}{g_*^2}\left (\hat \Gamma[g_*^2 M,\mst]+O(g_*^2/16\pi^2)\right )= \frac{1}{g_*^2}\left (\hat \Gamma[\bar M,\mst]+O(g_*^2/16\pi^2)\right )\,.
\eeq
This form for $\Gamma$  is consistent with $\langle \bar M\rangle\sim \mst^3$ as $N\to \infty$. As $\langle \bar M \rangle$ scales purely like mass\footnote{This result is also diagrammatically obvious: $\bar M=(16\pi^2/N) \bar q q$, so that the $1/N$ prefactor compensates  the $N$ growth   from the trace over colors.},
 the mass normalization is most suitable for power counting in the EFT. 
 
 For glueball operators like $S=G_{\mu\nu}^AG^{A\mu\nu}$, which are double trace, a similar discussion goes through, but with the role of the coupling played instead by $g_S=4\pi/N=g_*/\sqrt N$ (see \cite{Coleman:1985rnk} and also  Appendix A of
 \cite{Glioti:2024hye} for the discussion of mixed meson-glueball correlators).
 
One last comment concerns the role of the choice of normalization in the Operator Product Expansion (OPE). Again to illustrate the idea we can consider the OPE of single trace operators, for instance two mesons $M_{ij}$ and $M_{k\ell}$. In the coupling normalization
(where $M_{ij}=\bar q_iq_j$, $S=G_{\mu\nu}^AG^{A\mu\nu}$, $D_{ijkl}=M_{ij}M_{k\ell}$ with the elementary fields canonically normalized) a simple diagrammatic analysis gives
\beq
M_{ij} \otimes M_{k\ell}\sim w^{(1)} D_{ijk\ell} + \left (w^{(2)}_{ij}M_{k\ell}+w^{(3)}_{i\ell}M_{kj}+{\mathrm{perms}}\right)+\frac{1}{N}w^{(4)}_{ijk\ell} S+\dots
\eeq
with $w^{(i)}$ Wilson coefficients of order one. In the mass basis, where $\bar{M}_{ij}\equiv g_*^2 M_{ij}=
(16\pi^2/N) M_{ij}$, $\bar{D}_{ijk\ell}=g_*^4 {D}_{ijk\ell}=(16\pi^2/N)^2 D_{ijk\ell}$, $\bar S =g_S^2 S=16\pi^2/N^2 S$, this becomes
\beq
\bar M_{ij} \otimes \bar M_{k\ell}\sim w^{(1)} \bar D_{ijk\ell} + \frac {(4\pi)^2}{N}\left (w^{(2)}_{ij}\bar M_{k\ell}+w^{(3)}_{i\ell}\bar M_{kj}+{\mathrm{perms}}\right)+\frac{(4\pi)^2}{N}w^{(4)}_{ijk\ell} \bar S+\dots
\eeq
Notice that the OPE coefficient for the $M\otimes M\to D$ fusion remains $O(1)$ in both normalization. This is all that will matter in the analysis that follows.

\boldmath
\subsection{Partial compositeness and RG flow at large $N$}
\unboldmath
\label{ToyPartialComp}
In this section we want to illustrate the RG flow induced by partial compositeness.
Ours is just a concise review, focusing on bosonic operators for simplicity, of the results in ref. \cite{Contino:2004vy}.
Consider a ``large-$N$" CFT coupled through a single trace scalar operator ${\cal O}$ of dimension $\Delta$ to an elementary scalar $\varphi$.
Starting at some microscopic scale  $\LUV$, the Lagrangian takes the form
\beq
{\cal L}={\cal L}_{CFT}+\frac{1}{2}(\partial \varphi)^2+\lambda {\cal O}\varphi+{\cal L}^{(2)}_{CT}(\varphi,\LUV)\,,
\eeq
where the last term is a  kinetic counterterm  ensuring that the $\varphi$ propagator is roughly $1/p^2$  at momenta $\sim \LUV$ (see below). 
Working for convenience in Euclidean space and adopting the {\it {coupling normalization}} defined in \sec{sec:Large N}, we parametrize the two-point function of ${\cal O}$ as (see ref. \cite{Grinstein:2008qk})
\beq
\langle {\cal O}(p){\cal O}(-p)\rangle = c\, \frac{N}{(4\pi)^2}\frac{\Gamma(2-\Delta)}{\Gamma(\Delta)}\, (p^2)^{\Delta-2}\,,\eeq
with $c=O(1)$.

Consider now the effects of the mixing on the $\varphi$ propagator. The leading effects at large $N$ are captured by  formally working in the limit where $\lambda^2 N$ is kept finite as $N\to \infty$. In this way, one immediately sees that the leading effect is captured by resumming the geometric series, giving
\bea
\langle \varphi(p)\varphi(-p)\rangle &=& \frac{1}{p^2}+\lambda^2 \frac{1}{p^2}\langle {\cal O}(p){\cal O}(-p)\rangle\frac{1}{p^2}+\dots\\&=&\frac{1}{p^2-c\frac{\lambda^2N}{(4\pi)^2}\frac{\Gamma(2-\Delta)}{\Gamma(\Delta)} \left [(p^2)^{\Delta-2}-p^2(\LUV^2)^{\Delta-3}\right ]} \, ,
\eea
where the $\LUV$ dependent term is the contribution of the above-mentioned  counterterm. This counterterm is mainly added to eliminate the singularity  at $\Delta=3$.
Similarly, for the corrected $\langle {\cal O}\varphi\rangle$ correlator one has 
\beq
\langle {\cal{O}}(p)\varphi(-p)\rangle=c\, \frac{\lambda N}{(4\pi)^2}\frac{\Gamma(2-\Delta)}{\Gamma(\Delta)}\,
\frac{(p^2)^{\Delta-3}}{1-c\frac{\lambda^2N}{(4\pi)^2}\frac{\Gamma(2-\Delta)}{\Gamma(\Delta)} \left [(p^2)^{\Delta-3}-(\LUV^2)^{\Delta-3}\right ]} \,.
\eeq
A suitable definition of the running dimensionless coupling $\hlambda(p)$ is then given by
\begin{eqnarray}
\hlambda(p)&\equiv & \frac{1}{\sqrt{c\frac{N}{(4\pi)^2}\frac{\Gamma(2-\Delta)}{\Gamma(\Delta)}}}\frac{\langle {\cal{O}}(p)\varphi(-p)\rangle}{\sqrt{\langle {\cal O}(p){\cal O}(-p)\rangle}\sqrt {\langle \varphi(p)\varphi(-p)\rangle}}\\&=&
\frac{\lambda\,p^{\Delta-3}\,}{\sqrt{1-c\frac{\lambda^2N}{(4\pi)^2}\frac{\Gamma(2-\Delta)}{\Gamma(\Delta)} \left [(p^2)^{\Delta-3}-(\LUV^2)^{\Delta-3}\right ]}}\,.
\end{eqnarray}
In order to envisage the RG flow, it is convenient to focus on the near marginal case $\Delta=3-\epsilon$ with $|\epsilon|\ll 1$. Working at lowest order in $\epsilon$ we can approximate $\Gamma(2-\Delta)/\Gamma(\Delta) \approx -1/(2\epsilon)$. One has the following three possibilities for the flow at $p\ll\LUV$.
\begin{enumerate}
\item $\epsilon<0$ (Irrelevant):
\beq
\hlambda(p)=\frac{\lambda p^{-\epsilon}}{\sqrt{1-\frac{c}{\epsilon}\frac{\lambda^2 N}{2(4\pi)^2}\LUV^{-2\epsilon}}}\sim \lambda p^{-\epsilon} \, ,
\eeq
which corresponds to the case of positive anomalous dimension discussed in the main text.
\item $\epsilon=0$ (Marginal): 
\beq
\hlambda(p)=\frac{\lambda}{\sqrt{1+c\frac{\lambda^2 N}{(4\pi)^2}\ln (\LUV/p)}}\to \frac{4\pi}{\sqrt {cN}}\frac{1}{\ln(\LUV/p)}\sim \frac{g_*}{{\ln(\LUV/p)}}\, .
\eeq
\item $\epsilon>0$ (Relevant):
\beq
\hlambda(p)=\frac{\lambda p^{-\epsilon}}{\sqrt{1+c\frac{\lambda^2 N}{2(4\pi)^2\epsilon}p^{-2\epsilon}}}\to \frac{4\pi}{\sqrt {cN}}\sqrt{\epsilon}\sim g_*\sqrt{\epsilon}\, ,
\eeq
with $\Lambda_{flow}=(\lambda \sqrt{N}/(4\pi \sqrt{\epsilon}))^{1/\epsilon}$ representing the scale around which the UV CFT flows to  the IR CFT.
\end{enumerate}
In the first two cases, $\varphi$ decouples from the CFT in the IR, but in the third the system flows to a novel CFT where $\varphi$ appears as a primary of dimension $1+\epsilon$, while $\cal O $ is a descendant, as implied by the equation of motion of $\varphi$: $\lambda {\cal O}=\partial^2\varphi$ (the counterterm being irrelevant in the IR). Notice however that the coupling of $\varphi$ to the CFT, $\hlambda\sim g_*\sqrt \epsilon$, remains small for small $\epsilon$. Indeed one can also check that the three point function of the canonically normalized $\varphi$ is of order
$g_*(\epsilon)^{3/2}$, corresponding to a $\sqrt \epsilon$ composite fraction of $\varphi$. Full $O(1)$ compositeness is achieved however in the strongly relevant case $\epsilon =O(1)$. This is the situation that is normally considered for the right-handed component of the top quark.

As a final remark, notice that the results we have just shown strictly apply only in the $N\to \infty$ limit with $\lambda^2N$ fixed. For instance, in the case of $\epsilon>0$, one could in principle imagine that, at some finite order in the $1/N$ expansion, the IR fixed point is itself destabilized and the theory perhaps even be gapped. Indeed, the $N\to \infty$ IR fixed point is endowed with relevant operators like $\varphi$, or in-principle relevant ones like $\varphi^2$, with dimension $2+2\epsilon$ (relevant for $\epsilon<1$) and $\varphi^3$ with dimension $3+3\epsilon$ (relevant for $\epsilon < 1/3$). The case closest to the one of our interest, where $\varphi$ is replaced by a Weyl spinor, is one where some discrete symmetry forbids $\varphi$ and $\varphi^3$, so that we should only focus on $\varphi^2$. In that case, we can definitely conclude that at least for $\epsilon >1$, {\it i.e.}~$\Delta <2$, the IR fixed point is stable at any finite order in $1/N$.
On the basis of this somewhat qualitative discussion, we will envisage the three broad possibilities depicted above for the IR fate of partial compositeness. The  case of $\epsilon$ of order one and positive, with $\mst< \Lambda_{flow}$, corresponds then to the fully composite $t_R$ scenario much considered in the Composite Higgs literature.

\boldmath
\section{The SILSH construction}
\unboldmath
\label{sec:SILSH}

Following  the scheme illustrated in \fig{fig:Overall}, we  now describe our scenario step by step through the EFTs suitable for each successive energy range, between $\LUV$, $\mst$, $\mS$ and $m_{soft}$. As our procedure can be viewed as an extension to supersymmetry of the SILH~\cite{Giudice:2007fh} approach,  we find it  appropriate to name  it  Strongly Interacting Light Supersymmetric Higgs (SILSH).

\boldmath
\subsection{From $\LUV$ to $\mst$}
\unboldmath
\label{UVtostar}

Our basic assumption is that below some unspecified far UV scale 
 $\LUV$ there emerge two QFT sectors, weakly coupled to one another.
The first sector, which we call  $\SSMnH$, is the supersymmetric extension of the Higgsless SM. The $\SSMnH$  field content is described by   the $G_{\rm SM}=SU(3)\times SU(2)\times U(1)$ vector superfields $V_3,\,V_2,\, V_1$ and by the three families of matter chiral superfields $\Phi_i\equiv (\Phi_{L_i}; \Phi_{{\bar R}_i})\equiv (Q_i,L_i \, ; \,\bar U_i,\bar D_i, \bar E_i)$, $i=1,2,3$ (with $\Phi$ labelling the five  different species of matter multiplets).

We must ensure that our scenario respects the strong observational constraints on baryon and lepton number violation.
 At first sight just assuming $R$-parity, or equivalently matter parity, would seem sufficient, as this discrete symmetry forbids all relevant $B$ and $L$ violating interactions in the $\SSMnH$ superpotential. However, it does not forbid higher-dimensional $B$-violating interactions, such as $d$=5 superpotential quartics ($QQQL$ and $\bar U\bar U\bar D\bar E$) and $d$=6 K\"ahler quartics  ($QQ\bar U^\dagger \bar E^\dagger$, $Q\bar U^\dagger \bar D^\dagger L$, $Q^\dagger\bar D \bar D L$, $\bar D \bar D \bar D \bar E^\dagger$). The $d$=5 operators are typically problematic for proton decay even for  $\LUV \sim M_{ \rm Pl }$, while the $d$=6 terms tolerate $\LUV$ a bit below $10^{16}$ GeV. More symmetry
than just  $R$-parity is then necessary to safely consider lower $\LUV$. The most minimal option  is given by combined baryonic and leptonic matter parities:
  $P_B: \,(Q,\,L,\,\bar U,\, \bar D,\,\bar E)\to (-Q,\,L,\,-\bar U,\, -\bar D,\,\bar E)$ and $P_L: \,(Q,\,L,\,\bar U,\, \bar D,\,\bar E)\to (Q,\,-L,\,\bar U,\, \bar D,\,-\bar E)$. The combined constraints of $P_B\times P_L$ relegate  $B$ and  $L$ violation to operators of sufficiently high dimension for them to be safely neglected.
This seems the minimal requirement, as far as pure $\SSMnH$ sector is concerned.

The other sector of the theory is a strongly interacting and approximately conformal invariant sector which we simply dub CFT. This sector is charged under the SM gauge group and endowed with at least three families of chiral operators with complementary gauge quantum numbers to those of the SM matter fields. 
The mixing of these operators with the matter fields are the microscopic seeds that will generate the SM Yukawa interactions, through the mechanism of partial compositeness, as shown in the following.

Besides the above, we won't need to be too specific in the picturing of the CFT sector. However, our scenario has the ambition to offer a full-fledged theory of flavor and supersymmetry breaking at sufficiently low mass scales. This requires that the CFT respects baryon and lepton number, as well as CP. This is necessary to avoid that the mixing between the two sectors generates operators that contribute to proton decay at an unacceptable rate and to unrealistic neutrino masses. Similarly, an intrinsic violation of CP in the CFT sector would generically imply order-one phases in soft terms ($A$-terms, $B_\mu$ and gaugino masses), leading to a tension with experimental data on electric dipole moments. So our hypothesis is that CP is a symmetry of the CFT, broken only by the elementary-composite mixings that seed the Yukawa couplings. These hypotheses, while superficially {\it ad hoc}, do not seem implausible to us: $B$, $L$ and CP could simply arise accidentally. Technically, this corresponds to the absence of singlet scalar primaries of dimension $\le$ 4, and we know of such examples. The SM is one such golden example of  $B$ and $L$ accidental conservation, while CP is easily a symmetry of fixed points. In our scenario CP, along with the flavor symmetry of the Yukawaless SM, is broken by the mixing between elementary and composite sectors. While this structure can lead to a realistic CKM CP phase, it also generically implies unacceptably large $\theta_{\rm QCD}$. In view of that,
we implicitly invoke the axion solution to the strong CP problem.

The leading terms of the Lagrangian at the scale $\LUV$ are
\beq
{\cal L} = {\cal L}_{\SSMnH} +{\cal L}_{\rm CFT} + \int d^4 \theta \, \left( g_A V^A J^A_{\rm CFT} + \dots \right) +
\left[ \int d^2 \theta \, \left( {\O}_{{\bar \Phi}_a} \lambda_{\Phi_{ai}} \Phi_i   + \dots \right) +\hc \right] \, ,
\label{fundlag}
\eeq 
where ${\cal L}_{\SSMnH}$ and ${\cal L}_{\rm CFT}$ are the Lagrangians of the individual sectors, while the other terms describe the (weak) interaction between the two sectors. 
The index $A=1,2,3$ runs over the $G_{\rm SM}$ gauge group factors, with $g_A$, $V^A$,  and $J^A_{\rm CFT}$ respectively  the gauge coupling, vector superfield  and  CFT supercurrent (dropping the adjoint irrep indices). For simplicity, we have written explicitly only the linear term in $V^A$, hiding in the ellipsis mark the higher powers in $V^A$ which reconstruct the exponential $\exp(g_A V^A)$ that is covariant under gauge transformations.  The couplings $\lambda_{\Phi_{ai}}$ describe the linear mixing between the elementary matter fields $\Phi_i$ and the composite chiral operators $\O_{{\bar \Phi}_a}\equiv (\O_{{\bar Q}_a}, \O_{{\bar L}_a},\O_{{U}_a},\O_{{D}_a}, \O_{{E}_a})$, which belong to gauge representations conjugate to those of $\Phi$. They consist of 5 matrices, one for each multiplet type, $\lambda_{\Phi_{ai}}=(\lambda_{{Q}_{ai}}$, $\lambda_{{L}_{ai}}$, $\lambda_{{\bar U}_{ai}}$, $\lambda_{{\bar D}_{ai}}$, $\lambda_{{\bar E}_{ai}})$, with $i=1,2,3$ running over SM generations and $a=1,2,3$ running over the generations of CFT composite operators.
 In general, we expect an infinite  tower of such composites, with ever growing dimension, but we will assume it is enough to focus on the three families of lowest dimension. Our conclusions will not be qualitatively affected by considering the whole tower. 
 In \eq{fundlag}, the dots represent mixing of  SM matter to  the rest of the tower, along with all other higher dimensional operators that do not play any role in our analysis.  The only exception arises when considering smaller effects, such as neutrino masses or proton decay.
 
As we assume $\LUV\gg \mst$, the RG evolution between these two scales is important. As concerns the gauge couplings, the CFT sector simply adds a positive contribution to their $\beta$-functions.  More interesting is the RG effect on the matter mixing  parameters. Consider for definiteness the case where all $\lambda_\Phi$ are irrelevant , which occurs when all $\O_{{\bar \Phi}_a}$ scaling dimensions $\Delta_{\Phi_a}$ are larger than 2 or, equivalently, when the anomalous dimensions $\gamma_{\Phi_a}=\Delta_{\Phi_a}-2$ are positive.
Normalizing the coupling as 
\beq
\lambda_{\Phi_{ai}}\equiv \frac{\hlambda_{\Phi_{ai}}(\LUV)}{\LUV^{\gamma_{\Phi_a}}} \, ,
\eeq
the dimensionless $\hlambda_{\Phi_{ai}}(\LUV)$ then measures the strength of the interaction at the UV scale, while 
\beq
\hlambda_{\Phi_{ai}}(\muRG) \equiv  \lambda_{\Phi_{ai}} \,\muRG^{\gamma_{\Phi_a}}= \hlambda_{\Phi_{ai}}(\LUV) \left (\frac{\muRG}{\LUV}\right )^{\gamma_{\Phi_a}}
\label{running_lambda}
\eeq
represents the running coupling at the renormalization scale $\muRG$, which accounts only for the leading classical effects. Now, even when starting with a structureless or anarchic $\hlambda_{\Phi_{ai}}(\LUV)$, an $O(1)$ or smaller separation among the anomalous dimensions, together with a scale separation $\muRG \ll \LUV$, is sufficient to produce a hierarchy among the eigenvalues of the running coupling matrices $\hlambda_{\Phi_{ai}}(\muRG)$. This is the seed of the fermion mass hierarchy as it arises in partial compositeness.

The above mechanism, based on irrelevant couplings, works beautifully for all the fermions with the sole exception of the top quark,
whose Yukawa is not so small. For the top quark, the most plausible situation is when the mixing of one of the two chiralities,  either $\lambda_{Q_{a3}}$ or $\lambda_{{\bar U}_{a3}}$, is in fact  relevant.\footnote{In the non-supersymmetric composite Higgs scenario, the preferred option  is for such entry to correspond to the right-handed top $t_R$. This choice minimizes the effects of compositeness in the couplings of the left-handed bottom quark to the $Z$-boson, which are known to agree well with the SM. In our supersymmetric scenario, the constraints from $Z$ couplings do not play a significant role, as the compositeness scale will consistently turn out to be higher. Therefore, a negative or vanishing eigenvalue could equally well correspond to either $\gamma_{Q_3}$ or $\gamma_{{\bar U}_3}$.}
In this case, the deformation grows in the IR in such a way that the chirality involved in the mixing becomes {\it de facto} composite. At sufficiently large $N$,  as discussed in ref.~\cite{Contino:2004vy}
 and as reviewed in \sec{sec:Large N}, one can indeed show that the deformed theory flows to a new fixed point. At small enough $N$, it is also conceivable, but difficult  to control in general, that the deformed theory develops a gap, which can naturally be interpreted as the scale $\mst$ itself. In either situation, the chirality involved in the mixing becomes fully composite, and this is all that matters as concerns the power counting rules that determine the structure of the EFT below $\mst$.

The mechanism of partial compositeness has a simple interpretation in the dual extra-dimensional picture, where the energy scale separation is related to the size of the warped extra dimension and $\gamma_{\Phi_a}$ to the ratio of the corresponding 5D mass to the AdS curvature. In that case, the exponential hierarchy is visualized in terms of 5D geography, with the mostly-elementary fields localized far from the brane on which the composite Higgs is confined, and with the mostly-composite fields localized near the Higgs brane.

\boldmath
\subsection{From $\mst$ to $\mS$}
\label{sec:msts}
\unboldmath

As sketched in \fig{fig:Overall} a relevant deformation of the CFT gives rise to an IR threshold $\mst$ in the CFT sector. It is implicitly understood that the hierarchy $\mst\ll\LUV$  satisfies naturalness,  either by quasi-marginality or by (super)-symmetry. However, the precise mechanism has no direct consequences on the low-energy dynamics.

At the scale $\mst$, some of the CFT degrees of freedom are gapped. Below $\mst$ the CFT  flows to two separate sectors that interact with each other only through irrelevant couplings issued from the microscopic strong dynamics and through the (weaker) SM gauge interactions.

The first sector contains effectively supersymmetric Higgs fields and combines with $\SSMnH$ to give rise to a viable low-energy supersymmetric SM theory. At this stage, we don't specify the mechanism that keeps the Higgs composite massless. As already mentioned in \sec{sec:frameCFT}, supersymmetry offers a new perspective on the emergence of  massless composites. Either through chiral symmetries or more simply because of accidental reasons made robust by the non-renormalization property (or the holomorphy) of the superpotential, supersymmetry can consistently produce non-derivatively interacting massless scalars. That is unlike the case of the non-supersymmetric composite Higgs, where the same  mechanism, a shift symmetry, controls both the mass and the other interactions, in particular the quartic and the Yukawas. In the Super-Composite Higgs, the masslessness of the Higgses, amounts to assuming  $\mu=0$ in the supersymmetric limit, a choice that  does not prevent Higgs quartic couplings and Yukawa interactions. In the following, we will assume that $\mu=0$ in the effective theory below $\mst$, either for accidental reasons made robust by holomorphy or because of an approximate PQ-like symmetry\footnote{By ``PQ-like symmetry'' we mean a continuous or discrete symmetry, in principle even an $R$-symmetry, that forbids the bilinear $H_u H_d$ in the superpotential.}. In this paper we shall  focus on the minimal choice of just two Higgs doublets $H_{u,d}$, but our scenario invites  for the exploration of alternatives with  additional light composites. In particular, composite singlets or triplets can drastically change the dynamics underlying the generation of the Higgs quartic. We plan to explore such alternatives in a forthcoming study.

The second sector is an approximately conformal supersymmetric theory, denoted as $\CFTs$, which contains states charged under the SM gauge group and which will eventually break supersymmetry  originating the soft terms in  the SSM sector. 

Given  the above general picture, we are now ready to present the structure of the effective Lagrangian for the light states arising after integrating out the dynamics at $\mst$. We shall do so by 
adapting  the SILH hypotheses and methodology to the supersymmetric case, thus resulting  in the SILSH.
The basic assumption is that the strong sector is broadly described, besides the overall mass scale $\mst$, by a single coupling $g_*$. One could in principle roughly associate $g_*$ to the ``number" $N$ of degrees of freedom in the CFT: $g_*\sim 4\pi /\sqrt N$. 
In the holographic description of our scenario in warped compactifications, $\mst$ and $g_*$ correspond to the mass and coupling of the Kaluza-Klein resonances, respectively.

As indicated in \eq{fundlag}, besides the (strong) coupling $g_*$ of the CFT sector, our  model is described by 
the weak couplings $g_A$ and $\lambda_\Phi$, which also offer the main channels of communication between the SM and the CFT.  

The effective theory below the scale $\mst$ is described by an effective Lagrangian constructed according to the symmetry and power counting rules  of the SILSH
\beq
{\cal L} = {\cal L}_{\SSMnH} +{\cal L}_{\CFTs} + \int d^4 \theta \, {\cal K} 
+
\left( \int d^2 \theta \,  {\cal W}  + \hc \right)\, ,
\label{lsilsh}
\eeq
\beq
{\cal K}=\frac{\mst^2}{g_*^2}\, f_{\cal K}
\Big(  \frac{g_* H_{u,d}}{\mst}, \frac{\O}{\mst^{\Delta_{\O}}}, \frac{\R}{\mst^{\Delta_{\R}}},\frac{D_\alpha}{\mst^{1/2}},gV,\frac{gW_\alpha}{\mst^{3/2}}, \frac{\hlambda_\Phi \Phi}{\mst}\Big) \, ,
\label{genK}
\eeq
\beq
{\cal W}=\frac{\mst^3}{g_*^2}\, f_{\cal W} \Big(  \frac{g_* H_{u,d}}{\mst}, \frac{\O}{\mst^{\Delta_{\O}}}, \frac{D_\alpha}{\mst^{1/2}},\frac{gW_\alpha}{\mst^{3/2}}, \frac{\hlambda_\Phi \Phi}{\mst}\Big) \, .
\label{genW}
\eeq
Here ${\cal L}_{\SSMnH}$ and ${\cal L}_{\CFTs}$ are the Lagrangians of the two individual sectors.
All the other effective interactions between the SSM and the $\CFTs$ sector are contained in a ``K\"ahler potential" ${\cal K}$ and ``superpotential'' ${\cal W}$, expressed in terms of the generic functions of dimensionless quantities $f_{{\cal K},{\cal W}}$. Notice  that 
$\cal K$ and $\cal W$ involve arbitrary powers of the covariant derivatives $D_\alpha$, in such a way that the genuine K\"ahler and superpotential parts correspond  to just the leading terms in the derivative expansion.

In eqs.~(\ref{genK})--(\ref{genW}), $\O$ and $\R$ represent generic chiral and vector operators of the residual $\CFTs$ sector, respectively. There will also exist couplings to $\CFTs$  operators of higher spin, but we assume their dimension is sufficiently high to make their effect negligible. This hypothesis is cleanly realized in holographic incarnations,
where the dimension of these operators presents a gap controlled by a power of $N$. In the  parametrization of \eq{lsilsh},  $g_*H_{u,d}$, $\cal O$ and $\cal R$ are all taken in the mass normalization (see \sec{sec:allthat}), with $1$, 
 $\Delta_{\O}$ and  $\Delta_{\R}$  their respective scaling dimensions. Notice that, among the vector operators, we also count the weakly gauged conserved currents $J_{\CFTs}^A$ of $SU(3)\times SU(2)\times U(1)_Y$. Conservation laws nail their dimensions to be equal to $2$.

The couplings $g$ and $\hlambda_\Phi$ are here understood as running couplings evaluated at $\muRG=\mst$. In particular
$\hlambda_{\Phi_{ai}}(\mst)= \hlambda_{\Phi_{ai}} (\LUV)(\mst/\LUV)^{\gamma_{\Phi_a}}$ is hierarchical even for generic, or anarchic, $\hlambda_{\Phi_{ai}} (\LUV)$. Notice that the dependence of $\cal K$, $\cal W$ on the gauge vector $V$ and matter $\Phi$ superfields only occurs through the couplings $g$ and $\hlambda_\Phi$ respectively, in a way that implements the mechanism of partial compositeness. In particular,  any flavor non-invariant effect in the SSM sector arises through the portal couplings $\hlambda_\Phi$ in \eq{fundlag} upon integrating out the resonances of the gapped CFT at the scale $\mst$.  Moreover, according to our hypothesis that the SCFT respects CP on its own,  the $\hlambda_\Phi$ are  the only complex parameters in eqs.~(\ref{genK}) and (\ref{genW}).
For future convenience, we define
\beq
{\xi_\Phi}_{ai} \equiv \frac{\hlambda_{\Phi_{ai}}(\mst )}{\sqrt{g_*}}\,,\qquad  \qquad {\epsilon_\Phi}_{ai}  \equiv\frac{{\xi_\Phi}_{ai}}{\sqrt{g_*}}\equiv \frac{\hlambda_{\Phi_{ai}}(\mst )}{g_*}\, .
\label{def_xi}
\eeq
where $\xi_\Phi$ essentially corresponds to the square root of the Yukawa coupling, while $\epsilon_\Phi$ measures the composite fraction of SM matter field $\Phi$, according to standard notations \cite{Glioti:2024hye}.

\subsubsection*{Interactions: Matter and Higgs only}

\begin{figure}[t]
\begin{center}
\includegraphics[width=0.97\columnwidth]{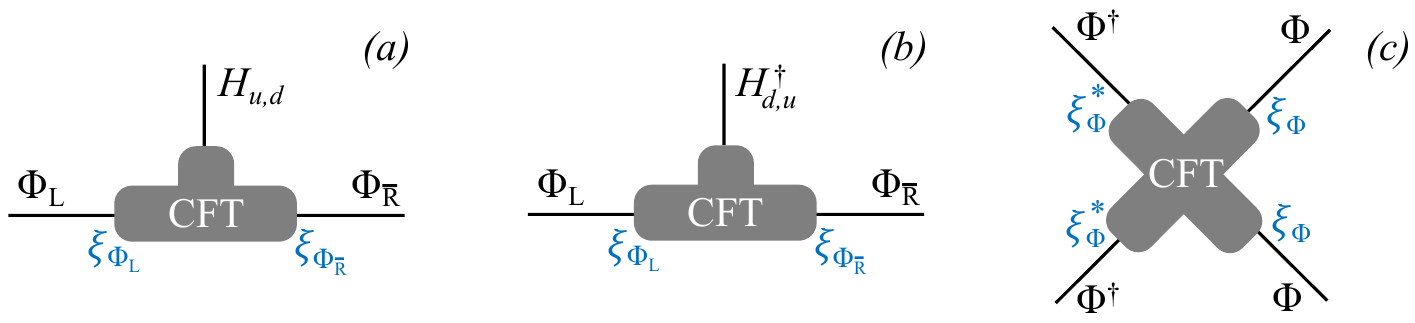}
\end{center}
\caption{Diagrams representing how the CFT dynamics generates {\it (a)} Yukawa interactions, {\it (b)} non-holomorphic trilinear interactions and {\it (c)} dimension-six contact interactions among SM fields. The power counting in partial-compositeness coefficients $\xi$ is shown in blue.}
\label{fig:feynCFT}
\end{figure}

The leading interactions involving only matter and Higgs fields are obtained by Taylor expanding the Lagrangian in \eq{lsilsh} 
\bea
\!\!\!\!\!\!\!\!\!\!\!\!\!\!\!\!{\cal K} &=& H_u^\dagger H_u +  H_d^\dagger H_d + c_\Phi\, \frac{\xi_\Phi^* \xi_\Phi}{g_*} \, \Phi^\dagger \Phi +
\frac{H_d^\dagger Q  {\hat y}_u {\bar U} }{\mst} + \frac{H_u^\dagger Q  {\hat y}_d {\bar D} }{\mst}  + \frac{H_u^\dagger L  {\hat y}_\ell {\bar E}}{\mst} 
\nonumber \\ \!\!\!\!\!\!\!\!\!\!\!\!\!\!\!\!&+&\!\!\!
c_{\Phi\Phi'} \, \xi_\Phi^* \xi_\Phi \xi_{\Phi^\prime}^* \xi_{\Phi^\prime} \, \frac{\Phi^\dagger  \Phi \, \Phi^{\prime\dagger}\Phi^\prime}{2\,\mst^2} +
 {c}_6 \, \xi_{Q}^* \xi_{{\bar D}}^* \xi_{L} \xi_{{\bar E}} \, \frac{Q^\dagger {\bar D}^{\dagger} L {\bar E}}{2\,\mst^2}
\nonumber \\ \!\!\!\!\!\!\!\!\!\!\!\!\!\!\!\!&+&\!\!\!
 \xi_\Phi^* \xi_\Phi \, \frac{\Phi^\dagger  \Phi}{\mst^2}\, {\cal H} +
 \frac{{\cal H}^2}{\mst^2}
+ \dots
\label{eccok}
\eea
\bea
{\cal W} &= &H_u \, Q \, y_u \,{\bar U}  + H_d \, Q  \, y_d \, {\bar D}  + H_d \, L  \, y_\ell \, {\bar E}  +c_4 \frac{g_*^2}{\mst} (H_u H_d)^2
\nonumber \\ &+&\!\!\!
 c_5 \, \xi_{Q} \xi_{{\bar U}} \xi_{L}  \xi_{{\bar E}} \frac{Q {\bar U} L  {\bar E}}{\mst} 
+  c_5^\prime \, \xi_{Q} \xi_{{\bar U}} \xi_{Q}  \xi_{{\bar D}} \frac{Q {\bar U} Q {\bar D}}{\mst} + \dots
\label{eccow}
\eea
Here and in the following, the dots indicate all possible higher dimensional term and higher order in the $\xi$'s as well as those enforced by the hermiticity of ${\cal K}$. All  $c$ coefficients in \eq{eccok} and \eq{eccow} are generic parameters of order one, which are  real because of  CP invariance of the SCFT. Conversely the   $\xi$'s are hierarchical  but otherwise freely complex. For simplicity, we have dropped the $i$ and $a$ indices labelling  generations of matter and composites, respectively. This choice makes formulae more readable, but hides the fact that
 most $c$ parameters are actually tensors
with $a$ indices. For instance, the second term in the second line of \eq{eccok} should be understood as
\beq
{(c_{\Phi\Phi'})}_{abcd}\, \, {\xi_\Phi^*}_{ai} \,{\xi_\Phi}_{bj} \,{\xi_{\Phi^\prime}^*}_{ck}\, {\xi_{\Phi^\prime}}_{d\ell} \, \Phi^\dagger_i  \Phi_j \, \Phi^{\prime\dagger}_k\Phi^\prime_\ell \, .
\eeq
In the third line of \eq{eccok}, we have defined ${\cal H} = g_*(H_u^\dagger H_u + H_d^\dagger H_d + H_u H_d)$, where each interaction term is implicitly multiplied by a different order-one constant. In the second line of \eq{eccow}, we have left implicit the fact that $c_5^\prime$ actually describes two independent sets of operators in which the color indices as saturated as products of two $SU(3)$ singlets or octets. 

All terms proportional to $\xi$ arise from integrating out the composite states at the scale $\mst$, as sketched in \fig{fig:feynCFT}. In particular, the Yukawa couplings (matrices in generation space) are
\beq
y_u =  \xi_{Q}^T c_u\, \xi_{{\bar U}}\, , ~~~~y_d = \xi_{Q}^T c_d \, \xi_{{\bar D}}\, , ~~~~y_\ell =  \xi_{L}^T c_\ell \, \xi_{{\bar E}}\, .
\label{yukint}
\eeq
Similarly, the coefficients of the trilinear terms in ${\cal K}$ are given by 
\beq
\hat y_u =  \xi_{Q}^T \hat c_u\, \xi_{{\bar U}}\, , ~~~~\hat y_d = \xi_{Q}^T \hat c_d \, \xi_{{\bar D}}\, , ~~~~\hat y_\ell =  \xi_{L}^T \hat c_\ell \, \xi_{{\bar E}}\, .
\eeq
 The $c_{u,d,\ell}$ (and $\hat c_{u,d,\ell}$) are expected to be featureless (or anarchic) matrices with $O(1)$ entries,  in line with the  absence of a flavor symmetry of  the CFT.  The Yukawa hierarchies are  encoded in the structure of the $\xi_\Phi$. They emerge purely from the dynamics of the RG flow, as shown by eqs.~(\ref{running_lambda}) and (\ref{def_xi}), rather than from selection rules of some {\emph{ad hoc}} approximate flavor symmetry. Notice that, by   \eq{yukint},  the $\xi_\Phi$'s can be thought of as ``square roots of Yukawa couplings".

It is convenient to perform field redefinitions to simplify the K\"ahler potential  so as to make the phenomenology more
manifest. First of all, one can perform a matrix rescaling of the matter fields in such a way that the term proportional to $c_\Phi$ in \eq{eccok}, together with the contribution from ${\cal L}_{\SSMnH}$, are put into a canonical form. This redefinition introduces a subleading non-holomorphic dependence on $\xi_\Phi^*$ in the superpotential, which should be kept in mind, though it turns out to be negligible for our considerations.
Secondly, the cubic terms proportional to $\hat y_{u,d,\ell}$ can be eliminated by redefining the Higgs superfields:
$H_u\to H_u-( Q \hat y_d \bar D+ L\hat y_\ell \bar E)/\mst$ and $H_d\to H_d- Q \hat y_u \bar U/\mst$. Apart from eliminating 
the trilinears, this transformation yields new interaction terms that can be absorbed in a redefinition of  $c_{\Phi\Phi'}$, $c_6$, $c_5$ and $c_5^\prime$. Notice that the elimination of the trilinears from ${\cal K}$ amounts to eliminating the linear terms in the K\"ahler metric,
and thus simply coincides with choosing Riemann normal coordinates.

In accordance with our hypotheses, the dimension-five operators in \eq{eccow} preserve $B$ and $L$, but violate flavor since the coefficients ${c}_5$ and ${\hat c}_5$ are generic tensors with 4 generation indices each. At low energy, they contribute to SM flavor-violating 4-fermion operators, once they are dressed at one loop, integrating out the supersymmetric particles.

The operators proportional to ${c}_{\Phi\Phi'}$ and ${ c}_6$ in \eq{eccok} give rise at low energy to SM flavor-violating 4-fermion operators. As implicit in our notation, $\Phi$ and $ \Phi'$ can be any pair of the chiral superfields $(Q,\,L,\,{\bar U}, \,{\bar D},\, \bar E)$, with all possible flavor indices.

According to our assumption that the Higgs is a light composite we did not include a supersymmetry-preserving $\mu$ term. 
Because of the non genericity of the superpotential, which is tied to holomorphy, this may not be the  consequence of a (chiral) symmetry. In the same spirit, we allowed in $\cal W$ higher-order terms starting with those controlled by $c_4$. We shall later discuss an example where the vanishing of $\mu$ is controlled by a symmetry.

\boldmath
\subsubsection*{Interactions: SSM  and $\CFTs$ }
\unboldmath
The interactions between the SSM  and $\CFTs$ are worked out by further expanding \eq{lsilsh} to include $\CFTs$ operators which, besides the gauge vector currents
$J_{\CFTs}^A$, consist of  chiral superfields $\O$ and vector superfields $\R$. In principle there will be a number of them, distinguished by their gauge quantum numbers  and possibly by their quantum numbers under some global (approximate) symmetry. The most relevant effects will be determined by gauge neutral operators that acquire vevs upon supersymmetry breaking. In the minimal realization of the scenario there will be just one chiral and one vector operator dominating all effects. We shall base our discussion of the phenomenology on such minimal hypothesis. However, in the following discussion, we shall remain more general and allow for different CFT operators to dominate the coupling to different operators involving matter and gauge fields.

To analyze the supersymmetry breaking effects it is convenient to classify the interactions in the following three classes.

\boldmath
\subsubsection*{I. Gauge sector}
\unboldmath
The interactions between gauge fields and operators in the $\CFTs$ are given by
\beq
{\cal K} =
\frac{g}{g_*^2}\, V_A J_{\CFTs}^A + \dots \, ,
\label{Kgaug}
\eeq
\beq
{\cal W} = \frac{g^2}{g_*^2} W^\alpha W_\alpha
\frac{\O_W}{\mst^{\Delta_{\O_W}}} \, .
\label{Wgaug}
\eeq
Equation (\ref{Kgaug}) describes the SM gauge interaction with the $\CFTs$ current, and, as before, we have left implicit higher orders in $V$. In \eq{Wgaug}, ${\O_W}$ indicates the lowest dimension chiral operator that couples to $W^\alpha W_\alpha$.

\boldmath
\subsubsection*{II. Higgs sector}
\unboldmath

The leading interactions between Higgs fields and $\CFTs$ operators  are given by
\beq
{\cal K} = H_u^\dagger H_u Z_u + H_d^\dagger H_d Z_d + \left( H_u H_d Z_\mu + \hc \right) \, ,
\label{kalH}
\eeq
with
\beq
Z_\alpha=  \frac{c_\alpha \O_\alpha+  \tilde c_\alpha \O_\alpha^\dagger}{\mst^{\Delta_{\O_\alpha}}} +  \hat c_\alpha\frac{\R_\alpha}{\mst^{\Delta_{\R_\alpha}}} \, , ~~~~
\alpha=u,d,\mu \, ,
\label{defZH}
\eeq
where hermiticity of ${\cal K}$ also implies $\tilde c_{u,d}=c_{u,d}^*$ and $\hat c_{u,d}=\hat c_{u,d}^*$.
 The chiral operator $\O_\mu$ gives no effect in the K\"ahler potential and can be omitted from the definition of $Z_\mu$. Since no quantum number distinguishes the leading lowest dimension operators appearing in $Z_u, Z_d$, we shall identify them:  $\O_u=\O_d\equiv \O$ and $\R_u= \R_d\equiv \R$. In the presence of a continuous or discrete PQ-like symmetry, $\O_\mu$ and $\R_\mu$ could be distinguished from $\O$ and $\R$.

Making use of the non-genericity of the superpotential and for the reasons stated above about the absence of a $\mu$ term at short distances, we do not include a term ${\cal W} = H_u H_d \, {\O_{\mu}}/{\mst^{\Delta_{\O_{\mu}}-1}}$.

\boldmath
\subsubsection*{III. Matter (\& Higgs) sector}
\unboldmath
Matter couples to $\CFTs$ via partial compositeness. At lowest order, the Taylor expansion of  \eq{lsilsh} 
gives
\bea
{\cal K} &=&\frac{ \xi_\Phi^* \xi_\Phi}{g_*} \, \Phi^\dagger \Phi Z_S + \frac{\xi_{\Phi_L} \xi_{\Phi_{\bar R}}}{\mst} \, H \Phi_L \Phi_{\bar R}  Z_{y}+
\frac{\xi_{\Phi_L} \xi_{\Phi_{\bar R}}}{\mst} \, H^\dagger \Phi_L \Phi_{\bar R}  Z_{\hat{y}}
\label{kal1}\\ &+&
\frac{\mst\, \xi_\Phi}{g_*^{3/2}} \, \Phi Z_{1K} + \frac{\xi_\Phi \xi_\Phi}{g_*} \, \Phi \Phi Z_{2K} + \dots   \, ,
\nonumber \\
{\cal W} &=& \frac{\xi_\Phi}{g_*^{3/2}} \, \Phi \, \frac{\O_1}{ \mst^{\Delta_{\O_1}-2}}+  \frac{\xi_\Phi \xi_\Phi}{g_*} \, \Phi \Phi \, \frac{\O_2}{\mst^{\Delta_{\O_2}-1}}
+   \xi_{\Phi_L} \xi_{{\bar \Phi}_R} \, H \Phi_L \Phi_{\bar R}  \, \frac{\O_{3}}{\mst^{\Delta_{\O_3}}}+\dots\, 
\label{pot1}
\eea
Here we have defined the combinations of chiral and real operators as in \eq{defZH}.
However, now $\alpha=S,y, {\hat y}, 1K,2K$ and $c_\alpha, \tilde c_\alpha, \hat c_\alpha$  are tensors in CFT generation space with order-one entries   (again with 
$\tilde c_\alpha= c_\alpha^\dagger$ only where dictated by hermiticity). Analogous coefficients control the terms in ${\cal W}$, although we omit them for simplicity.
In eqs.~(\ref{kal1})--(\ref{pot1}), $H$ denotes the Higgs superfield,  $H_u$ or $H_d$, entering the suitable holomorphic  (involving ${\O_3}$ and $Z_y$) or non-holomorphic trilinear interactions (involving $Z_{\hat{y}}$). $\O_y$ gives no effect in the K\"ahler potential and can be omitted from the definition of $Z_y$. 

The  term involving  ${\O_3}$ generates dangerous flavor or CP-violating effects since it is not exactly proportional to Yukawa couplings. We will thus work under the assumption, made consistent by the non-genericity of the superpotential, that this term is indeed absent.  As we shall see, the trilinear proportional to $Z_{\hat y}$, even if 
suppressed by an extra power of $\mst$, also implies strong bounds, but it can be controlled by symmetries.

The operators ${\cal O}_S$ and ${\cal R}_S$ cannot be distinguished by symmetry from the analogous operators 
that coupled to the Higgs bilinears. We will thus work under the assumption $\O_S\equiv  \O$, $\R_S\equiv  \R$, as discussed below \eq{defZH}.

Finally, let us comment on the interactions involving $\O_{1,2}$ and $Z_{1K,2K}$. They exist only if the $\CFTs$ sector possesses operators that match the gauge quantum numbers of the $\Phi$'s or of their bilinears. Indeed, it is also possible that, even when allowed by gauge invariance, such terms are barred by residual accidental global symmetries ($B$, $L$, PQ or their discrete subgroups). The impact of these terms  depends on the dimension of the composites,  and, as commented in \sec{sec:flavor}, one can consistently assume they are negligible. We also remark that the $\O_1$ interaction in \eq{pot1}, although having the same form as the partial-compositeness coupling in \eq{fundlag}, is completely independent, as $\O_1$ belongs to the $\CFTs$ sector below $\mst$, while $\O_{\bar \Phi}$ belongs to the CFT above $\mst$. In the AdS$_5$ language, $\O_1$ is related to fields that live between the IR$_*$ and IR$_S$ branes, while $\O_{\bar \Phi}$ is related to fields that live between the UV and IR$_*$ branes.

\boldmath
\subsection{From $\mS$ to $m_{\rm soft}$}
\unboldmath
\label{sec:fromstosoft}

The next stage of the dynamical tumbling process occurs when the $\CFTs$  sector breaks supersymmetry at a mass scale $\mS$, parametrically smaller than $\mst$ (but not necessarily much smaller). We will assume   the simplest  scenario defined as follows.
\begin{itemize}
\item  At the scale $\mS$, all  
$\CFTs$ states are gapped, apart from the mandatory gapless Goldstino $\chi$. As $\CFTs$ is charged under the SM, the gapped states automatically act as messengers of supersymmetry breaking {\it {\`a la}} gauge mediation. A variant of this minimal scenario would involve another scale $m_M>\mS$ controlling the mass of the mediators. In \sec{sec:robust}, we will make some brief comments on the implications of such variant.
\item $\CFTs$ is characterized down to the scale $\mS$ by the same coupling strength $g_*$ (or by the same large $N\sim 16\pi^2/g_*^2$)
of  the parent theory at the $\mst$ threshold. In the holographic realization of our construction it is straightforward to implement the alternative scenario where $\CFTs$ is described by a different coupling strength $g_*^\prime$ (plausibly larger than $g_*$): all is needed is to tweak the coupling in the AdS slice between the Higgs and supersymmetry breaking branes. We leave this variant for future study.
\end{itemize} 

Under the above hypotheses, the effect of supersymmetry breaking is derived by iterating at $\mS$  the same EFT construction already worked out at $\mst$.
The only residual $\CFTs$ field,  the Goldstino, can be encapsulated in a dimensionless chiral superfield ${X}$ that satisfies the constraint $X^2=0$~\cite{Komargodski:2009rz}. This removes from $X$ the scalar fields, leaving the Goldstino as the only propagating field and realizing supersymmetry non-linearly.  This field reflects the most infrared description of the supersymmetry breaking $F_{\cal X}$ in the 5D picture. In the mass normalization (see \sec{sec:allthat}), $X$ takes the form $X=(\sqrt \mS \theta_\alpha+\chi_\alpha/ \sqrt {2 \mS})^2$, which evidently satisfies $X^2=0$. The Goldstino EFT can be  straightforwardly derived from this expression of ${X}$, but we will be mainly concerned with the $F$-component of $X$, as it sources the soft terms 
\beq
\langle X\rangle =\mS\, \theta^2 \, .
\label{vevX}
\eeq
In the EFT below $\mS$, terms involving both $X$ and the SSM fields control
supersymmetry breaking effects. These terms originate by integrating out the $\CFTs$ dynamics at $\mS$ and can
be formally expressed in terms of correlators of $\CFTs$ operators of increasing order. In practice, however,
it will suffice to consider just the one- and two-point functions, as they cover all relevant effects.

Accounting for one-point functions simply amounts to matching   the SM neutral composites in ${\cal K}$ and ${\cal W}$
to their Goldstino EFT form as dictated by generic power counting
\beq
\O_a \to \mS^{\Delta_{\O_a}} \, c_{\raisebox{-1.0pt}{$\scriptstyle {{\O}_a}$}} X \, , ~~~~
\R_a \to \mS^{\Delta_{\R_a}} \left( {\hat c}_{\raisebox{-1.0pt}{$\scriptstyle {{\R}_a}$}} X + {\hat c}^*_{\raisebox{-1.0pt}{$\scriptstyle {{\R}_a}$}} X^\dagger + c_{\raisebox{-1.0pt}{$\scriptstyle {{\R}_a}$}} XX^\dagger \right) \, ,
\label{replac}
\eeq
where $c$ are unknown dimensionless parameters of order unity.  
Since we are considering the most general K\"ahler potential compatible with symmetries, $\R_a$ always appears in the combination $Z_a$, given in \eq{defZH}. Under the plausible assumption $\Delta_{\O_a} < \Delta_{\R_a}$, the effect of ${\hat c}_{\raisebox{-1.0pt}{$\scriptstyle {{\R}_a}$}}$ can be reabsorbed in $c_{\raisebox{-1.0pt}{$\scriptstyle {{\O}_a}$}}$. Therefore, in the following, we will neglect ${\hat c}_{\raisebox{-1.0pt}{$\scriptstyle {{\R}_a}$}}$. 

The second class of effects arises from two-point functions, and is genuinely associated to the exchange of states with mass $\sim \mS$. In particular,
\eq{Kgaug}, through the current two-point function $\langle V_{\CFTs}^A
V_{\CFTs}^B\rangle$, will give rise to effective interactions among $X$ and the SSM fields, which realize gauge mediated supersymmetry breaking. Similarly, exchange of $Z_{1K}$ and $Z_{2K}$ in \eq{kal1} will produce flavor breaking soft masses for the matter fields.

The effects induced by the interactions of class I and II of the previous subsection  basically amount to the standard supersymmetry breaking soft terms and are characterized by an overall scale $\msoft$. These effects are flavor preserving and determine the main features of the spectrum. Instead, the effects induced by the interactions of class III are flavor breaking and, as we shall explain, not even fully described by the standard soft terms.

With the above general picture in mind, we can now study the supersymmetry breaking effects induced by the interactions in classes I, II and III of the previous subsection.

\subsubsection*{I. Gauge mediated effects}
We consider here the supersymmetry breaking effects induced by \eq{Kgaug}. We anticipate that these will essentially 
amount to  the soft terms of standard gauge mediation.

Let us first focus on the contribution of  one-point functions, which would arise by applying  \eq{replac} to the gauge currents  superfields $J_{\CFTs}^A$ in \eq{Kgaug}.  By color and weak charge conservation  only the hypercharge current $J_{\CFTs}^Y$ could in principle have non-zero vev, {\it i.e.}~non-zero $c$'s  in \eq{replac}. That would give rise to an  $O(\mS^2)$   contribution to soft masses via the hypercharge $D$-term. This contribution is non-positive definite and, if present, would dominate the soft masses leading to a non-realistic vacuum structure. Luckily, however, one can imagine that $\langle J_{\CFTs}^Y\rangle =0$ due to an accidental charge conjugation  of  $\CFTs$,  under which $J_{\CFTs}^Y\to -J_{\CFTs}^Y$.
We will henceforth work under this assumption. 

\begin{figure}[t]
\begin{center}
\includegraphics[width=0.97 \columnwidth]{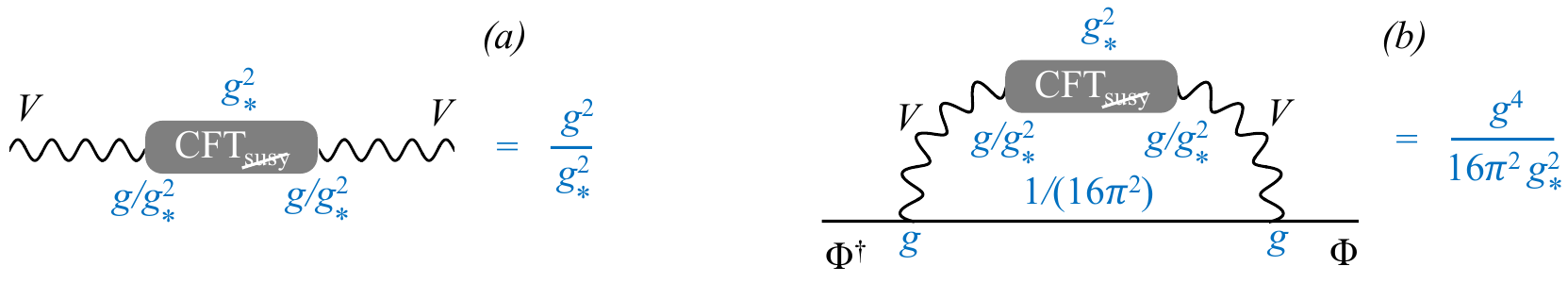}
\end{center}
\caption{Diagrams representing how the $\CFTs$ dynamics generates {\it (a)} gaugino masses and {\it (b)} squark and slepton masses. The power counting in couplings and loop factors is shown in blue. }
\label{fig:feynCFTs}
\end{figure}

The diagrams in  \fig{fig:feynCFTs} describe the leading contribution  controlled by the $\CFTs$ current two-point function.
While the gauge sector is  directly affected, matter fields are only affected through light-field quantum fluctuations, at one-loop.
 The corresponding operators in the Goldstino + SSM EFT are easily ``power counted", including a calculable IR log in diagram (b),
\bea
\frac{1}{g_*^2}&&\!\!\!\!\!\!\!\!  \int d^2\theta  \sum_A (a_0^A+a_1^A X)g_A^2 W_\alpha^A W^{\alpha A}+ \dots
\label{gaugemed_gauge} \\
\frac{1}{g_*^2}&&\!\!\!\!\!\!\!\! \int d^4\theta 
\sum_A \frac{g_A^4 C^A}{16\pi^2}\left \{c_0^A+\left [(c_1^A +2 a_1^A \ln\frac{\muRG}{\mS})X+{\mathrm {h.c.}}\right ]+c_2^A X X^\dagger\right \} \Phi^\dagger \Phi+\dots
\label{gaugemed_matter}
\eea
where $a_{0,1}^A$ and $c_{0,1,2}^A$ are incalculable $O(1)$ coefficients, $C^A$ is the corresponding Casimir and
the dots represent higher derivative terms suppressed by inverse powers of $\mS$,
along with  higher orders in the SM gauge couplings. By using \eq{vevX}, we obtain the gaugino and scalar masses 
\be\label{softm}
M_{\lambda}^A =  \frac{a_1^Ag_A^2}{g_*^2} \, \mS\sim \frac{ g_A^2 N}{16\pi^2} \, \mS\,,\qquad
{\tilde m}^2_\Phi= \sum_A   \frac{c_2^AC_\Phi^Ag_A^4}{16 \pi^2 g_*^2} \, \mS^2\sim  \sum_A \frac{ C_\Phi^A g_A^4  N}{(16\pi^2)^2}\, \mS^2\,.
\ee
With  the replacement $g_*=4\pi/\sqrt{N}$,  \eq{softm} shows that the mass relations are (up to unknown order-one factors) identical to the familiar gauge mediation expressions, where gaugino masses are generated at one loop and scalar masses at two loops and where $N$ plays the role of the number of messenger species. 

The term  linear in $X$ in \eq{gaugemed_matter}  involves a threshold effect proportional to $c_1^A$ and a dominant IR log (where $\muRG$ is the renormalization scale) associated with the gaugino mass. Saturating  the log at the scale $\msoft$ leads to  a universal contribution proportional to $g_A^2/16\pi^2 M_{\lambda}^A\ln (\mS/\msoft) $ to 
 $A$-terms. Because of the moderate hierarchy between $\mS$ and $\msoft$, in spite of the log enhancement,  this contribution remains smaller than both the gaugino masses and the 
 $A$ and $B$-terms induced by Higgs compositeness, which will be discussed later.

The effects we just discussed constitute the leading contribution to gaugino and sfermion masses. As concerns gauginos, subleading contributions to their masses arise  for instance from  \eq{Wgaug}, once $\O_W$ is matched to the Goldstino EFT. In a sense also this contribution is gauge mediated, but associated with dynamics at $\mst$. The result   is  suppressed by a factor $(\mS/\mst)^{\Delta_{O_W}}$ with respect to \eq{softm}, and is therefore unimportant.
The situation with the sfermion masses is more delicate, as the subdominat contributions violate flavor, and are part of a constellation of flavor and CP violating effects, not all classifiable as {\it soft}. We shall offer an overview of these effects in subsection III and study them in detail  in \sec{sec:flavor}.

It must also be stressed that, besides these soft masses, there also arise hard breaking effects controlled by  $g^2/g_*^2$. Indeed, the vev of $X$ and the derivative expansion of eq.~\eqref{gaugemed_gauge}  are controlled by the same mass scale $m_S$. So, for instance, the higher derivative term
\be
\label{hardgauge}
\frac{g^2}{g_*^2} \int d^4 \theta \frac{XX^\dagger}{m_S^2} \left (D_\alpha W_\beta+D_\beta W_\alpha\right )\left (D^\alpha W^\beta+D^\beta W^\alpha\right )
\ee
will purely induce an $O(g^2/g_*^2)$ renormalization of the  gauge kinetic term $F_{\mu\nu}F^{\mu\nu}$, without affecting the gaugino and auxiliary field quadratic action. This will induce a mismatch of the same order among  the gauge couplings  as defined by the covariant derivative and by the gaugino coupling to matter. While this effect is small, \ie~formally one-loop when identifying $N/16\pi^2= 1/g_*^2$, it nonetheless represents a hard form of supersymmetry breaking. This is because the scale of supersymmetry breaking and the scale of its mediation through gauge interaction coalesce in our scenario into a single scale.

It is interesting to compare the results as seen from the two perspectives, 5D warped compactification and 4D CFT. In the 4D  picture, gaugino masses come from a mixing between SM gauginos and composite fermion adjoints of the strong dynamics, as shown in \fig{fig:feynCFTs}{\it (a)}, with mixing measured by  $g/g_*$.\footnote {In the figure we use the  mass normalization, but scaling back to canonical normalization makes clear that the elementary-composite mixing is $O(g/g_*)$.} This can be pictured as  partial compositeness of SM gauge superfields. The result is $\propto N$,  representing the number of fundamental constituents of the composite adjoints. Gaugino and scalar masses have a rough ratio
\beq
\frac{{\tilde m}_\Phi^2}{M_{\lambda}^2}\sim \frac{g_*^2}{16\pi^2} \sim \frac1N\, .
\label{masrat}
\eeq
This shows that the presence of a strong coupling constant is essential to explain why all supersymmetric particle masses are at roughly the same scale, which we call $\msoft$. As $N$ grows, the coupling $g_*$ gets weaker and a certain mass separation is generated.

In the 5D picture, gauginos directly feel supersymmetry breaking, since gauge fields live everywhere in the extra dimension and have a full overlap with the brane where supersymmetry breaking is localized. 
Effectively, this is the picture of gaugino mediation, where scalar masses are generated through one-loop diagrams involving the zero-modes and the excited KK-modes of the 5D gauginos. 
However, there is a crucial difference in the parametric expressions of scalar masses between flat and warped spaces.

In the case of flat extra dimensions, both the zero and excited KK-modes of the gauge fields are delocalized and therefore couple in a similar way to the supersymmetry-breaking source. As a result, the ratio between scalar and gluino squared masses evaluated at the compactification scale $M_{ \rm KK }$ is parametrically given by
\beq
 \frac{ \tilde{m}_\Phi^2 (M_{ \rm KK } )}{ M_{\lambda}^2 (M_{ \rm KK } ) }  \sim  \frac{ g^2 }{ 16 \pi^2 } ~~~~~{\rm (flat~extra~dimension)} \, .
\label{flat_scalar_mass}
\eeq

In warped spaces, the excited KK-modes of the gauge fields are strongly localized near the supersymmetry-breaking brane, see \fig{fig:susy}. Hence, they couple to the supersymmetry-breaking source with a strength $g_{ KK }$ exceeding  the  strength $g$ with which the zero mode couples. This enhances   the  excited KK-modes contribution over the flat case
\beq
 \frac{ \tilde{m}_\Phi^2 (M_{ \rm KK } )}{ M_{\lambda}^2 (M_{ \rm KK } ) }  \sim  \frac{ g_{ KK }^2 }{ 16 \pi^2 } ~~~~~{\rm (warped~extra~dimension)} \, .
\label{warped_scalar_mass}
\eeq

Of course, the contributions from RG running below $M_{KK}\sim \mS$ are the same in the two cases of space geometry. According to the AdS/CFT dictionary, KK-modes localized near the supersymmetry breaking IR brane are interpreted as composites in the CFT picture. This implies the identifications
\beq
\mS~~ \leftrightarrow ~~   M_{KK} \qquad\qquad g_*  ~~ \leftrightarrow ~~ g_{ \rm KK }  \, ,
\eeq
with $g_{ \rm KK }$ controlling  the coupling among excited gauge KK-modes localized near the IR brane.
Thus, \eq{warped_scalar_mass} fully matches \eq{masrat}.

\subsubsection*{II. Supersymmetry breaking  from  Higgs compositeness}
The Higgs sector, like all fields charged under the SM gauge group, is affected by gauge mediation. However, compositeness too gives a contribution, which originates at lowest order  from \eq{kalH}. These are also standard soft terms, consisting of the higgsino mass\footnote{We find it convenient to treat $\mu$ as a soft  supersymmetry breaking effect as it arises according to the mechanism of ref.~\cite{Giudice:1988yz}.}
\beq
{\cal L}_\mu = \mu \tilde h_u\tilde h_d+{\mathrm{h.c.}}
\eeq
and of the scalar potential, which we parametrize as 
\bea
&V& = m_1^2 |H_d|^2 + m_2^2 |H_u|^2 +\left( B_\mu H_u H_d + \hc \right)
\nonumber\\
&+&\left(  H_u \,{\tilde \varphi}_Q A_u \,{\tilde \varphi}_{\bar U} +  H_d \,{\tilde \varphi}_Q A_d \,{\tilde \varphi}_{\bar D} + H_d \,{\tilde \varphi}_L A_\ell \,{\tilde \varphi}_{\bar E} + \hc \right) + \dots \, ,
\eea
\beq
m_{1,2}^2 = {\tilde m}_{H_{d,u}}^2 + \mu^2 \, .
\label{defm12}
\eeq
To study these effects we shall focus on the simplest and most generic case, where  
\begin{enumerate}
\item no additional global symmetry controls the appearance of the   composites $\O_a$ and $\R_a$;
\item   a finite gap in dimension allows to collapse all the $\O_a$ and $\R_a$ neutral under the SM to the single pair  of  lowest dimension, which we shall indicate by $\O$ and $\R$.
\end{enumerate}
Under these hypotheses and after the replacements in \eq{vevX} dictated by \eq{replac},  the supersymmetry breaking part of the kinetic coefficients in \eq{kalH} becomes
\beq
\langle Z_{u,d,\mu} \rangle =  \left( c_{\raisebox{-1.0pt}{$\scriptstyle {{\O}_{u,d,\mu}}$}}
 \frac{\mS^{\Delta_\O +1}}{\mst^{\Delta_\O}}\, \theta^2 +\hc \right) + 
 c_{\raisebox{-1.0pt}{$\scriptstyle {{\R}_{u,d,\mu}}$}} \frac{\mS^{\Delta_\R +2}}{\mst^{\Delta_\R}}\, \theta^4\, .
\label{renZ}
\eeq
The computation of the soft masses by inserting \eq{renZ} in \eq{kalH} encounters a  known technical subtlety,  associated with double insertions of $\O$, which in general do not factorize. This affects $X X^\dagger$ terms arising at second order in the Wilson coefficients of $\O$. 
As pointed out in \cite{Perez:2008ng,Murayama:2007ge,Craig:2009rk}, the result of such double insertions are effectively controlled by the OPE at $\mst$: $\O^\dagger \times \O \sim \mst^{2\Delta_\O-\Delta_\R} \R$. In practice, this means that the renormalization of  $\O^\dagger \O$ is controlled by  $\R$ 
\beq
\langle \O_{u,d,\mu}^\dagger \O_{u,d,\mu}\rangle = a|c_{\raisebox{-1.0pt}{$\scriptstyle {{\O}_{u,d,\mu}}$}} |^2 \, \frac{\mS^{\Delta_\R +2}}{\mst^{\Delta_\R}}\, \theta^4\, ,
\label{renO}
\eeq
with $a$ an $O(1)$  fudge factor that depends on the detailed dynamics at $\mst$.
At linear order in $X$, we then find
\beq
\mu = c^*_{\raisebox{-1.0pt}{$\scriptstyle {{\O}_{\mu}}$}}
 \frac{\mS^{\Delta_\O +1}}{\mst^{\Delta_\O}} \, , ~~~~
A_{u,d}= c_{\raisebox{-1.0pt}{$\scriptstyle {{\O}_{u,d}}$}}
 \, y_{u,d}\frac{\mS^{\Delta_\O +1}}{\mst^{\Delta_\O}}\, , ~~~~A_\ell =
c_{\raisebox{-1.0pt}{$\scriptstyle {{\O}_{d}}$}}\,  y_{\ell}\frac{\mS^{\Delta_\O +1}}{\mst^{\Delta_\O}}\, .
\label{eqmu}
\eeq
The trilinear coupling  matrix $A$ is of order $\mu$ and exactly proportional to the corresponding Yukawa coupling. This is because this term is purely determined by  the Higgs leg in the vertex and therefore cannot affect the flavor structure. 

At order  $X^\dagger X$, after replacing the composite operators in \eq{kalH} according to eqs.~(\ref{renZ})--(\ref{renO}), we find the Higgs mass terms 
\beq
m_{1,2}^2= 
\frac{c_{1,2}\, g^4 N\mS^2}{{(16\pi^2)}^2}+
( a |
c_{\raisebox{-1.0pt}{$\scriptstyle {{\O}_{d,u}}$}}
|^2 + a|
c_{\raisebox{-1.0pt}{$\scriptstyle {{\O}_{\mu}}$}}
|^2- 
c_{\raisebox{-1.0pt}{$\scriptstyle {{\R}_{d,u}}$}}
)\, \frac{\mS^{\Delta_\R +2}}{\mst^{\Delta_\R}}\, ,
\label{eqhso}
\eeq
\beq
B_\mu = [a 
c^*_{\raisebox{-1.0pt}{$\scriptstyle {{\O}_{\mu}}$}}
(
c_{\raisebox{-1.0pt}{$\scriptstyle {{\O}_{u}}$}}
+ 
c_{\raisebox{-1.0pt}{$\scriptstyle {{\O}_{d}}$}}
)- 
c_{\raisebox{-1.0pt}{$\scriptstyle {{\R}_{\mu}}$}}
]\, \frac{\mS^{\Delta_\R +2}}{\mst^{\Delta_\R}}\, ,
\label{eqbmu}
\eeq
where we have added the gauge-mediated  contribution to ${\tilde m}_{H_{u,d}}^2$ given by \eq{softm} multiplied  by unknown order-one parameters 
$c_{1,2}$.

Incidentally, eqs.~(\ref{eqhso})--(\ref{eqbmu}) are in accordance with the scalar sequestering phenomenon, which dictates that the Higgs mass matrix has a vanishing boundary condition $m_{1,2}^2(\mS)=B_\mu (\mS)=0$, if the following two assumptions are satisfied~\cite{Roy:2007nz,Murayama:2007ge,Perez:2008ng}: {\it (i)} gauge mediated effects are absent; {\it (ii)} vector superfields have a significant dimension gap, {\it i.e.} ~$(\mS/\mst)^{\Delta_\R}\to 0$.

\boldmath
\subsubsection*{The solution to the $\mu$--$B_\mu$ problem}
\unboldmath

Comparing eqs.~(\ref{eqmu}) and (\ref{eqbmu}), we find
\beq
\frac{B_\mu }{ \mu^2} \approx \left(\frac{\mS}{\mst}\right)^{\Delta_\R-2\Delta_\O}\, . 
\label{eccobbmu}
\eeq
Therefore, the $\mu$--$B_\mu$ problem is solved if $\Delta_\R \gsim 2\Delta_\O$, with the case $\Delta_\R \simeq \Delta_\O$ singled out
as a specially intriguing situation. Interestingly, this condition is automatically realized in large $N$ theories, where we find $\Delta_\R = 2 \Delta_\O + O(1/N)$. Indeed, following the discussion in \sec{sec:Large N}, consider taking $\O$ as a single trace operator. The double trace vector operator $\O(x)\O^\dagger(x)$ has,  at leading order in $1/N$, a two-point function equalling the square of $\langle \O^\dagger (x)\O(0)\rangle$. Then, $\R$ coincides with $\O\O^\dagger$ at leading order and  $\Delta_\R = 2 \Delta_\O + O(1/N)$ follows.

Thus, even at moderately large $N$, we find $B_\mu \sim \mu^2$, while all entries in the Higgs mass matrix are automatically of order $\msoft$, as long as
\beq
\left(\frac{\mS}{\mst}\right)^{\Delta_\O} \sim \frac{g^2}{16 \pi^2}\, .
\label{ratformu}
\eeq
We must assume that \eq{ratformu}  holds accidentally. However, note that  \eq{ratformu} does not represent a technically unnatural tuning, but only a numerical coincidence. Moreover, the tumbling mechanism is expected to generate only mild hierarchies in $\mS/\mst$ and, from theoretical considerations, we expect $\Delta_\O$ to be of order unity. Therefore, the two quantities in \eq{ratformu} are generically in the same ballpark. The degree of coincidence in their equality can be estimated by using the criterion of ref.~\cite{Barbieri:1987fn}, which measures the parameter sensitivity in \eq{ratformu}
\beq
\delta_{\rm FT}=\max_{a\in\{\mS/\mst,\Delta_\O\}} \left| \frac{\partial \ln (\mS/\mst)^{\Delta_\O}}{\ln a}\right|=\ln \left( \frac{g^2}{16 \pi^2}\right) \, .
\eeq
This procedure gives a degree of coincidence in one part over $\delta_{\rm FT}$, which is fairly mild in the parameter range of interest. Coincidences of this kind occur in nature. As a playful example, note that \eq{ratformu} has a parametric resemblance to the accidental coincidence $(m_u-m_d)/\Lambda_{\textrm{QCD}}\sim \alpha_{\textrm{QED}}/4\pi$, which is crucial to make the neutron slightly heavier than the proton, a condition vital for the existence of chemical elements.

We would also like to emphasize that the need for an accidental coincidence is a common feature of all calculable mechanisms for the origin of  $\mu$ in gauge mediation.  Typically, one must assume that the coupling constant of the interaction generating $\mu$ at one loop is approximately equal to the gauge coupling constant. This assumption represents a coincidence that is as unexplained as \eq{ratformu}. However, while \eq{ratformu} itself may not raise any particular concerns, it is part of a larger issue: the unexplained little-hierarchy tuning required for EW symmetry breaking.

Besides these considerations, the important implication of \eq{ratformu} is that the scales $\mst >\mS >\msoft$ cannot be widely separated, a tight prediction of our scenario. This result has significant phenomenological implications, as we will discuss in sects.~\ref{sec:Minimal} and \ref{sec:flavor}.

One way to better appreciate the physical meaning of our solution to the $\mu$--$B_\mu$ problem is to consider an alternative model where $H_{u,d}$, like matter,  are partially composite. Let us indicate by $\lambda_{u,d}$  the couplings controlling their partial compositeness and by $\epsilon_{u,d}= \lambda_{u,d}/g_*$
their composite fraction, in full analogy with \eq{def_xi}. Eq.~\eqref{kalH} would then be   modified by  powers of 
$\epsilon_{u,d}$. In particular, $Z_\mu$ would get an overall factor $\epsilon_{u}\epsilon_{d}$. Assuming  $\Delta_\R \approx 2\Delta_\O$, we obtain 
\beq
\frac{\mu^2}{B_\mu} \approx \epsilon_u\epsilon_d \approx
\frac{N\,\lambda_u\lambda_d}{16\pi^2}  \, ,
\label{solpermu}
\eeq
where we have used $g_*=4\pi/\sqrt{N}$.
Equation~(\ref{solpermu}) gives a transparent physical explanation of why the solution to the $\mu$--$B_\mu$ problem works and how it is linked to the composite nature of the Higgs. Only in the case in which the Higgs is fully composite ($\epsilon_{u,d}=1$), the problem is solved. If the Higgs were only partially composite ($\epsilon_{u,d}\ll1$), we would find that $\mu^2/B_\mu$ suffers from the usual loop suppression factor that one encounters in the simplest variants of gauge mediation, in which both $\mu$ and $B_\mu$ are generated at one loop~\cite{Dvali:1996cu}. 

It is useful to interpret our $\mu$--$B_\mu$ solution in the language of warped compactifications. As illustrated in \fig{fig:susy}, the extra-dimensional setup of the Super-Composite Higgs is such that the Higgs fields and the supersymmetry-breaking source are localized on two different branes, one at  position  $z_*=1/\mst$ and the other at $z_S=1/\mS$. This means that the Higgs does not  feel  supersymmetry-breaking head on, but only through  a tail extending to  $z_*$ from $z_S$ through the profile of some auxiliary massive bulk field. Calling $\epsilon_S$ the resulting dimensionless dilution factor, the effective supersymmetry-breaking $F$-term felt by the Higgs fields is measured by $\epsilon_S \mS^2$.
Since two supersymmetry-breaking insertions, evaluated on the Higgs brane, are needed for $B_\mu$ and $m_{1,2}^2$, and only one for $\mu$, we obtain
\beq
\mu \sim   \frac{  \epsilon_S \mS^2 }{ \mst  } \, , ~~~~
B_\mu \sim   \left(\frac{  \epsilon_S \mS^2 }{ \mst  }\right)^2 \, , ~~~~
m_{1,2}^2 \sim   \left(\frac{  \epsilon_S \mS^2 }{ \mst  }\right)^2\, .
\eeq
This manifestly shows the right power counting for a solution to the $\mu$--$B_\mu$ problem and parametrically agrees with eqs.~(\ref{eqhso}) and (\ref{eccobbmu}) upon making the identification $\epsilon_S \sim (\mS/\mst)^{\Delta_\O-1}=(z_*/z_S)^{\Delta_\O-1}$.

In contrast, one of the original gaugino mediation models in flat space 
\cite{Chacko:1999mi} had the Higgs fields delocalized in the bulk, with supersymmetry breaking localized on a brane. In this case,  $B_\mu$ and $\mu$ have similar volume suppressions, hence giving too large $B_\mu / \mu^2$, which is problematic for obtaining correct EW breaking. 
Instead, in ref.~\cite{Okada:2011ed}, both the Higgs fields and supersymmetry breaking are localized at 
the same brane, which gives $B_\mu / \mu^2 \sim 1$, apparently as desired. However, the delocalization of gauginos in the bulk makes their masses volume-suppressed 
with respect to $\mu$, which again is phenomenologically unacceptable.

The solution to the $\mu$--$B_\mu$ problem offered by the Super-Composite Higgs is new with respect to the several other solutions previously proposed in the context of gauge mediation. One class of these solutions relies on the Higgs doublets acting as pseudo-Goldstone bosons of an approximate global symmetry~\cite{Dvali:1996cu}. Others  involve the addition of a gauge singlet superfield, whose vev  generates $\mu$~\cite{deGouvea:1997cx,Chacko:2001km,Delgado:2007rz}, or rely on the logarithmic dependence of the K\"ahler interaction on messenger thresholds~\cite{Giudice:2007ca}, or a hierarchical structure of the Higgs mass soft terms~\cite{Csaki:2008sr,DeSimone:2011va}, or a generalized form of gauge mediation~\cite{Komargodski:2008ax}. All these solutions are viable and interesting possibilities. However, they involve some ad-hoc structure, while the advantage of the Super-Composite Higgs solution is that the dynamical elements required to obtain $\mu^2 \sim B_\mu$ are already built in the theory.

A $\mu$--$B_\mu$ solution close in spirit to that  of the Super-Composite Higgs is found in  the scalar sequestering scenario
of refs.~\cite{Roy:2007nz,Murayama:2007ge,Perez:2008ng}, where $B_\mu$ is set to $0$  in the UV through a dimension gap analogous to $\Delta_\R> 2\Delta_\O$, and then regenerated by IR running. However, the phenomenological consequences of the two approaches are quite different. Scalar-sequestering  is based on  a relatively large mediation scale implying boundary conditions $m_{1,2}^2 ={\tilde m}_{q,\ell}^2=0$ and $B_\mu /\mu^2 =0$. Instead, the Super-Composite Higgs  relies on a low mediation scale and on $\Delta_\R\simeq 2\Delta_\O$ which at the boundary imply  $m_{1,2}^2$, ${\tilde m}_{q,\ell}^2\ne 0$ and $B_\mu /\mu^2 =\O (1)$. The phenomenological implications are obviously quite distinct.

A construction closely related to ours has been proposed in ref.~\cite{Okada:2011ed}. The crucial differences with our approach
concern the $\mu$--$B_\mu$ problem and the possibility to take $m_S$ and $m_*$ arbitrarily large. Loosely speaking, the scenario in ref.~\cite{Okada:2011ed}
corresponds to the case in which $\CFTs$ is neutral under the SM. As a result, the masses  in eqs.~(\ref{softm}), (\ref{eqmu}) and (\ref{eqbmu}) are controlled by the same leading composite operator $\O$, but $\msoft$ is diluted by $g^2/g_*^2$,
\be
\msoft \sim \frac{g^2}{g_*^2} \frac{m_S^{\Delta_\O+1}}{m_*^{\Delta_\O}}\,,\qquad\mu  \sim  c_\mu \frac{m_S^{\Delta_\O+1}}{m_*^{\Delta_\O}}\qquad B_\mu \sim c_{B_\mu}\frac{m_S^{2\Delta_\O+2}}{m_*^{2\Delta_\O}}\,.
\label{okada}
\ee
Obtaining $\mu$ and $B_\mu$ at the right scale requires two different unexplained small numbers $c_\mu \sim g^2/g_*^2$ and $c_{B_\mu}\sim g^4/g_*^4$. Moreover, it does not imply any upper bound on $\mst / \mS$.

\subsubsection*{III. Effects of partial matter compositeness}

The effects described in the last two subsections determine the basic features of the sparticle spectrum and the dynamics of EW symmetry breaking. In other words, they control the phenomenology underlying  direct searches. However, there are additional effects associated with partial compositeness. These do not significantly affect  the mass spectrum 
but crucially control the phenomenology of  indirect searches, as they  violate flavor and CP. The discussion of these effects is a bit more involved as they are not even properly described by soft terms. The reason for that is the relatively low value of the scale of compositeness $m_*$. This implies the phenomenological importance of a plethora of effects that would otherwise decouple when formally taking the limit $m_*\to \infty$ with $\msoft$ fixed.
We will refer to such effects as {\it Beyond Soft}. To be more explicit, we define Beyond Soft any supersymmetry breaking term that decouples in the   formal triple scaling limit $m_*, m_S\to \infty$, $g_{SM}\to 0$ with  the soft masses  in eqs.~(\ref{softm}) and  (\ref{eqmu})--(\ref{eqhso}) held fixed.\footnote{Indeed, as exemplified by eq.~\eqref{hardgauge}, there are  $O(g_{SM}^2/g_*^2)$ hard breakings of supersymmetry below the messenger scale. The breaking thus reduces to purely soft terms in the limit $g_{SM}\to 0$ with $g_{SM}^2 m_S$ fixed.}

Some interesting examples of Beyond Soft are the interactions originating from eqs. (\ref{kal1}) and (\ref{pot1}), which exhibit a subtle interplay between supersymmetry breaking and compositeness. As we shall now see, these interactions lead to renormalizable operators in the low energy EFT, but with coefficients controlled by $\msoft/m_*$ and with structures that do not necessarily correspond to ordinary soft terms.

The full  study of the phenomenological implications  of partial compositeness unavoidably involves the  complexity of flavor physics, and we leave it for \sec{sec:flavor}.  Here  we shall instead limit ourselves to classifying the different operators that are generated upon supersymmetry breaking  from the terms in eqs.~(\ref{kal1})--(\ref{pot1}). This will also render explicit the class of Beyond Soft effects.

Supersymmetry breaking can be distinguished, as before, between one- and two-point function contributions. The effects of the first class are easily worked out by  using \eq{replac} in eqs.~(\ref{kal1})--(\ref{pot1}). 
Sfermion masses receive contributions from the first term in \eq{kal1}, which, upon  using \eq{eqbmu}, become
\beq
{{\tilde m}}^2_{\Phi_{ij}} \approx {c}_{ab}\frac{{\xi}_{\Phi_{ai}}^* {\xi}_{\Phi_{bj}}}{g_*} \left( \frac{\mS}{\mst} \right)^{\Delta_\R} \! \mS^2 
\, \approx\, {c}_{ab} \frac{  {\xi}_{\Phi_{ai}}^* {\xi}_{\Phi_{bj}}\sqrt{N}}{4\pi} \,B_\mu
\, , 
\label{flavs1}
\eeq
where $c_{ab}$ are  $O(1)$ CFT flavor tensors. As they are controlled by $B_\mu$, these are genuine soft masses and with  a pattern analogous to Flavorful Supersymmetry~\cite{Nomura:2007ap}. The structure is not precisely controlled by the Yukawas and induces effects beyond minimal flavor violation. However, the flavor mixings are tamed by two powers of the compositeness fraction $\epsilon_i\equiv \xi_i/\sqrt g_*$, as an additional gift from partial compositeness. This contribution 
is unavoidable in the minimal incarnation of the Super-Composite Higgs, 
since the same universal operators $\O$ and $\R$ contribute to both $Z_S$ and $Z_\mu$. At large $N$, we have   $B_\mu\sim \mu^2$, so that the overall scale of \eq{flavs1} is mainly controlled by the Yukawas ($y\sim \xi^2$), by the higgsino mass $\mu$, and mildly by $N$.

When operators with the  suitable SM quantum numbers exist, also the two-point function of $Z_{1K}$ in \eq{kal1} affects
sfermion masses. These effects  are saturated by intermediate states  of  virtuality $\sim \mS$. Focusing on the contribution of ${\cal O}_{1K}$, simple power counting gives
\beq
{\tilde m}^2_{\Phi_{ij}} \approx  {b}_{ab}\frac{{\xi}_{\Phi_{ai}}^* {\xi}_{\Phi_{bj}}}{g_*}\left( \frac{\mS}{\mst} \right)^{2(\Delta_{\O_{1K}}-1)} \! \mS^2 
\, 
\label{flavs2}
\eeq 
where, again, $b_{ab}$ are generic matrices with order-one entries.
Exchange of ${\cal R}_{{1K}}$ gives the same result with $\Delta_{\O_{{1K}}}\to \Delta_{\R_{{1K}}}$. These effects have precisely the same form as \eq{flavs1}. However, while before $\Delta_{\R}$ was controlled by $B_\mu$, now $\Delta_{\O_{1k}}$ and $ \Delta_{\R_{1k}}$
could be in principle large enough to make \eq{flavs2} subdominant.
That will henceforth be our assumption. Similar considerations also apply to the terms generated by the two point function of $\O_1$ in \eq{pot1}. 

Let us now consider the trilinear terms. Holomorphic $A$-terms
are generated by the vevs of $Z_S$, $Z_y$ in \eq{kal1} and $\O_3$ in \eq{pot1}, each giving  distinct effects.
The vev of $Z_S$, upon using \eq{eqmu}, gives
\beq
A_{I} \sim \frac{y_{\raisebox{-1.5pt}{$\scriptstyle I$}}}{g_*} \left( 
\xi^*_{\Phi_{L}} \xi_{\Phi_{L}} +
\xi^*_{\Phi_{\bar R}}\xi_{\Phi_{\bar R}} \right) 
\mu \, ,~~~~~~I=u,d,\ell \, .
\label{Apartc}
\eeq
This contribution  is  flavor non-universal and 
genuinely soft, as it is purely controlled by $\mu$.   However it is suppressed by four powers of $\xi$, which greatly reduces its relevance, even with respect to higher dimensional operators. 

Next, consider the effect of $Z_y$ in \eq{kal1}. Using \eq {eqbmu}, we find
\beq
A_{I} \sim 
 \xi_{\Phi_{L}} \xi_{\Phi_{\bar R}}\, \frac{B_\mu}{\mst} \, ,~~~~~~I=u,d,\ell \, .
\label{Apartc2}
\eeq
This is an instance of Beyond Soft effect, as it decouples in the  limit $\mst\to \infty$ with $B_\mu$ fixed. Yet, in our scenario \eq{ratformu}, with the unitarity condition $\Delta_\O\geq 1$,  does not permit to take that limit. The consequence is that  \eq{Apartc2}, which involves only two powers of $\xi$,  is more important than the soft contribution in \eq{Apartc}, which is $O(\xi^4)$.

Notice in passing that the $\bar \theta^2$ component of $Z_y$ produces  an extra contribution to the Yukawas in \eq{eccow},  with a suppression $\sim \mu/\mst$, but with the same structure. This effect has no practical consequences.

Finally, consider the effects of $\O_3$ in \eq{pot1}. In our minimal scenario where $\O_3=\O$,  \eq{ratformu} implies
\beq
A_{I} \sim 
 \xi_{\Phi_{L}} \xi_{\Phi_{\bar R}}\, \mu \, ,~~~~~~I=u,d,\ell \, .
\label{Apartc3}
\eeq
This is a genuine soft and flavor/CP breaking trilinear, which would place significant lower bounds on the sparticle masses. Holomorphy and non-renormalization theorems, however, render non-compulsory the presence of the parent term in \eq{pot1}. We will thus work under the consistent hypothesis that this term is absent.

The effects we discussed so far have the same operator structure as standard soft terms (sfermion masses and A-terms), even though \eq{Apartc2} already indicates the presence of Beyond Soft, as it strictly decouples for $\mst\to \infty$ with $\msoft$ fixed.
The terms  with coefficient $Z_{\hat y}$ in \eq{kal1}, however, generate relevant Beyond Soft effects structurally different from soft terms, as we now turn to discuss.

To analyze the effects of the non-holomorphic trilinear terms, let us focus for simplicity only on the leptonic part, as the extension to $u$ and $d$ quarks is completely straightforward. For completeness, we reinstate here the generation indices. The relevant K\"ahler term in \eq{kal1} is
\beq
{\cal K} = \frac{\xi_{L_{ai}}\xi_{{\bar E}_{bj}}}{\mst}\, H_u^\dagger L_i {\bar E}_jZ_{{\hat y}_{ab}}\, ,
\label{Klep}
\eeq
with the replacement implied by supersymmetry breaking
\beq
Z_{{\hat y}_{ab}}\to\frac{\mu}{\mS}\left( c_{ab}X+{\tilde c}_{ab}X^\dagger \right) + \frac{B_\mu}{\mS^2}\, {\hat c}_{ab} XX^\dagger \, , ~~~~~~X\to \mS\, \theta^2 \, .
\eeq
The effects of the insertion linear in $X$ in \eq{Klep} (originating from the chiral operator $\O_{\hat y}$) are best understood by making the field redefinition
\beq
H_u \to H_u -\frac{\mu}{\mst}\, c_{ab}\, \xi_{L_{ai}}\xi_{{\bar E}_{bj}}L_i {\bar E}_j\, \frac{X}{\mS} \, ,
\eeq
which shifts the leading contribution to the superpotential. After supersymmetry breaking, this yields two effects. One is a trilinear term 
\beq
{A}_{\ell_{ij}}\, H_d \, \scalar{L_i} \scalar{{\bar R}_j} \, ,
~~~~~~\textrm{with}~~A_{\ell_{ij}}=c_{ab}\, \xi_{L_{ai}}\xi_{{\bar E}_{bj}}\, \frac{\mu^2}{\mst} \, ,
\label{holhol}
\eeq
which is parametrically equal to \eq{Apartc2}. Here and in the following, we define $\scalar{\Phi}$ as the scalar component of the chiral superfield $\Phi$. The second effect is the quartic scalar interaction
\beq
\frac{\mu}{\mst}\, c_{ab}\, \xi_{L_{ai}}\xi_{{\bar E}_{bj}}y_{u_{k\ell}}\, \scalar{L_i}\scalar{{\bar E}_j} \scalar{Q_k} \scalar{{\bar U}_\ell} \, .
\eeq
This new scalar interaction has limited phenomenological relevance because its coupling constant is highly suppressed, being $O(\xi^4\mu/\mst)$.

The $XX^\dagger$ insertion in \eq{Klep} (originating from the real operator ${\cal R}_{\hat y}$) gives a non-holomorphic scalar trilinear interaction
\beq
{\tilde A}_{\ell_{ij}}\, H_u^\dagger \scalar{L_i} \scalar{{\bar E}_j} \, ,
~~~~~~\textrm{with}~~
{\tilde A}_{\ell_{ij}}={\hat c}_{ab}\, \xi_{L_{ai}}\xi_{{\bar E}_{bj}}\, \frac{\mu^2}{\mst} \, ,
\label{nonhol}
\eeq
which has the same structure as its holomorphic counterpart in \eq{holhol}, but cannot be assimilated by any ordinary soft term.

The $X^\dagger$ insertion in \eq{Klep} (originating from the chiral operator $\O^\dagger_{\hat y}$) leads to quartic scalar interactions
\beq
\frac{\mu}{\mst}\, {\tilde c}_{ab} \xi_{L_{ai}}\xi_{{\bar E}_{bj}}\, 
H_u^\dagger H_d^* \left( y^*_{\ell_{kj}} 
\scalar{L_k}^* \scalar{L_i} 
+ y^*_{\ell_{ik}} 
\scalar{{\bar E}_k}^*
\scalar{{\bar E}_j}
\right) \, ,
\eeq
whose coupling constant is, again, $O(\xi^4\mu/\mst)$ suppressed. More importantly, the $X^\dagger$ insertion leads to a modified Yukawa interaction, which includes a ``wrong-Higgs" contribution
\beq
{\bar e}_{R_j} \ell_{L_i} \left( y_{\ell_{ij}}H_d+ \frac{\mu}{\mst}\, {\tilde c}_{ab} \xi_{L_{ai}}\xi_{{\bar E}_{bj}}\, 
H_u^\dagger \right) \, .
\label{wronghiggs}
\eeq
Eqs.\eqref{nonhol} and \eqref{wronghiggs} are examples of phenomenologically relevant Beyond Soft effects that have no counterpart among ordinary soft terms.
As we will discuss in \sec{sec:flavor}, the ``wrong-Higgs" Yukawa interaction, if present, gives the strongest constraint on the compositeness scale $\mst$.

\subsection{Supersymmetric particle masses in Super-Composite Higgs}

\renewcommand{\arraystretch}{1.5}
\begin{table}[t]
\centering
\begin{tabular}{|c|c|c|c|c|}
\hline
\multicolumn{2}{|c|}{\bf Supersymmetric particle} & I&II&III  \vspace{-0.3cm}\\
\multicolumn{2}{|c|}{\bf mass contributions in}  & Gauge  & Higgs  & Partial matter  \vspace{-0.3cm} \\ 
\multicolumn{2}{|c|}{\bf Super-Composite Higgs} & mediation & compositeness  &  compositeness \\
\hline
gaugino masses& ${M_{\lambda}}/{\mS}$ & $\frac{g^2\, N}{16 \pi^2}$ & {}&\\
\hline
squark \& slepton masses & ${{\tilde m}_\Phi^2}/{\mS^2}$ & $\frac{g^4\, C_\Phi N}{(16 \pi^2)^2}${\footnotesize$\, \id$}&{}&$\frac{\xi_\Phi^*\xi_\Phi \sqrt{N}}{4\pi} \big( \frac{\mS}{\mst}\big)^{\Delta_\R}$ \\ \hline
Higgs masses & $m_{1,2}^2/\mS^2$ & $\frac{g^4\, C_{H} N}{(16 \pi^2)^2}$&$\big( \frac{\mS}{\mst}\big)^{\Delta_\R}$&{} \\ \hline
$\mu$ term & $\mu /\mS$ & {}&$\big( \frac{\mS}{\mst}\big)^{\Delta_\O}$&{} \\ \hline
$B_\mu$ term & $B_\mu /\mS^2$ & {}&$\big( \frac{\mS}{\mst}\big)^{\Delta_\R}$&{} \\ 
 \hline
 holomorphic trilinears  & $A/\mS$ & {}&$y\, \big( \frac{\mS}{\mst}\big)^{\Delta_\O}$ & {\footnotesize $\xi_{\Phi_L} \xi_{\Phi_{\bar R}}$}$\, \big( \frac{\mS}{\mst}\big)^{\Delta_\R +1}$\\ \hline
non-holomorphic trilinears  & $\tilde A/\mS$ & {}&{} & {\footnotesize $\xi_{\Phi_L} \xi_{\Phi_{\bar R}}$}$\, \big( \frac{\mS}{\mst}\big)^{\Delta_\R +1}$\\ \hline
\end{tabular}
\renewcommand{\arraystretch}{1}
\caption{Parametric dependence of the contributions to supersymmetric particle masses in the Super-Composite Higgs in units of the scale $\mS$. Here $C_{\Phi ,H}$ denote the Casimir of the gauge representations for the matter superfields $\Phi$ and Higgs superfields $H_{u,d}$, while $\id$ is the identity matrix in generation space. Because of strong-dynamics effects, each contribution is implicitly multiplied by an unknown order-one coefficient.}
\label{tab:soft}
\end{table}

For the ease of the reader, we summarize in table~\ref{tab:soft} the main  contributions to the supersymmetric particle masses from effects of class I, II and III, as defined  in \sec{sec:fromstosoft}. In the table we have kept generic $\Delta_\O$ and $\Delta_\R$ for a quicker identification of the effect's origin. However, to study the phenomenology we will specify to $\Delta_\R =2 \Delta_\O+O(1/N) \approx 2 \Delta_\O$.

 Here we comment on each line of table~\ref{tab:soft}.

\begin{itemize}
\item 
Gaugino masses are  produced by a variant of gauge mediation via the current two-point function in the $\CFTs$ sector.  Equivalently, their origin can be traced back to the mixing between gauginos and  composite  adjoint fermions, in full analogy with the origin of Yukawa couplings in partial compositeness.  The 5D description of this mechanism is instead just a variant of gaugino mediation. Since gaugino masses describe the primary source of supersymmetry breaking in the SM sector, they establish the typical soft scale, denoted as $\msoft$.

\item Squark and slepton soft masses receive contributions of class I and III.
Those belonging to class I, see \fig{fig:feynCFTs}, consist of
a one-loop dressing of the two-point function  of the $\CFTs$ gauge current. The result is flavor universal, as befits gauge interactions. Moreover, as shown in \eq{softm} after identifying  $g_*\sim 4\pi/\sqrt N$, this contribution, along with gaugino masses,
parametrically reproduces the result of gauge mediated supersymmetry breaking with $N$ families of messengers.

The contribution of class III
is associated with the expectation value $\langle \R\rangle$ of a $\CFTs$ operator and is suppressed by $(\mS /\mst)^{\Delta_\R} $. Although subleading in size, this contribution is significant because, unlike the gauge mediated one, it does not respect minimal flavor violation. Consequently, it can lead to observable effects in flavor physics. This phenomenon is a realization of the scheme known as flavorful supersymmetry~\cite{Nomura:2007ap}.

\item The  Higgs soft masses receive contributions of class I and II. The second kind is controlled by the same ratio $(\mS /\mst)^{\Delta_\R} $ that controls flavorful corrections to sfermion masses.

\item The $\mu$ term is an effect of Higgs compositeness and comes entirely from the expectation value $\langle \O \rangle$ of a $\CFTs$ chiral operator. For  $\mu$ to be of order $\msoft$ a numerical coincidence is required, which we cannot justify dynamically: $(\mS/\mst)^{\Delta_\O}$ must be roughly equal to a SM gauge loop factor, see \eq{ratformu}. However, once we accept this as a working hypothesis,  we can derive the interesting conclusion that the compositeness scale must be somewhat above the supersymmetry-breaking scale. This is welcome because a certain separation of scales helps taming the strong flavor constraints, which pose the main difficulty for traditional compositeness. Moreover, since unitarity requires ${\Delta_\O}\geq 1$, this separation cannot be excessively large, so that compositeness remains a relatively low-energy phenomenon, meaning that its effects could potentially be observed in flavor experiments.

\item $B_\mu$ is generated by the same dynamics that produces $\mu$. This often leads to the problematic relation $B_\mu / \mu = \mS$, known as the
 $\mu$--$B_\mu$ problem that plagues nearly all approaches for dynamically generating these parameters. Interestingly, the Super-Composite Higgs, coupled to a moderately large $N$, provides a novel and surprisingly straightforward dynamical explanation for the issue. Indeed the leading large $N$ expectation $\Delta_\R =2 \Delta_\O+O(1/N) \approx 2 \Delta_\O$, naturally implies the successful relation $B_\mu \sim \mu^2$, see \eq{eccobbmu}. 

\item Holomorphic trilinear  terms receive contributions of class II and III. Those belonging to class II, purely due to Higgs compositeness, satisfy minimal flavor violation, since they are proportional to the corresponding Yukawa coupling. They are sizable, as they are controlled by $\mu$. The existence of sizable trilinear couplings marks a significant departure from gauge mediation, where $A$-terms are loop suppressed and negligible. This distinction is crucial because, as we will explain in \sec{sec:Minimal}, having stop $A$-terms of order $\msoft$ is important for the Super-Composite Higgs to reproduce the physical Higgs mass, while maintaining a particle spectrum within current or future collider searches.

The class III contributions present instead a flavorful structure, but are 
suppressed with respect to the class II universal contributions by a $B_\mu/(\mu \mst)$ factor. Yet, they remain important because they violate flavor.

\item Non-holomorphic trilinear  terms arise as class  III effects. Their structure parallels precisely that of the  flavorful  holomorphic trilinears we just mentioned above. 
It must be added that class III  trilinears, with their extra suppression by $\msoft/\mst$ fall into a class of non-soft effects that will be discussed in \sec{sec:sourcesFCP}.

\end{itemize}

\boldmath
\section{Super-Composite Higgs: parameters and mass scales}
\unboldmath
\label{sec:Minimal}

\subsection{Input parameters}
\label{sec:param}

Now that we have established the structure of the theory and the form of the soft terms, we are ready to present its features and phenomenological implications. The goal of this section is to discuss the role of the input parameters that describe the theory. 

In building the Super-Composite Higgs model, we have encountered two fundamental mass scales (the compositeness scale $\mst$ and the scale of supersymmetry breaking $\mS$) and two fundamental quantities associated with the $\CFTs$ sector (the index $N$ and $\Delta_\O$, the scaling dimension of the chiral operator whose supersymmetry-breaking vev is responsible for the generation of $\mu$). These are the structural parameters of our theory. The phenomenology is also controlled by two derived
scales (the overall scale of sparticle masses $\msoft$ and the $\mu$ parameter) and  by one quantity associated with the Higgs sector ($\tan\beta$). These 7 parameters are not independent because of several constraints intrinsic to the theory. Let us analyze each of these quantities, studying their ranges of variability and their mutual relations.

\boldmath
\subsubsection*{$\bullet$ $\msoft$}
\unboldmath

The gauge-mediation mass spectrum is quite broad. Therefore, to proceed with the discussion, we must specify what we mean precisely by $\msoft$. The most convenient choice is to take the squark mass ${\tilde m}_q$ as input. This is because the squark mass is a central parameter to determine {\it (i)} the Higgs mass $m_h$, {\it (ii)} the EW breaking condition, {\it (iii)} the experimental observability at hadron colliders, since squarks are the lightest colored sparticles, especially at large $N$. Once we fix ${\tilde m}_q$, the entire sparticle spectrum is determined, according to gauge mediation, up to order-one coefficients.

In the following, we will use ${\tilde m}_q =$ 3 and 10 TeV, as our reference points. Using exact gauge-mediation mass relations, the choice ${\tilde m}_q =$ 3 TeV is already ruled out (for $N>5$) by experimental limits on slepton masses, as we will discuss in \sec{sec:cosmo}, but could be saved by relatively small modifications of the mass spectrum due to strong-dynamics effects. Moreover, this choice requires a certain adjustment among parameters to obtain the correct Higgs mass. Altogether, ${\tilde m}_q \sim $ 3 TeV roughly characterizes the lower bound of the mass spectrum. The choice ${\tilde m}_q = $ 10 TeV is representative of a heavier spectrum, which more easily satisfies the various phenomenological constraints.
 
 \boldmath
\subsubsection*{$\bullet$ $\mS$}
\unboldmath

Once ${\tilde m}_q$ is fixed, the scale of supersymmetry breaking $\mS$ is no longer a free parameter, and we determine it from the gauge-mediation mass relation,
\beq
{\tilde m}_{ q_{\rm (GM)}}^{2} = 2\,\mS^2\sum_{A=1}^3 \frac{C_A \,\alpha_A^2(m_S)\, N}{(4\pi)^2}\left[ 1+\frac{N}{b_A}\left( 1-\frac{\alpha_A^2({\tilde m}_q)}{\alpha_A^2(\mS)}\right)\right] \, ,
\label{eqq2}
\eeq
up to incalculable factors describing strong-dynamics effects. In \eq{eqq2}, the gauge coupling constant have GUT normalization, $C_A=(N^2-1)/2N$ for SU($N$) gauge groups, $C_1=(3/5)\,Y^2$ for U(1)$_Y$. Finally,
\beq
\frac{\alpha_A(\mS)}{\alpha_A({\tilde m}_q)}=1+\frac{b_A\,\alpha_A(\mS)}{2\pi}\, \ln \frac{\mS}{{\tilde m}_q}\, ,~~~~b=(33/5,1,-3) \, .
\label{betafun}
\eeq

 \boldmath
\subsubsection*{$\bullet$ $\mu$}
\unboldmath

The Higgs $\mu$ parameter originates from Higgs compositeness, and not from gauge mediation, but is tied to $\msoft$ by the condition of reproducing the Higgs vacuum expectation value $v$. Using the notations of \eq{defm12} for the Higgs mass parameters, the conditions for EW breaking are
\beq
m_1^2 \approx m_2^2 \tan^2\beta \, ,~~~~
\tan\beta + \frac{1}{\tan\beta} = \frac{m_1^2+m_2^2}{B_\mu}\, ,~~~~
m_1^2 + m_2^2 > 2|B_\mu | \, .
\label{condb}
\eeq
In the first equation in (\ref{condb}), we have dropped terms $\O (M_Z^2/\msoft^2)$, which are completely negligible in our context. The inequality in (\ref{condb}) is the condition for the scalar potential to be bounded from below which, with the help of the two equations in (\ref{condb}) can be simply rewritten as $m_1^2 >0$, irrespectively of the value of $\tan\beta$. In the limit of large $\tan\beta$, the EW breaking condition is simply $m_2^2\approx 0$.

Under the simplifying assumption of omitting the contributions from compositeness to the Higgs soft masses and the coefficient $A$ of the top trilinear term, the EW breaking condition can be approximated as
\beq
\mu^2 \approx \frac{3 \,y_t^2\, {\tilde m}_t^2}{4\pi^2}\,\ln \frac{\mS}{{\tilde m}_t} 
-{\tilde m}^2_{H_u}
\, ,
\label{EEWW}
\eeq
where $y_t$ is the top quark Yukawa and ${\tilde m}^2_{H_u}$ is the gauge-mediation expression for the Higgs soft mass. In the spirit of this paper, which aims at deriving only order-of-magnitude estimates, we will use \eq{EEWW} as a proxy for EW breaking, encapsulating all theoretical uncertainties in a coefficient that multiplies $\mu$. A more realistic analysis may lead to significant modifications of \eq{EEWW}, though at the price of introducing further unknown coefficients.

\boldmath
\subsubsection*{$\bullet$ $\tan \beta$}
\unboldmath

The ratio of Higgs vevs $\tan\beta = \langle H^0_u\rangle / \langle H^0_d\rangle$ plays a crucial role in the determination of the physical Higgs mass $m_h$. In minimal low-energy supersymmetry with large $\tan\beta$, the measured value $m_h=$ 125 GeV is obtained for stop masses of about 2 TeV, when the stop mixing parameter $X_t \equiv (A_t/y_t - \mu/\tan \beta )/{\tilde m}_t$ reaches a critical value
of about 2. For $X_t \approx 1$, the Higgs mass requires stop masses in the range 6--8 TeV and, for vanishing $X_t$, stop masses in the range 10--14 TeV~\cite{Slavich:2020zjv}. The result is very sensitive to $\tan\beta$ because any reduction of the tree-level contribution to $m_h$ must be compensated by an increase of the quantum correction, which grows only logarithmically with ${\tilde m}_t$. For instance, for $\tan\beta =5$, one needs stop masses of about 6 TeV (for the critical value of $X_t$),  20 TeV (for $X_t \approx 1$), and 40 TeV (for $X_t \approx 0$).

In order to keep the supersymmetric spectrum within experimental reach, it is beneficial to have a large, or moderately large, $\tan\beta$. Therefore, we assume this condition to hold, although we don't have sufficient theoretical control over the Higgs potential to compute $\tan\beta$. Having a relatively large $\tan\beta$ requires a tuning of parameters at the level of $1/\tan\beta$. This is a generic problem of low-energy supersymmetric models and, in this respect, there is nothing new in the Super-Composite Higgs. 

The novelty of the Super-Composite Higgs, with respect to gauge mediation, is the presence of $A$-terms at the matching scale $\mS$, which are helpful for the Higgs mass as they can lead to $X_t$ of order unity, and possibly close to the critical value. Since the $A$-terms are generated from Higgs compositeness, they are expected to be $O(\mu)$. Using \eq{EEWW} to relate $\mu$ to the stop mass, we would find a suppressed contribution to $X_t$. However, the inclusion of the Higgs soft masses from compositeness could easily modify \eq{EEWW} making $\mu \sim A \sim {\tilde m}_t$, which leads to $X_t = O(1)$. Furthermore, the soft parameter $A$ receives an additive one-loop quantum correction $\Delta A$ proportional to the gluino mass, whose leading-log expression is
\beq
\Delta A = \frac{8\,\alpha_sM_{\tilde g}}{3\pi}\, \ln \frac{\mS}{M_{\tilde g}} \, .
\eeq
This can give a sizable contribution to $X_t$, especially at large $N$.

In conclusion, the Super-Composite Higgs can account for the correct Higgs mass, although it requires a favorable adjustment among various unknown parameters and incalculable coefficients from strongly-coupled dynamics.

\begin{figure}[t]
\begin{center}
\includegraphics[width=0.6\columnwidth]{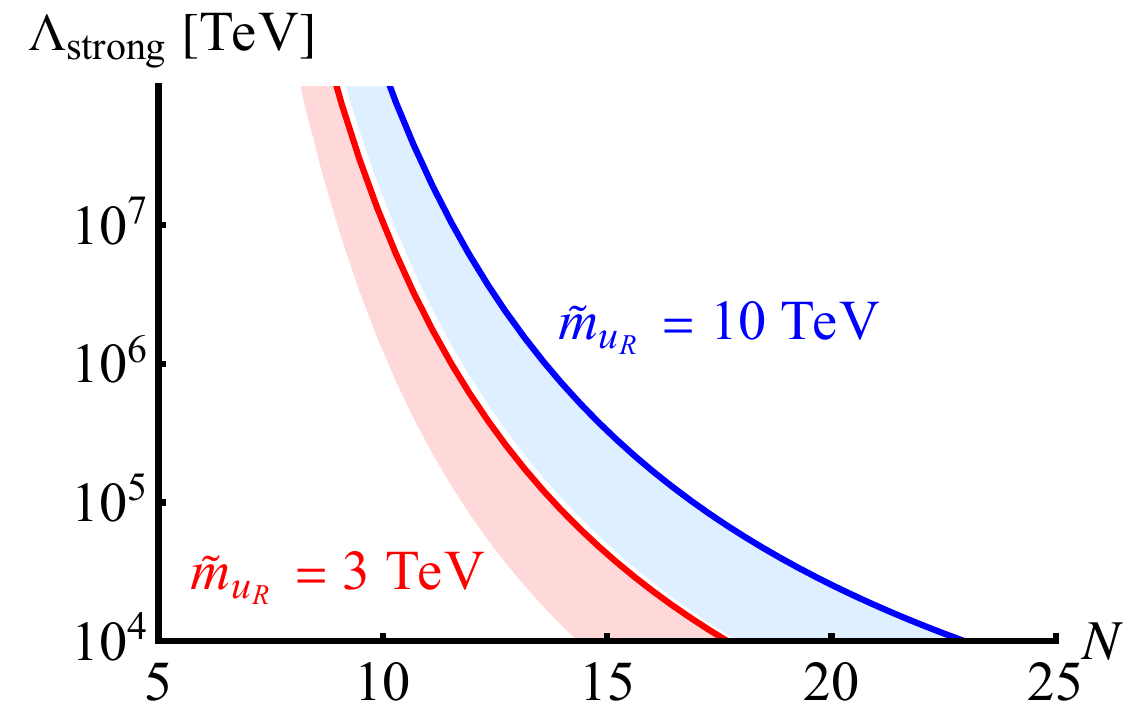}
\end{center}
\caption{The energy scale $\Lambda_{\textrm{strong}}$ at which the QCD gauge coupling becomes comparable to $g_*$, based on a one-loop RG analysis. The solid lines refer to the condition $g_3 (\Lambda_{\textrm{strong}}) =g_*$, where $g_3$ is the QCD coupling constant and $g_*=4\pi/\sqrt{N}$. The colored bands correspond to parameter regions in which $\nicefrac12 <g_3/g_*<1$.}
\label{fig:landau}
\end{figure}

 \boldmath
\subsubsection*{$\bullet$ $N$}
\unboldmath

The index $N$ is a parameter pertaining to the theory describing the $\CFTs$ sector, which is defined through its relation with the strong coupling constant $g_* =4\pi/\sqrt{N}$ (see the discussion in section \ref{sec:allthat}). To model the $\CFTs$ sector, we have focused on the regime of large $N$. There are several reasons to prefer large $N$: {\it (i)} it allows  for a weakly coupled holographic realization 
with coupling $g_*\sim 4\pi/\sqrt{N}$; {\it (ii)} it seems more likely to support  the complex dynamical  requirements of the composite sector, since there are probably very few theories at small $N$; {\it (iii)} it adds some quantitative control, like leading-order factorization and additivity of operator dimensions, and can consistently allow for parameter choices that help the plausibility of the model, for instance, as those discussed in \sec{sec:robust}.

While we choose $N$ to be large, we have to respect the upper bound on $N$ coming from the requirement that the QCD gauge coupling doesn't prematurely exceed $g_*$, or else the SM becomes part of the strongly-coupled sector. To estimate this bound, we evolve SM gauge couplings with one-loop RG running, under the simplified assumption that the QCD $\beta$-function above the scale $\mS$ is such that $b=b_{3} +N$, where $b_{3}=-3$ is the $\beta$-function coefficient of the minimal supersymmetric model, following the notations of \eq{betafun}. Generically, the QCD $\beta$-function is further modified above the threshold $\mst$ but, for simplicity, we assume that $b$ remains the same. We also emphasize that the value of $N$ that appears in the $\beta$-function, counting the number of colored degrees of freedom in the $\CFTs$ sector, is related to the index $N$ that appears in the strong coupling $g_*$ but is not necessarily equal. An unknown order-one factor between the two definitions of $N$ is implicit.

Our results are presented in \fig{fig:landau}, where we have expressed $\mS$ in terms of the squark mass, using the gauge-mediation relation in \eq{eqq2}. In the figure, the solid lines refer to the condition $g_3 (\Lambda_{\textrm{strong}}) =4 \pi/\sqrt{N}$, with $g_3$ being the QCD coupling constant.
 The colored bands correspond to regions in which $\nicefrac12 <g_3/g_*<1$ and give an indication of the theoretical uncertainty in the bound. The regions to the right of the solid lines correspond to situations in which the SM is embedded in the strongly-coupled sector and our analysis is no longer valid. Requiring that SM gauge interactions remain in a controllable perturbative regime up to tens of PeV implies that $N$ cannot exceed about 15. Therefore, our preferred range for $N$ is
\beq
3\lsim N \lsim 15 \, .
\label{N515}
\eeq

\boldmath
\subsubsection*{$\bullet$ $\Delta_\O$}
\unboldmath

The scaling dimension of the supersymmetry-breaking chiral operator generating $\mu$ is a critical parameter because it enters \eq{eqmu} exponentially. In a supersymmetric CFT, unitarity constrains the scaling dimension of a chiral scalar primary to be $\geq 1$, with the lower bound corresponding to a free field. We will thus impose $\Delta_\O >1$.
Theoretical consistency does not impose a strict upper bound on $\Delta_\O $, but  structural plausibility does. Indeed the gapping of $\CFTs$  at $m_S$ must arise from a relevant or approximately marginal deformation. The only possibility for that to happen is the existence of a chiral scalar primary $\O_{gap}$ with dimension $\lsim 3$. In view of that, if we had  $\Delta_\O >3$,
the dominance of $\O$ in the mediation of supersymmetry breaking effects could be achieved only with the assumption that all couplings to $\O_{gap}$ are suppressed compared to those to $\O$. In the absence of additional control symmetries, which is our assumption in the minimal construction, this would be unnatural. We thus conclude that\footnote{$\Delta_\O\gsim 2$ is also  in principle another critical threshold. That is due to the existence of vector superfields of the form $J^2$, with  $J$  any SM gauge current operator in the CFT. In the large-$N$ limit, the operators $J^2$ have dimensions $4+O(1/N)$. For $\Delta_\O>2$, these operators have dimension lower than   $\R=\O\O^\dagger$, and in principle can dominate the generation of $B_\mu$, ruining our solution to the $\mu$--$B_\mu$ problem. However, at leading order in the $1/N$ expansion (or equivalently at tree level in the AdS$_5$ description), the vev of $J^2$ factorizes and, since $\langle J\rangle =0$, the vev of $J^2$ must be suppressed by some power of $1/N$. This makes the effects of this class of operators negligible.} 
\beq
1<\Delta_\O \lsim 3 \, .
\label{D12}
\eeq

As we will show in the following, see \eq{stima2}, for $\Delta_\O$ much larger than 1, the ratio $\mS/\mst$ rapidly converges towards 1. 
Therefore, the exact numerical choice for the upper bound on $\Delta_\O$ is unimportant for our considerations. Just for illustrative purposes,  in our plots we will often take $\Delta_\O \le 2$.

\boldmath
\subsubsection*{$\bullet$ $\mst$}
\unboldmath

Using \eq{eqmu}, we can determine $\mst$ as
\beq
\mst \approx \mS\left( \frac{ \mS}{\mu} \right)^{1/\Delta_{\cal O}} \, ,
\label{uffamu}
\eeq
up to an order-one unknown factor. Since both $\mS$ and $\mu$ can be estimated from ${\tilde m}_q$, we infer that the compositeness scale is calculable, up to an order-one coefficient, with $\Delta_\O$ and $N$ as the only unknowns. The remarkable result is that the unitarity bound $\Delta_\O >1$ sets an absolute upper limit on $\mst$ given by
\beq
\mst \lsim \frac{ \mS^2}{\mu} \, ,
\label{unitbou}
\eeq
which is calculable in terms of ${\tilde m}_q$ and $N$. We will present the numerical estimate of this bound in \sec{sec:masss}. Here we only want to emphasize the conceptual importance of the result. In the Super-Composite Higgs, $\mst$ is not directly related to the Higgs naturalness problem. Nonetheless, the theoretical structure imposes a rigid upper bound on the compositeness scale.

\bigskip

In summary, the physical requirements of reproducing the Higgs mass $m_h$ and vev $v$, together with the constraints inherent in our theoretical construction, impose several relations among the parameters of the Super-Composite Higgs, such that only 3 parameters remain independent. In our analysis, we will use ${\tilde m}_q$, $N$ and $\Delta_\O$ as independent parameters, with their ranges of variability limited by eqs.~(\ref{N515})--(\ref{D12}). Once a choice for these 3 parameters is made, everything else in the theory is calculable, up to unavoidable order-one uncertainties.

\subsection{The spectrum of mass scales}
\label{sec:masss}

\begin{figure}[t]
\begin{center}
\includegraphics[width=0.49\columnwidth]{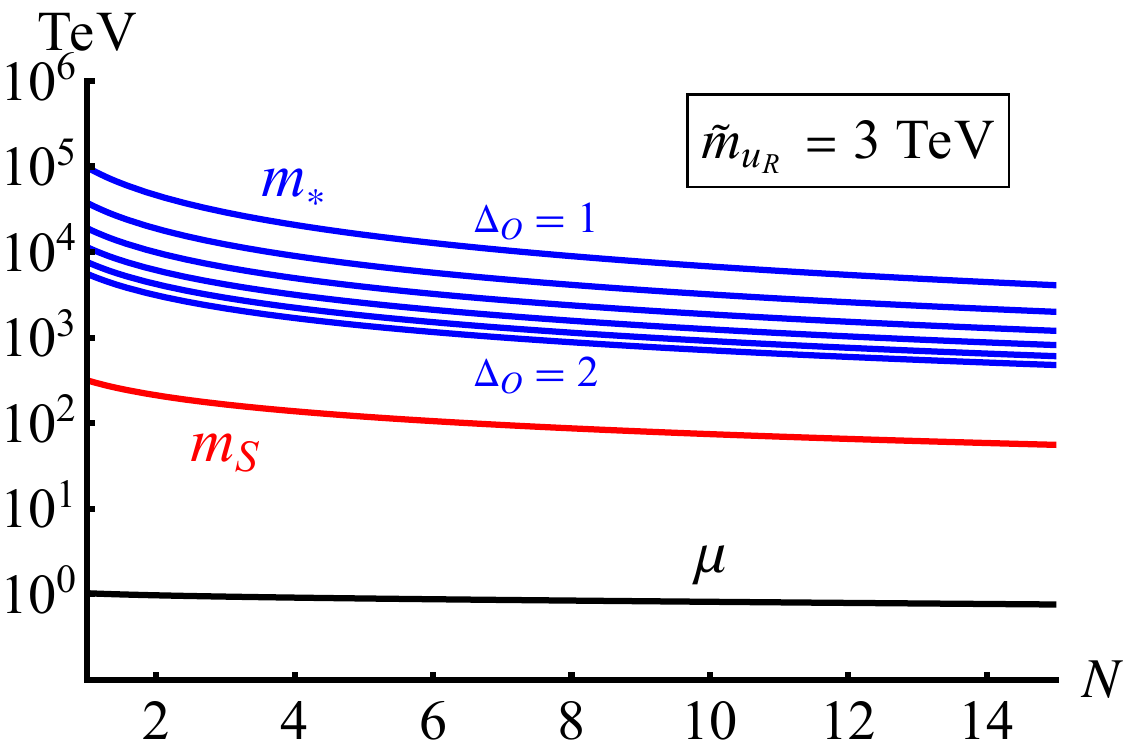}\hfill
\includegraphics[width=0.49\columnwidth]{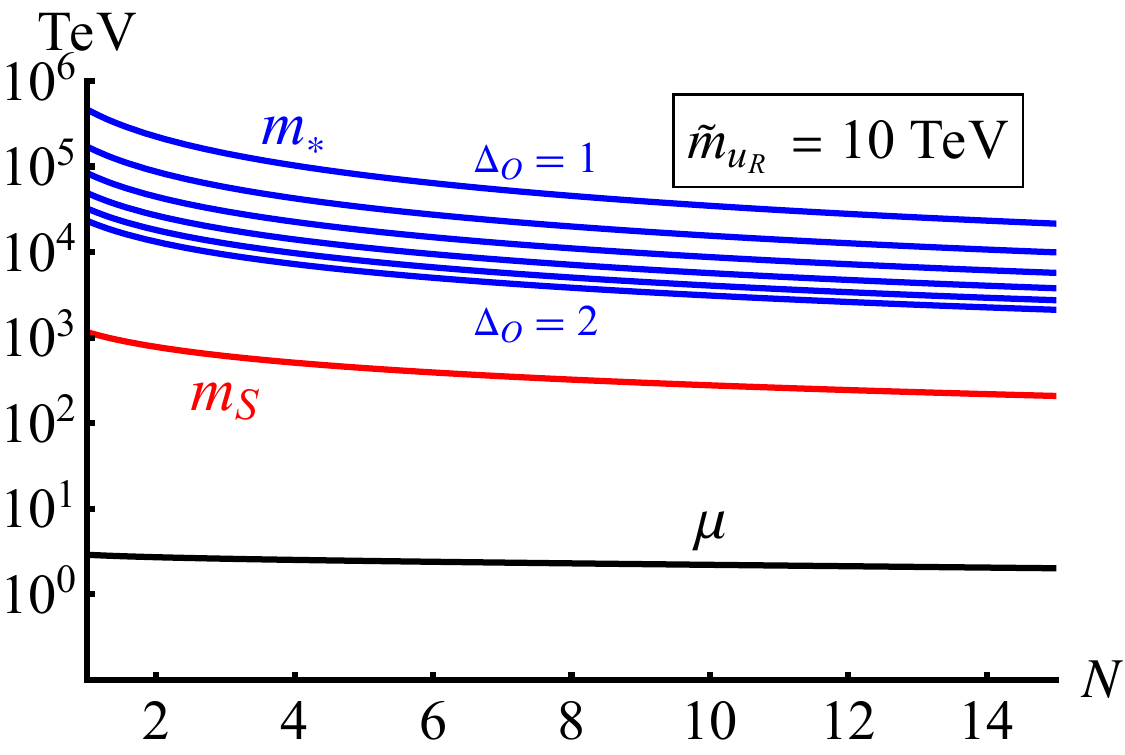}
\end{center}
\caption{The spectrum of mass scales in the Super-Composite Higgs, based on a one-loop RG improved calculation with gauge mediation mass relations and radiative EW breaking. The blue lines show the predictions for $\mst$ corresponding to $\Delta_\O=$ 1, 1.2, 1.4, 1.6, 1.8, 2 (from top to bottom).}
\label{fig:mu}
\end{figure}

In \fig{fig:mu} we show the Super-Composite Higgs prediction for $\mu$, $\mS$ and $\mst$ in terms of ${\tilde m}_q$, $N$ and $\Delta_\O$, following the logic explained in \sec{sec:param}. The calculation is based on gauge mediation mass relations with one-loop RG evolution and radiative EW breaking. 

A clear hierarchical pattern emerges with $\mS$ between 50 and 500 TeV, and $\mst$ between 1 and 100 PeV, with order-one uncertainties. The dependence of $\mS$ on $N$ is relatively mild since, as $N$ varies from 3 to 15, $\mS$ decreases only by a factor of 3, which is within the theoretical uncertainty. The compositeness scale $\mst$ has a more pronounced dependence on $N$, especially for $\Delta_\O$ close to 1. More significant is the dependence on $\Delta_\O$ (which is exponential) since $\mst$ changes by one order of magnitude, as $\Delta_\O$ ranges from 1 to 2. 

The most important result is the upper limit on $\mst$, given in \eq{unitbou}, which is obtained from the unitarity bound $\Delta_\O>1$. The choice $\Delta_\O \approx 1$ is not realistic, since it corresponds to a nearly free field, but it acts as a limiting case of plausible CFTs. The upper bound on $\mst$ depends on ${\tilde m}_q$ and $N$ and we find that $\mst$ cannot exceed 140 PeV (20 PeV) for $N=3$ ($N=15$), when ${\tilde m}_q=10$ TeV, without accounting for the uncertainties due to strong dynamics. This result has direct phenomenological consequences because it implies that compositeness effects can never be completely decoupled. We will discuss these effects in \sec{sec:flavor}.

\begin{figure}[t]
\begin{center}
\includegraphics[width=0.6\columnwidth]{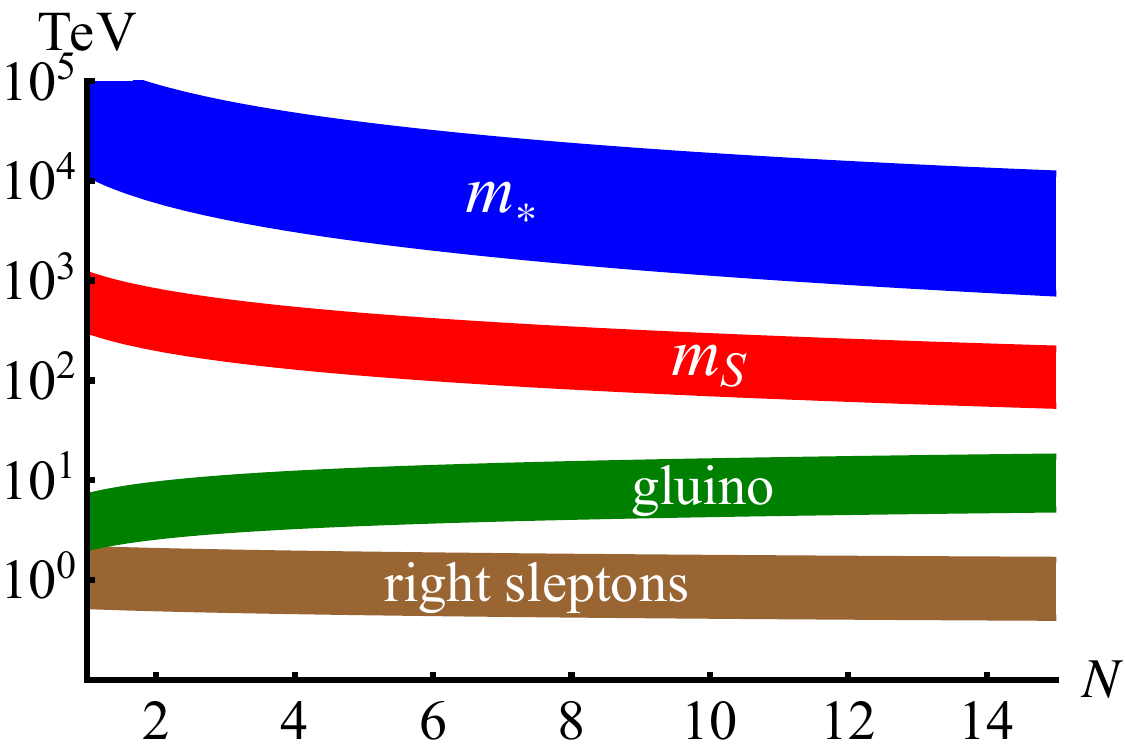}
\end{center}
\caption{The preferred range of the mass spectrum in the Super-Composite Higgs, obtained by varying the squark mass in the range 3--10 TeV and $\Delta_\O$ in the range 1.2--1.6.
Gauge mediation mass relations and radiative EW breaking are assumed.}
\label{fig:Spectrum}
\end{figure}

We can get an approximate analytical understanding of the Super-Composite Higgs spectrum by keeping only the leading effects in the mass relations, which yield the hierarchical pattern in terms of only two quantities
\beq
R_\mu^{-1/\Delta_\O}\, \mst ~\approx~ \mS ~\approx~ R\, {\tilde m}_q ~\approx~ R_\mu \, \mu \, ,
\label{stima1}
\eeq
\beq
R\approx \frac{\sqrt{6}\, \pi}{\alpha_s \sqrt{N}}\approx 30\, \sqrt{\frac{10}{N}} \,,~~~~~
R_\mu \approx \frac{2\pi R}{y_t\sqrt{3\ln R}} \approx 100\, \sqrt{\frac{10}{N}} \, .
\label{stima2}
\eeq
These simple expressions give an adequate approximation of the full result presented in \fig{fig:mu}, and are useful for simple estimates. For instance, the supersymmetry breaking scale $\mS$ and the theoretical upper bound on $\mst$ can be written as
\bea
\mS &\approx& R\, {\tilde m}_q ~\approx \left( \frac{{\tilde m}_q}{10~\textrm{TeV}}\right)
\sqrt{ \frac{10}{N}}~~ 300~\textrm{TeV} \, , \\
\mst &\lsim& R_\mu R\, {\tilde m}_q ~\approx \left( \frac{{\tilde m}_q}{10~\textrm{TeV}}\right)
\left( \frac{10}{N}\right)\,\, 30~\textrm{PeV} \, ,
\eea
which approximately explain the behavior with $N$ and agree reasonably well with the full numerical calculation. We warn the reader that the expressions in eqs.~(\ref{stima1})--(\ref{stima2}) are adequate for order-of-magnitude estimates, but do not precisely capture the dependence on $N$ because they ignore the effect of gauginos in the RG running, which contaminates scalar masses with a different functional dependence on $N$. 

To visualize the Super-Composite Higgs spectrum in a more compact form, we present it in \fig{fig:Spectrum} by varying squark masses 
 in the interval 3--10 TeV and $\Delta_\O$ in the interval 1.2--1.6. To illustrate the mass pattern, in \fig{fig:Spectrum} we have chosen the gluino (which typically lies at the top of the sparticle spectrum) and the right sleptons (which typically lie at the bottom), as well as $\mS$ and $\mst$. The sizes of the bands in \fig{fig:Spectrum} are merely indicative, as they include some parametric dependence, but ignore the corrections due to strong dynamics. Nevertheless, this figure highlights several key points.

\begin{itemize}

\item It is remarkable how well we can outline the mass pattern of the Super-Composite Higgs, despite our inability to resolve the strong dynamics that governs the theory. While we must accept unknown order-a-few factors in front of any prediction, a clear picture emerges. Starting from supersymmetric particles that populate the region 1--10 TeV, we encounter the supersymmetry breaking sector at the scale $\mS$ in the range of about 50--500 TeV, and then the composite sector at the scale $\mst$ in the range of about 1--100 PeV.

\item The proximity of all these different mass scales in the theory indicates that our EFT approach with multiple thresholds can only serve as a rough approximation at best. In reality, the Super-Composite Higgs features complex and intertwined dynamics, with multiple overlapping sectors influencing one another.

\item The Super-Composite Higgs couldn't be further apart from the picture of the Big Desert. Above the weak scale, there exists a complex structure that spans from a few TeV to multi PeV. Furthermore, it is possible that such a theory will encounter additional thresholds on the way to the far UV, perhaps needed to resolve Landau poles that may arise in QCD given the large amount of colored matter implied by the CFT sector. Indeed the traditional picture of supersymmetric grand unification will need to be re-thought. In the 
 Super-Composite Higgs framework there is still much to explore and discover beyond our current knowledge of particle physics.

\item The Super-Composite Higgs offers an exciting prospect for future searches of supersymmetric particles in the region 1--10 TeV. This is a particularly interesting mass range for future collider projects, as we will further comment in \sec{sec:collider}.

\item The edifice of the Super-Composite Higgs yields an upper bound on $\mst$ of about 100 PeV. This implies that flavor and CP violating effects originating from the compositeness sector are a necessary and inevitable feature of the theory. These effects will be discussed in the next section.
\end{itemize}

\boldmath
\section{Flavor effects}
\unboldmath
\label{sec:flavor}

\subsection{The flavor structure}
\label{sec:flavstruc}

To complete the last step in the journey throughout the energy scales, following the stack of effective theories illustrated in \fig{fig:Overall}, we integrate out the supersymmetric particles at the scale $\msoft$. The resulting effective theory is the SM endowed with a tower of higher-dimensional operators. Starting from the dynamical superfields describing matter, $\Phi =(\scalar{\Phi}, \psi_\Phi)$, Higgs, $H_{u,d}=(H_{u,d},{\tilde h}_{u,d})$, and gauge, $V_A=({\lambda}_A,A_A^\mu)$,  we integrate out the squarks and sleptons $\scalar{\Phi}$, the higgsinos ${\tilde h}_{u,d}$, the gauginos ${\lambda}_A$ and the heavy Higgs doublet, while retaining in the effective theory the chiral components of SM fermions $\psi_\Phi$, the SM Higgs doublet $h= H_d \cos\beta +i \sigma_2 H_u^* \sin \beta$, and the SM gauge bosons $A_A^\mu$. Among all higher-dimensional operators, the most important operators for constraining the mass scales of the Super-Composite Higgs are those that lead to flavor and CP violation. In this section, we will focus on this class of operators. The first step in view of this goal is the study of the Yukawa couplings.

The hypothesis that the hierarchical structure of the Yukawa couplings derives from partial compositeness implies that the mixing matrices $\xi_{\Phi_{ai}}$ take the form
\beq
\xi_{ai} = \xi_a \, A_{ai} \, ,
\label{partz}
\eeq
where, to keep the notation light, we drop the matter superfield index $\Phi$. Here $\xi_a$ are real numbers exhibiting a substantial hierarchy $\xi_{a>b}>\xi_b$, which reflects the mild hierarchy in the dimensions of the composite operators $\Delta_{{\cal O}_a}$. The matrix $A$ has generic complex entries of order unity.

Using a biunitary rotation, we can express \eq{partz} as $\xi_{ai} =U_{ab}\xi_b V_{bi}$, where $U$ and $V$ are unitary matrices, and we have reabsorbed in each $\xi_b$ an overall ${\mathcal O}(1)$ coefficient that multiplies the $b$-th row of $A$, so as to identify $\xi_b$ with the corresponding eigenvalue. The matrix $V$ can be eliminated through a unitary rotation of the elementary fields in generation space, $\Phi \to V^\dagger \Phi$. On the other hand, a unitary rotation of the composite operators to eliminate $U$ is undesirable because it would contaminate the CP-conserving CFT sector with spurious phases. Therefore, working at the leading order in $\xi$ ratios, we obtain
\beq
\xi_{ai} = U_{ai}\xi_i \, , ~~~~U= \begin{pmatrix}
1 & a_{12}\, {\xi_1}/{\xi_2} & (a_{13}+a_{12}a_{23})\, {\xi_1}/{\xi_3} \\
-a_{12}^*\, {\xi_1}/{\xi_2}& 1 & a_{23}\, {\xi_2}/{\xi_3}\\
-a_{13}^*\, {\xi_1}/{\xi_3} & -a_{23}^*\, {\xi_2}/{\xi_3} & 1
\end{pmatrix} \, ,
\label{matphas}
\eeq
where $a_{ij}$ are generic $O(1)$ complex coefficients. In particular ${\mathrm{arg}}(a_{ij})=O(1)$.

The Yukawa couplings of \eq{yukint} become
\beq
y_{ij} =\xi_{L_i} C_{ij} \xi_{{\bar R}_j} \, ,~~~~C_{ij}\equiv U_{L_{ia}} c_{ab} U_{{\bar R}_{bj}} \, ,
\label{yukintclever}
\eeq
where the indices $L,{\bar R}$ refer to the matter superfields $\Phi_{L,{\bar R}}$ that participate in the Yukawa interaction, and $c_{ab}$ are ${\mathcal O}(1)$ coefficients which are real as a consequence of CP invariance in the CFT sector.

The genericity of $c_{ab}$ implies that
\beq
{\rm Re}\, C_{ij} \approx 1 ~~\forall i,j \, .
\eeq
As $c_{ab}$ are real, any imaginary part of $C_{ij}$ must originate from phases in $U_{L,{\bar R}}$ which, as shown in \eq{matphas}, always appear with a suppression factor $\xi_i /\xi_{j>i}$. Since $c_{ab}$ are fully generic, with no price associated with transitions in generation space, the only suppression in ${\rm Im}\, C_{ij}$ is governed by the off-diagonal terms of $U_{L_{ia}}$ and $U_{{\bar R}_{bj}}$. One can easily check that this results in
\beq
{\rm Im}\, C_{ij} \approx {\text{max}}\left( \chi_{L_i} \, ,\, \chi_{{\bar R}_j} \right) \, ,
\label{ImC}
\eeq
\beq
\chi_{A_1} = {\text{max}}\left( \frac{\xi_{A_1}}{\xi_{A_2}}\, ,\, \frac{\xi_{A_1}}{\xi_{A_3}} \right),~~~
\chi_{A_2} = {\text{max}}\left( \frac{\xi_{A_1}}{\xi_{A_2}}\, ,\, \frac{\xi_{A_2}}{\xi_{A_3}} \right),~~~
\chi_{A_3} = {\text{max}}\left( \frac{\xi_{A_1}}{\xi_{A_3}}\, ,\, \frac{\xi_{A_2}}{\xi_{A_3}} \right),~~~A=L,{\bar R}  .
\nonumber 
\eeq

Employing the basis (\ref{yukintclever}) for the Yukawas matrices in \eq{yukint},  one readily estimates, up to $O(1)$ coefficients,  the eigenvalues and  CKM entries to be
\beq
y_{u_i} \approx   \xi_{Qi}\xi_{\bar U i}\,,\qquad y_{d_i} \approx  \xi_{Qi}\xi_{\bar D i}\,,\qquad y_{\ell_i} \approx  \xi_{Li}\xi_{\bar E i}\,,
\label{eigenvaluesY}
\eeq
\beq
V_{us} \approx \frac{\xi_{Q_1}}{\xi_{Q_2}} e^{i\varphi_{12} } \,,~~~~
V_{cb} \approx \frac{\xi_{Q_2}}{\xi_{Q_3}} e^{i\varphi_{23} }\,,~~~~
V_{ub} \approx \frac{\xi_{Q_1}}{\xi_{Q_3}} e^{i\varphi_{13} } \, ,
\label{CKMel}
\eeq
with $\varphi_{12},\,\varphi_{23},\,\varphi_{13}$ all controlled by the biggest phases in the $C$ matrices of \eq{yukintclever} for the $u$ and $d$ quark sector:
\beq
\varphi_{12},\,\varphi_{23},\,\varphi_{13}\, \approx \,\Max\limits_{i<j}\,(\xi_{Qi}/ \xi_{Qj},\, \xi_{\bar U i}/ \xi_{\bar U j},\, \xi_{\bar D i}/ \xi_{\bar D j})\,.
\label{phi_ij}
\eeq
Of course the phases in \eq{CKMel} are redundant  and  the only observable CP breaking effect is controlled by the Jarlskog invariant
\beq
J= {\mathrm {Im}}(V_{us}V_{tb}V_{ub}^*V_{td}^*)\approx \frac{\xi_{Q_1}^2}{\xi_{Q_3}^2} \sin(\varphi_{12}+\varphi_{23}+\varphi_{13}) \approx |V_{ub}V_{td}| \sin(\varphi_{12}+\varphi_{23}+\varphi_{13})\, .
\eeq
Formally, with strictly hierarchical mixings, $\xi_{\Phi_{i>j}}\gg \xi_{\Phi_j}$, the CKM phase $\varphi_{CKM}\equiv (\varphi_{12}+\varphi_{23}+\varphi_{13})$ is suppressed. 
However, as we will show later, see \eq{maxang},  $\xi_{\bar D_2}/ \xi_{\bar D_3}$ turns out to be sizable,
which in turn implies $\varphi_{CKM}\approx {\xi_{\bar D_2}}/{ \xi_{\bar D_3}}\approx 0.5$, in qualitative agreement with its measured $O(1)$ value. Notice that the numerical proximity between $\xi_{\bar D_2}$ and $\xi_{\bar D_3}$ corresponds, according to \eq{running_lambda}, to $\gamma_{{\bar D}_2}\simeq \gamma_{{\bar D}_3}$. We view this as an acceptable numerical coincidence.

The values of the $\xi_\Phi$'s are determined by matching \eq{eigenvaluesY} and \eq{CKMel}
 to the observed fermion masses and  CKM angles. In the quark sector, these  fix all the $\xi_\Phi$'s up to one combination, which we choose to be the
ratio 
\beq
\kappa_t \equiv \sqrt{\frac{\xi_{Q_3}}{\xi_{{\bar U}_3}}} \, ,
\eeq
which measures the relative degree of compositeness of $t_L$ and  $t_R$.  Equation~(\ref{def_xi}) and the additional constraints 
$\hlambda_{Q,{\bar U}}\lsim g_*$, imply
\beq
\sqrt{\frac{y_t}{g_*}} < \kappa_t < \sqrt{\frac{g_*}{y_t}}\, , 
\eeq
with the lower and upper limits corresponding respectively to fully composite $t_R$ and fully composite $t_L$. 
When $\kappa_t =1$, there is an equal degree of compositeness for $t_R$ and $t_L$.
For the lepton sector, we assume, for simplicity, that the degree of compositeness is equally shared between left and right, but the generalization to other cases is straightforward. This completely determines the values of the $\xi_\Phi$'s to be
\beq
\xi_{Q_1}\approx \frac{V_{us} V_{cb} \, \kappa_t\sqrt{m_t}}{\sqrt{v\, \sin \beta}} \,,~~~~
\xi_{Q_2}\approx \frac{V_{cb} \, \kappa_t\sqrt{m_t}}{\sqrt{v\, \sin \beta}} \,,~~~~
\xi_{Q_3}\approx \frac{ \kappa_t\sqrt{m_t}}{\sqrt{v\, \sin \beta}} \, ,
\label{xi1}
\eeq
\beq
\xi_{\bar U_1}\approx \frac{m_u}{V_{us} V_{cb} \, \kappa_t  \sqrt{m_t \, v\, \sin\beta}} \,,~~~~
\xi_{\bar U_2}\approx \frac{m_c}{V_{cb} \, \kappa_t \sqrt{m_t \, v\, \sin\beta}} \,,~~~~
\xi_{\bar U_3}\approx \frac{\sqrt{m_t}}{\kappa_t \sqrt{v\, \sin\beta}} \, ,
\label{xi2}
\eeq
\beq
\xi_{\bar D_1}\approx \frac{m_d\, \tan\beta}{V_{us} V_{cb} \, \kappa_t \sqrt{m_t \, v\, \sin\beta}}  \,,~~~~
\xi_{\bar D_2}\approx \frac{m_s\, \tan\beta}{V_{cb} \, \kappa_t \sqrt{m_t \, v\, \sin\beta}}  \,,~~~~
\xi_{\bar D_3}\approx \frac{m_b\, \tan\beta}{ \kappa_t \sqrt{m_t \, v\, \sin\beta}}  \, ,
\label{xi3}
\eeq
\beq
\xi_{L_1}\approx \xi_{\bar E_1} \approx \sqrt{\frac{m_e}{v \cos\beta}} \,,~~~~
\xi_{L_2}\approx \xi_{\bar E_2} \approx \sqrt{\frac{m_\mu}{v \cos\beta}} \,,~~~~
\xi_{L_3}\approx \xi_{\bar E_3} \approx \sqrt{\frac{m_\tau}{v \cos\beta}} \, .
\label{xi4}
\eeq
The mass parameters in eqs.~(\ref{xi1})--(\ref{xi4}) are running masses to be evaluated at the relevant scale $\muRG$. In table~\ref{tab:run} we give the input values of the running quark masses, derived from ref.~\cite{Huang:2020hdv} and evolved with two-loop SM $\beta$-functions. For $\muRG >\msoft$, the evolution must be performed with supersymmetric $\beta$-functions and, for $\muRG > \mS$, the contribution from the strongly-coupled sector must be included as well.

\renewcommand{\arraystretch}{1.2}
\begin{table}[t]
\centering
\begin{tabular}{|c|c|c|c|c|c|c|}
\hline
& $m_u$  & $m_d$  & $m_s$  & $m_c$  & $m_b$  & $m_t$  \\
$\muRG$ & [MeV] & [MeV] &  [MeV] &  [GeV] &  [GeV] &  [GeV] \\
\hline
$M_t$ & 1.2 & 2.6 & 51 & 0.59 & 2.7 & 162 \\
3 TeV & 1.0 & 2.2 & 45 & 0.51 & 2.3 & 144 \\
10 TeV & 1.0 & 2.1 & 42 & 0.48 & 2.1 &  137 \\
100 TeV & 0.9 & 1.9 & 37 & 0.43 & 1.9 & 124 \\
\hline
\end{tabular}
\renewcommand{\arraystretch}{1}
\caption{SM quark running masses, derived from the input values of ref.~\cite{Huang:2020hdv} and evaluated at the energy scale $\muRG$. Here $M_t= 172.4$~GeV is the pole top mass.}
 \label{tab:run}
 \end{table}

We are now ready to study the impact on the phenomenology of flavor and CP violation from the terms associated with compositeness in the effective action of eqs.~(\ref{eccok})--(\ref{eccow}) and
(\ref{kal1})--(\ref{pot1}). These terms are controlled by tensor coefficients ($c_\Phi$, $c_{\Phi \Phi'}$, \dots ) carrying    indices ($a,b,\dots$)  in the space of composite operators ${\cal O}_{\bar \Phi}$. In analogy with \eq{yukintclever}, it is convenient to define $C$ tensors for which  the $U_\Phi$ are absorbed in  the original  $c$-tensors. For instance for the correction to the matter kinetic matrix  controlled by $c_\Phi$ in \eq{eccok} we have
\beq
(c_\Phi)_{ab}\, \xi^*_{\Phi_{ai}}\xi_{\Phi_{bj}}=(c_\Phi)_{ab}\, U^*_{\Phi_{ai}} U_{\Phi_{bj}} \xi_{\Phi_i}\xi_{\Phi_j}
\equiv (C_\Phi)_{ij}\, \xi_{\Phi_i} \xi_{\Phi_j}\,.
\eeq
The $C$'s are still anarchic with  $O(1)$ entries, but now with phases  of order $\xi_{\Phi i}/\xi_{\Phi j>i}$ in an obvious generalization of the pattern of  \eq{ImC}.
In this final form, the size of the flavor and CP violating effects are readily estimated, in analogy with what we did above for the Yukawa couplings. In practice,  in each sector $\Phi$, CP phases will be controlled by the largest $\xi_{\Phi i}/\xi_{\Phi j>i}$. We will indicate such dominant phase as $\chi_\Phi$. From the comparison with the observed masses and  mixings we then have 
\beq
\chi_\Phi =
\left\{ \begin{array}{ll}
{|\xi_{Q_1}}/{\xi_{Q_2}}|\approx
V_{us}=0.22 & ~~~\text{for}~\Phi=Q \\
{|\xi_{\bar U_2}}/{\xi_{\bar U_3}|}\approx
m_c/(V_{cb} m_t)=0.09 & ~~~\text{for}~\Phi={\bar U} \\
{|\xi_{\bar D_2}}/{\xi_{\bar D_3}|}\approx
m_s/(V_{cb}m_b)=0.48 & ~~~\text{for}~\Phi={\bar D} \\
{|\xi_{L_2,\bar E_2}}/{\xi_{L_3,\bar E_3}|}\approx
\sqrt{m_\mu / m_\tau} =0.24 & ~~~\text{for}~\Phi=L,{\bar E} 
\end{array}
\right. 
\label{maxang}
\eeq

In summary, a mechanism reminiscent of the SM Jarlskog suppression operates in the Super-Composite Higgs. Although CP is maximally violated by the $\xi_\Phi$ couplings, the effect percolates into the SM sector only through mixing angles suppressed by the hierarchical structure of $\xi_\Phi$. However, because of the maximal flavor violation in the CFT sector, the transmission of CP violation doesn't have to go through all the mixing angles of the theory---as in the case of the SM CKM matrix---but only through a single mixing. This can be identified with the relevant mixing, in the case of a flavor transition, or the largest mixing, in the case of flavor-diagonal observables.  As shown in \eq{maxang}, the largest mixings are not particularly small and, therefore, the suppression of CP-violating effective operators is not very pronounced, even for flavor-diagonal observables.  This will become more clear case by case in the discussion that follows.

\subsection{Sources of flavor and CP violation}
\label{sec:sourcesFCP}

We are now ready to present the flavor-violating higher-dimensional operators involving SM fields only, which we separate in the following different classes.

\boldmath
\subsubsection*{K\"ahler dim-6 matter contact interactions}
\unboldmath

The K\"ahler terms proportional to $c_{\Phi \Phi^\prime}$ and $c_6$ in \eq{eccok} yield the 4-fermion contact interactions in the SM Lagrangian at dimension six, using the decomposition
\bea
\frac{1}{2}\, \int d^4\theta \,\Phi_{L_i}^\dagger \Phi_{L_j}\Phi_{L_k}^\dagger \Phi_{L_\ell} &\longrightarrow &  {\bar \psi}_L^i \gamma^\mu \psi_L^j \, {\bar \psi}_L^k \gamma^\mu \psi_L^\ell  \, ,\\
\frac{1}{2}\, \int d^4\theta \,\Phi_{{\bar R}_i}^\dagger \Phi_{{\bar R}_j}\Phi_{{\bar R}_k}^\dagger \Phi_{{\bar R}_\ell} &\longrightarrow &  {\bar \psi}_R^j \gamma^\mu \psi_R^i \, {\bar \psi}_R^\ell \gamma^\mu \psi_R^k  \, ,\\
\frac{1}{2}\, \int d^4\theta \,\Phi_{L_i}^\dagger \Phi_{L_j}\Phi_{{\bar R}_k}^\dagger \Phi_{{\bar R}_\ell}  &\longrightarrow & 2\, {\bar \psi}_L^i  \psi_R^k \, {\bar \psi}_R^\ell  \psi_L^j  \, .
\eea
The interactions proportional to $c_{\Phi \Phi^\prime}$ describe the product of all combinations of SM currents with arbitrary flavor indices. We also leave implicit that  different contractions of gauge indices lead to independent operators: for SU(2), the product of the two currents can be in the singlet-singlet or triplet-triplet channels and, for SU(3), in the singlet-singlet or octet-octet channels.

\renewcommand{\arraystretch}{1.5}
\begin{table}[t]
\centering
\begin{tabular}{|c|c|c|r|}
\hline
Process & Operator & Limit on & \multicolumn{1}{|c|}{Limit on} \\
 & & $\Lambda$ [TeV] & \multicolumn{1}{|c|}{$\mst$}  \\
 \hline
 $\varepsilon_K$ & ${\text{Im}}(Q_2^\dagger Q_1Q_2^\dagger Q_1)$ & $3 \times 10^4$ & $ \frac{\kappa_t^2} {\sin\beta}~ 4~\text{TeV}$\\
 $\varepsilon_K$ & ${\text{Im}}(\bar D_1^{\dagger}\bar D_2\bar D_1^{\dagger}\bar D_2)$ & $3 \times 10^4$ & $\big(\frac{\tan \beta}{5}\big)^2\frac{1}{\kappa_t^2 \sin\beta}~ 5~\text{TeV}$\\
 $\varepsilon_K$ & ${\text{Im}}(Q_2^\dagger Q_1\bar D_1^{\dagger}\bar D_2)$ & $5 \times 10^5$ & $\big(\frac{1/5}{\cos\beta}\big)~ 130~\text{TeV}$\\
 $\Delta a_{\text{CP}}$ & ${\text{Im}}(Q_2^\dagger Q_1Q_2^\dagger Q_1)$ & $1 \times 10^4$ & $ \frac{\kappa_t^2} {\sin\beta}~ 1~\text{TeV}$\\
 $\Delta a_{\text{CP}}$ & ${\text{Im}}(\bar U_1^{\dagger}\bar U_2\bar U_1^{\dagger}\bar U_2)$ & $1 \times 10^4$ & $ \frac{1}{\kappa_t^2\sin\beta}~ 0.2~\text{TeV}$\\
 $B_d$-${\bar B}_d$ mixing & $Q_3^\dagger Q_1Q_3^\dagger Q_1$ & $1 \times 10^3$ & $ \frac{\kappa_t^2} {\sin\beta}~ 7~\text{TeV}$\\
 $B_d$-${\bar B}_d$ mixing & $\bar D_1^{\dagger}\bar D_3 \bar D_1^{\dagger}\bar D_3$ & $1 \times 10^3$ & $\big(\frac{\tan \beta}{5}\big)^2\frac{1}{\kappa_t^2 \sin\beta}~ 0.5~\text{TeV}$\\
 $B_s$-${\bar B}_s$ mixing & $Q_3^\dagger Q_2Q_3^\dagger Q_2$ & $3 \times 10^2$ & $ \frac{\kappa_t^2} {\sin\beta}~ 10~\text{TeV}$\\
 $B_s$-${\bar B}_s$ mixing & $\bar D_2^{\dagger}\bar D_3 \bar D_2^{\dagger}\bar D_3$ & $3 \times 10^2$ & $\big(\frac{\tan \beta}{5}\big)^2\frac{1}{\kappa_t^2 \sin\beta}~ 0.7~\text{TeV}$\\
 $\mu \to 3e$ & $L_2^\dagger L_1 L_1^\dagger L_1$ & $2 \times 10^2$ & $\big(\frac{1/5}{\cos\beta}\big)~ 10~\text{GeV}$\\
 $\tau \to 3\mu$ & $L_3^\dagger L_2 L_2^\dagger L_2$ & $1 \times 10^1$ & $\big(\frac{1/5}{\cos\beta}\big)~ 60~\text{GeV}$\\
  $\mu \,{\rm Au} \to e \,{\rm Au}$ & $Q_1^\dagger Q_1 L_2^\dagger L_1$ &$7 \times 10^2$&  $\sqrt{\frac{\tan \beta}{5}}\frac{\kappa_t}{ \sin\beta}~ 80~\text{GeV}$\\
 \hline
 \end{tabular}
 \renewcommand{\arraystretch}{1}
 \caption{Limits on the scale $\Lambda$ of flavor-violating four-fermion operators, as defined in \eq{4ferd6}. The limits have been derived from the results in ref.~\cite{Bona:2024bue} (for $\Delta F =2$ processes) and ref.~\cite{Raidal:2008jk} (updating some of the limits on lepton number violation). The constraints from lepton-number violating processes apply to all operators obtained with the exchange $L \leftrightarrow {\bar E}$ and $Q \leftrightarrow {\bar U}, {\bar D}$. In the last column, we translate the limits on $\Lambda$ into limits on the compositeness scale $\mst$ by using \eq{transmst} and the expressions for the $\xi$-coefficients in eqs.~(\ref{xi1})--(\ref{xi4}). The bounds on $\mst$ are subjected to order-one uncertainties due to incalculable strong-dynamics effects.}
 \label{tab:4ferm}
 \end{table}
 
In \tab{tab:4ferm} we show the bounds from flavor-violating processes on the scale $\Lambda$ of each operator in the effective Lagrangian  
\beq
{\cal L}^{\rm SM}_6=
 \sum_{\alpha=L,R}\frac{1}{\Lambda^2}\, {\bar \psi}^{i}_{\alpha} \gamma^\mu {\psi}^j_{\alpha}\,
{\bar \psi}^{\prime k}_{\alpha} \gamma_\mu {\psi}^{\prime \ell}_{\alpha}+
\frac{1}{\Lambda^2}\, {\bar \psi}_L^i  \psi_R^{\prime k }\, {\bar \psi}_R^{\prime \ell}  \psi_L^j  + \hc 
\label{4ferd6}
\eeq
We then translate the bounds on $\Lambda$ into bounds on the compositeness scale $\mst$ through the relation 
\beq
\mst = \sqrt{\xi_i^* \xi_j \xi_k^* \xi_\ell \, \kappa_\chi \, \kappa_{\textrm CP}}\, \Lambda
\, ,
\label{transmst}
\eeq
where $\kappa_\chi =1$ for $LLLL$ or $RRRR$ operators and $\kappa_\chi =2$ for $LLRR$ operators, while $\kappa_{\textrm CP}=1$ for CP-conserving observables and $\kappa_{\textrm CP}=\chi_\Phi$ for CP-violating observables, with $\chi_\Phi$ defined in \eq{maxang}.

The strongest bound on $\mst$ comes from $\varepsilon_K$ through the operator $Q_2^\dagger Q_1\bar D_1^{\dagger}\bar D_2$, whose coefficient is simply $2\chi_{\bar D} y_d y_s/\mst^2$, thus featuring the same chiral suppression of the corresponding Yukawa entries and benefitting from a $\tan\beta^2$ enhancement. The corresponding limit on $\mst$ is above $100$ TeV  for large $\tan \beta$. 
Rare lepton decays and $\mu$-$e$ conversion are completely ineffective in setting any constraint.

\boldmath
\subsubsection*{Superpotential dim-5 matter contact interactions}
\unboldmath

\begin{figure}[t]
\begin{center}
\includegraphics[width=0.35\columnwidth]{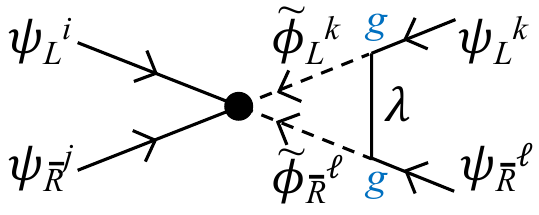}
\end{center}
\caption{Dressing of the dim-5 matter contact interaction (denoted by the black dot) via a gauge loop with gaugino ($\lambda$) exchange, leading to a 4-fermion contact interaction.}
\label{fig:feynFive}
\end{figure}

The superpotential terms proportional to $c_5$ and $c_5^\prime$ in \eq{eccow} yield two-fermions/two-scalars contact interactions. These can be turned into SM four-fermion interactions through the loop diagram involving squarks/sleptons and gauginos, shown in \fig{fig:feynFive}. The result of the loop integration is
\beq
\frac{g^2 \, C_G}{16\pi^2 M_{\lambda}} \, f({\tilde m_1}^2/ M_{\lambda}^2,{\tilde m_2}^2/M_{\lambda}^2) \, ,
\label{loopg}
\eeq
where $g$, $C_G$ and $M_{\lambda}$ are respectively the coupling, the quadratic Casimir and the gaugino mass of the corresponding SM gauge interaction, $\tilde m_{1,2}$ are the masses of the two sfermions. Using the expression of the gaugino mass given in \eq{softm}, the coefficient in front of \eq{loopg} becomes, up to an unknown order-one coefficient, equal to $1/(N \mS)$ and therefore independent of gauge couplings. The loop function 
\beq
f(x,y) =\frac{2[x(1-y)\ln x-y(1-x)\ln y]}{(x-1)(y-1)(y-x)}
\eeq
is normalized such that $f(1,1)=1$. In the limit of one heavy sfermion, the loop function is $f(x\!\!\gg\!\!1,1)\simeq 2(\ln x-1)/x$, and therefore the exchange of squarks and weak gauginos is suppressed by a factor of $\sim \alpha_2/\alpha_3$. As a result, the leading effect comes from the exchange of virtual colored sparticles and the resulting effective four-fermion interactions in the SM Lagrangian at dimension six are 
\beq
{\cal L}^{\rm SM}_6=
\frac{\xi_i \xi_j \xi_k \xi_\ell}{N \mS \mst}\left( {\bar u}^i_R q^j_L \, {\bar e}^k_R \ell^\ell_L +   {\bar u}^i_R q_L^j \, {\bar d}^k_R q^\ell_L\right) + \hc
\label{4ferd5}
\eeq 
Again the Wilson coefficient features the same chiral suppression of the corresponding Yukawas. The overall coefficient is modified 
corrected with respect to the K\"ahler dim-6 operators in \eq{4ferd6}  by a factor $\mst/(N \, \mS)$. For $N\sim 5$ and $\Delta_{\cal O} \to 1$,  this represents an enhancement of order a few. However, the enhancement evaporates for larger $N$ and $\Delta_{\cal O}$.
The interactions in \eq{4ferd5} do not contribute to neutral meson mixing nor to the chirally-violating decays $B^0, K^0 \to \ell^i \ell^j$. Instead, they lead to the decay $D^0 \to \ell^i \ell^j$, whose branching ratio in the muon channel has been studied by LHCb~\cite{LHCb:2022jaa} and CMS~\cite{CMS:2025fmx}, resulting in an upper bound of $3\times 10^{-9}$. In the Super-Composite Higgs, we find
\bea
{\rm BR} (D^0 \to \mu^+ \mu^-) &=& \frac{M_D^5 f_D^2}{128 \pi \Gamma_D m_c^2} \left( \frac{\xi_{{\bar U}_2}\xi_{Q_1} \xi_{{\bar E}_2}\xi_{L_2}}{N\mS \mst}\right)^2 
\left( 1-4\frac{m_\mu^2}{M_D}^2 \right)^{3/2}
\nonumber \\
&\approx& \frac{4\times 10^{-16}}{\sin^2 2\beta} \left( \frac{ \rm TeV}{\sqrt{{N\mS \mst}}} \right)^4 \, ,
\eea
which does not yield any competitive constraint. 

\boldmath
\subsubsection*{Dipole contact interactions}
\unboldmath

It is well known that supersymmetry prohibits dipole interactions~\cite{Ferrara:1974wb}, which therefore can
only be generated by operators involving ${\cal O}$ and/or ${\cal R}$ in \eq{pot1}, once supersymmetry is broken.
The leading direct source  of dipoles  is the K\"ahler interaction 
\beq
{\cal K}_{dipole} =  g\, 
\xi_{\Phi_L} \xi_{\Phi_{\bar R}} \, H_u^{(\dagger)} \Phi_L { D}^{\alpha}\Phi_{\bar R}
  \,{ W}_\alpha \frac{\O^\dagger}{\mst^{3+\Delta_{\O}}} + \hc \, ,
\label{odd}
\eeq
where $H_u^{(\dagger)}$ stands for $H_u$ when $\Phi_L \Phi_{\bar R} = Q {\bar U}$, and $H_u^\dagger$ when $\Phi_L \Phi_{\bar R} = Q {\bar D}$ or $L {\bar E}$, and where we stick to the simplest scenario where a single composite ${\cal O}$ dominates all supersymmetry breaking effects induced by compositeness.
As explained in \sec{sec:fromstosoft}, below the scale $\mS$ we can effectively replace $\langle \O\rangle\to \mS^{1+\Delta_{\O}}\theta^2$. When inserted in \eq{odd}, this leads to a contact electromagnetic dipole interaction in the dimension-six SM Lagrangian which, up to an unknown overall factor of order one, is given by
\beq
{\cal L}^{\rm SM}_6=
\frac{e}{ \Lambda_{d_{ij}}^2}\, {\bar \psi}_R^i \sigma_{\mu \nu} \psi_L^j \, H F^{\mu \nu} + \hc  \, ,
\label{dipolec}
\eeq
\beq
\frac{1}{\Lambda_{d_{ij}}^2}\approx {\xi_{\Phi_{{\bar R}_i}} \xi_{\Phi_{L_j}}}\, \frac{\mu\sin\beta}{\mst^3} \, ,
\label{deffL}
\eeq
where $e$ is the electric charge, $H$ is the SM Higgs doublet with $\langle H\rangle =v$. Notice that, besides \eq{odd} there is also the term with $D^\alpha$ acting on $\Phi_L $. That also leads to a dipole, although with opposite sign, consistently with the fact that when $D^\alpha$ acts on $(\Phi_L \Phi_{\bar R})$,  integration by parts gives a vanishing   dipole. However, we generically expect a dipole to be generated.

The matrix $1/{\Lambda_{d_{ij}}^2}$ in \eq{deffL} has the same pattern of entries and phases as the corresponding Yukawa coupling matrix. To work out the physical effects of the former, one must choose the diagonal basis for the latter. In such basis, the structure of the entries of $1/{\Lambda_{d_{ij}}^2}$ remain the same, but  all its entries, without exception, generically acquire a phase 
of the order of the corresponding $\chi_\Phi$, see \eq{maxang}.

Below the scale of EW breaking, \eq{dipolec} contributes to three kinds of observables. The first is the electric dipole moment (EDM) of the electron $d_e$, which is given by
\beq
{\cal L} = -\frac{i\,d_e}{2}\, {\bar e} \sigma_{\mu \nu} \gamma_5 e \, F^{\mu \nu} \, , 
\eeq
\bea
d_e &\approx& \frac{2 e \, \mu\tan\beta\, m_e}{\mst^3} \sqrt{\frac{m_\mu}{m_\tau}}
\nonumber
 \\
&=& \frac{\tan\beta}{5} \, \,  \frac{\mu}{\rm TeV}\left( \frac{180~{\rm TeV}}{\mst}\right)^3 4\times 10^{-30}~ e\,{\rm cm} \, ,
\label{EDMel}
\eea
where we have used \eq{xi4} for the $\xi$ coefficients and took into account the leptonic CP phase $\chi_L=\chi_{\bar E}= \sqrt{m_\mu/m_\tau}$ of \eq{maxang}. 

The second observable is the anomalous magnetic moment of the muon $a_\mu$
\beq
{\cal L} = \frac{e\, a_\mu}{4m_\mu}\, {\bar \mu} \sigma_{\mu \nu} \mu\, F^{\mu \nu} \, ,
\eeq
\beq
a_\mu \approx \frac{4  \mu\tan\beta\, m_\mu^2}{\mst^3} 
= \frac{\tan\beta}{5} \, \frac{\mu}{\rm TeV}\left( \frac{180~{\rm TeV}}{\mst}\right)^3 4\times 10^{-14}
 \, .
 \label{ecamu}
 \eeq
 The expression for $a_\tau$ is simply obtained from \eq{ecamu} by replacing $m_\mu$ with $m_\tau$.

The third class of observables are lepton-family transition dipole interactions
\beq
{\cal L} = \frac{e\, \mu\tan\beta \sqrt{m_i m_j}}{\mst^3} \, {\bar \ell^i_R} \sigma_{\mu \nu}  \ell^j_L F^{\mu \nu} + \hc \, , 
\eeq
which contribute to various rare lepton-family violating processes.

\begin{table}[t]
\centering
\begin{tabular}{|c|c|}
\hline
Process & Estimate in units of the electron EDM \\
\hline
Muon anomalous magnetic moment &$a_\mu \approx \left(\frac{d_e}{4\times 10^{-30}~ e\,{\rm cm}}\right) \, 4 \times 10^{-14}$ \\
Tau anomalous magnetic moment & $a_\tau \approx \left(\frac{d_e}{4\times 10^{-30}~ e\,{\rm cm}}\right) \, 1 \times 10^{-11}$ \\
Muon electric dipole moment &$d_\mu \approx \left(\frac{d_e}{4\times 10^{-30}~ e\,{\rm cm}}\right) \, 1 \times 10^{-27}$\\
Rare muon decays & BR$(\mu \to e \gamma) \approx \left(\frac{d_e}{4\times 10^{-30}~ e\,{\rm cm}}\right)^2 \, 1 \times 10^{-15}$\\
 & BR$(\mu \to 3 e ) \approx \left(\frac{d_e}{4\times 10^{-30}~ e\,{\rm cm}}\right)^2 \, 6 \times 10^{-18}$\\
 Rare tau decays & BR$(\tau \to \mu \gamma) \approx \left(\frac{d_e}{4\times 10^{-30}~ e\,{\rm cm}}\right)^2 \, 4 \times 10^{-12}$\\
 & BR$(\tau \to e \gamma) \approx \left(\frac{d_e}{4\times 10^{-30}~ e\,{\rm cm}}\right)^2 \, 2 \times 10^{-14}$\\
 $\mu$--$e$ conversion & B$(\mu N \to e N) \approx \left(\frac{d_e}{4\times 10^{-30}~ e\,{\rm cm}}\right)^2 B_{AZ} \, 2 \times 10^{-18}$\\
\hline
\end{tabular}
\caption{Estimates of various leptonic processes based on the flavor structure of the Super-Composite Higgs and expressed in units of the electron EDM. B$(\mu N \to e N)$ is the rate of $\mu$--$e$ conversion in nuclei in units of the nuclear capture rate and $B_{AZ}$ is an effective nuclear coefficient~\cite{Haxton:2024amf}, which is typically of order one for elements heavier than aluminum.} 
\label{tab:edm}
\end{table} 

The current experimental limit on the electron EDM is $4.1 \times 10^{-30}~e\,$cm~\cite{Roussy:2022cmp}, which, using \eq{EDMel}, sets a lower bound on the compositeness scale $\mst$ of about 200~TeV for $\tan\beta \sim 5$, with unavoidable uncertainties due to unknown order-one factors in our estimates. Future experiments aim at improving the sensitivity on the electron EDM by several orders of magnitude~\cite{Alarcon:2022ero} and these searches are evidently very effective in probing the Super-Composite Higgs.

The muon $g$--2 and lepton-family transitions provide less stringent limits than the electron EDM. However, it is interesting that our framework predicts robust correlations among a variety of leptonic processes, which can be expressed in terms of the value of the electron EDM, as show in \tab{tab:edm}. These correlations depend uniquely on the flavor structure of the Super-Composite Higgs and therefore hold for all kinds of contributions to the dipole operators that we will encounter in this section. In the lucky event of a measurement of the electron EDM, the correlations in \tab{tab:edm} will provide a target for future experimental searches of rare phenomena in the leptonic sector. 

\boldmath
\subsubsection*{Dipoles from Beyond Soft non-holomorphic $A$-terms}
\unboldmath

As discussed in \sec{sec:fromstosoft}, the operators proportional to $Z_y$ and $Z_{\hat y}$ in \eq{kal1} lead, after supersymmetry breaking, to Beyond Soft holomorphic and non-holomorphic $A$-terms. In the leptonic case, they are shown in eqs.~({\ref{holhol}) and ({\ref{nonhol}), and analogous expressions hold for quarks as well. The coefficients $A_{\ell_{ij}}$ 
and ${\tilde A}_{\ell_{ij}}$ in eqs.~({\ref{holhol}) and ({\ref{nonhol}) are not aligned with Yukawa couplings and carry CP-violating phases. After EW breaking, these $A$-terms contribute to slepton left-right mixings. Particularly important is the non-holomorphic $A$-term as it leads to a $\tan\beta$ enhancement in the contribution to an effective electromagnetic dipole through the diagram in \fig{fig:feynDip2} and to an electron EDM of size
\bea
d_e &\approx& \frac{e\,g^{\prime 2}  \tan\beta   \, m_e}{192 \pi^2 M_{\tilde B}\, \mst}\sqrt{\frac{m_\mu}{m_\tau}} \, F_{\tilde B}
\nonumber \\
&=& \left(\frac{\tan \beta}{5}\right) \left( \frac{\rm TeV}{M_{\tilde B}}\right)\, \left( \frac{200~{\rm TeV}}{\mst}\right) \, F_{\tilde B}~4\times 10^{-30} ~e\,{\rm cm}\, ,
\label{EDM2}
\eea
\beq
f_{\tilde B}(x) =\frac{3(1-x^2+2x\ln x)}{(1-x)^3} \, ,~~~~
F_{\tilde B}=  
\frac{
2\mu^2\,\Big[ f_{\tilde B}\Big( \frac{{\tilde m}_{e_R}^2}{M_{\tilde B}^2}\Big)-
f_{\tilde B}\Big( \frac{{\tilde m}_{\ell_L}^2}{M_{\tilde B}^2}\Big)\Big]}
{{\tilde m}_{\ell_L}^2-{\tilde m}_{e_R}^2} \, ,
\label{FWB}
\eeq
where $f_{\tilde B}(1) =1$ and $F_{\tilde B}=1$ when all mass ratios are equal to 1.

\begin{figure}[t]
\begin{center}
\includegraphics[width=0.26\columnwidth]{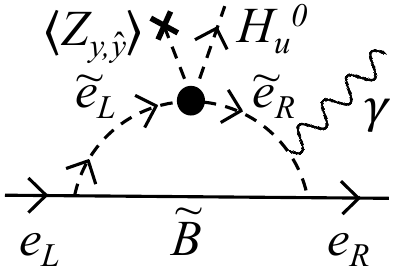}
\end{center}
\caption{Feynman diagrams representing the one-loop contribution to an effective electron dipole interaction from Beyond Soft non-holomorphic $A$-terms (denoted by the black dot).}
\label{fig:feynDip2}
\end{figure}

\begin{figure}[t]
\begin{center}
\includegraphics[width=0.55\columnwidth]{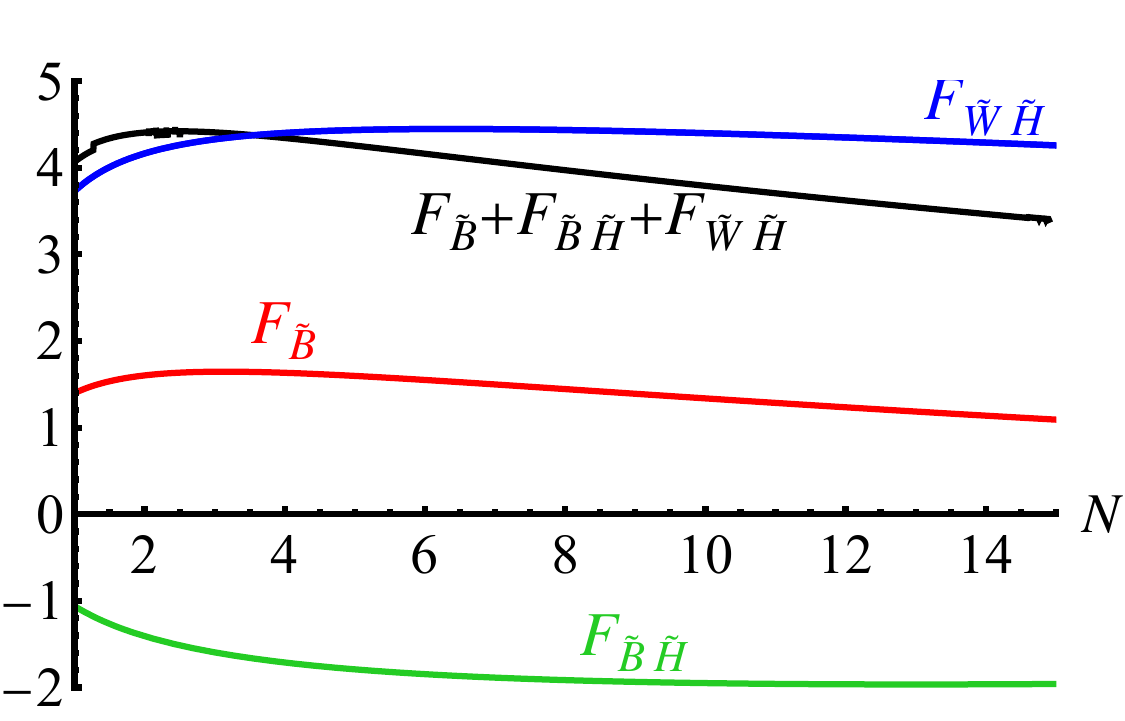}
\end{center}
\caption{The loop functions $F_{\tilde B}$, $F_{\tilde B \tilde H}$ and $F_{\tilde W \tilde H}$, as well as their sum, under the assumption of gauge-mediation mass ratios and the EW breaking condition for $\mu$ given by \eq{EEWW}. We have chosen $\mS =100$ TeV.}
\label{fig:FLoop}
\end{figure}

To be more quantitative, we determine the mass ratios of supersymmetric particles using the gauge-mediation expressions
\beq
\frac{{\tilde m}_{e_R}^2}{M^2_{\tilde B}}=\frac{6\, g^{\prime 4}(\mS)}{5\, g^{\prime 4}(M_{\tilde B})}
\left[ \frac1N +\frac{5}{33}\Big( 1-\frac{g^{\prime 4}({\tilde m}_{e_R})}{g^{\prime 4}(\mS)}\Big) \right] \, ,
\label{GM1}
\eeq
\beq
\frac{{\tilde m}_{{ \ell}_L}^2}{M^2_{\tilde B}}=\frac{27\, g^{ 4}(\mS)}{50\, g^{\prime 4}(M_{\tilde B})}
\left[ \frac1N +\Big( 1-\frac{g^{ 4}({\tilde m}_{{ \ell}_L})}{g^{ 4}(\mS)}\Big) \right] 
+\frac{3\, g^{\prime 4}(\mS)}{10\, g^{\prime 4}(M_{\tilde B})}
\left[ \frac1N +
\frac{5}{33}\Big( 1-\frac{g^{\prime 4}({\tilde m}_{{ \ell}_L})}{g^{\prime 4}(\mS)}\Big) \right] 
\label{GM2}
\eeq
and the EW breaking condition for $\mu$, see \eq{EEWW} and \fig{fig:mu}. The corresponding value of $F_{\tilde B}$ is shown in \fig{fig:FLoop}.

We conclude that the contribution to the electron EDM from Beyond Soft trilinears gives a lower bound on the compositeness scale
\beq
\mst \gsim \left(\frac{\tan \beta}{5}\right) \left( \frac{\rm TeV}{M_{\tilde B}}\right)\, 200~{\rm TeV}\, ,
\label{limEDMtr}
\eeq
with large uncertainties from strong dynamics.

\subsubsection*{Dipoles from Beyond Soft wrong-Higgs Yukawa interactions}

A prominent example of Beyond Soft effects is the wrong-Higgs Yukawa interaction in \eq{wronghiggs}, which modifies the relation between electron mass and Yukawa coupling
\beq
\frac{m_e}{v} = y_e \left( e^{i\phi}\cos\beta +c\, \sin\beta\, \frac{\mu}{m_*}\,\right) \, .
\label{wHrel}
\eeq
Here $y_e e^{i\phi}$ (with $y_e$ real) is the electron Yukawa coupling that appears in the superpotential term. Recalling the discussion in \sec{sec:flavstruc}, we expect the constant $c$ to have a real part of order unity and an imaginary part of size $\chi_{L,\bar E} \sim \sqrt{m_\mu /m_\tau}$. Therefore, in the basis in which $m_e$ is real, the phase $\phi$ is given, up to an order-one factor, by
\beq
\phi \approx \tan\beta \, \frac{\mu}{m_*}\, \sqrt{\frac{m_\mu}{m_\tau}}.
\eeq

Because of the Beyond Soft effect in \eq{wHrel}, there is a mismatch between the Yukawa coupling that enters sparticle interactions and the value of the Yukawa measured by the lepton mass. As a result, diagrams with exchange of Bino (${\tilde B}$), Bino/higgsino (${\tilde B}{\tilde H}$) and Wino/higgsino (${\tilde W}{\tilde H}$) lead to an electron EDM
\bea
d_e &\approx& \frac{e\,g^{\prime 2}  \tan^2\beta \, m_e}{192 \pi^2 M_{\tilde B}\, \mst}\sqrt{\frac{m_\mu}{m_\tau}}\, (F_{\tilde B}+ F_{\tilde B \tilde H} +F_{\tilde W \tilde H})
\nonumber \\
&=& \left(\frac{\tan \beta}{5}\right)^2  \left( \frac{\rm TeV}{M_{\tilde B}}\right)\, \left( \frac{{\rm PeV}}{\mst}\right) \, (F_{\tilde B}+ F_{\tilde B \tilde H} +F_{\tilde W \tilde H})~4\times 10^{-30} ~e\,{\rm cm}\, ,
\label{EDM3}
\eea
\bea
F_{{\tilde B}{\tilde H}}\!\!\!&=&\!\!\!\frac{M_{\tilde B}^2
 \Big[ 
2\, f_{\tilde B}\Big( \frac{{\tilde m}_{e_R}^2}{\mu^2}\Big)-
f_{\tilde B}\Big( \frac{{\tilde m}_{\ell_L}^2}{\mu^2}\Big) \Big] -
\mu^2 \Big[ 
2\, f_{\tilde B}\Big( \frac{{\tilde m}_{e_R}^2}{M_{\tilde B}^2}\Big)-
f_{\tilde B}\Big( \frac{{\tilde m}_{\ell_L}^2}{M_{\tilde B}^2}\Big) \Big] }
{\mu^2- M_{\tilde B}^2} \, ,
\\
F_{{\tilde W}{\tilde H}}\!\!\!&=&\!\!\! \bigg( \frac{3M_{\tilde B}}{M_{\tilde W}\tan^2\theta_W}\bigg)\,\frac{\Big[
\mu^2  \,
f_{\tilde W}\Big( \frac{{\tilde m}_{\ell_L}^2}{M_{\tilde W}^2}\Big) 
- M_{\tilde W}^2 \,
 f_{\tilde W}\Big( \frac{{\tilde m}_{\ell_L}^2}{\mu^2}\Big)\Big]}
{(\mu^2- M_{\tilde W}^2)} \, , 
\eea
\beq
 f_{\tilde W}(x) =\frac{1-8x+7x^2-2x(1+2x)\ln x}{(1-x)^3} 
\, , 
\eeq
where $f_{\tilde W}(1) =1$, and the functions $f_{\tilde B}$ and $F_{\tilde B}$ are given in \eq{FWB}.

Taking the gauge mediation mass relations $M_{\tilde B}/M_{\tilde W}=(5/3)\tan^2\theta_W$ and eqs.~(\ref{GM1})--(\ref{GM2}), together with the EW breaking condition in \eq{EEWW}, we obtain the loop functions shown in \fig{fig:FLoop}. In \eq{EDM3}, the contribution from $F_{\tilde B}$ is identical to the one in \eq{EDM2}, but for an additional power of $\tan\beta$. The contribution from $F_{\tilde B \tilde H}$ nearly cancels the one from $F_{\tilde B}$. The contribution from $F_{\tilde W \tilde H}$ gives the leading effect (as long as one relies on the EW breaking condition) partly because it comes from SU(2) gauge interactions and is therefore enhanced by a $1/\tan^2\theta_W$ factor, and partly because of accidental coefficients in the loop functions.

From \eq{EDM3} and the results in \fig{fig:FLoop}, we conclude that the contribution to the electron EDM from wrong-Higgs Yukawa interaction gives a lower bound on the compositeness scale in the range
\beq
\mst \gsim \left(\frac{\tan \beta}{5}\right)^2 \left( \frac{\rm TeV}{M_{\tilde B}}\right)\, 4~{\rm PeV}\, .
\label{limEDMw}
\eeq
This is a factor of $0.8\,\tan\beta\, (3M_{\tilde B})/(M_{\tilde W}\tan^2\theta_W)\sim 20\, (\tan\beta /5)$ times larger than the result from non-holomorphic $A$-terms in \eq{limEDMtr}.

\begin{figure}[t]
\begin{center}
\includegraphics[width=0.65\columnwidth]{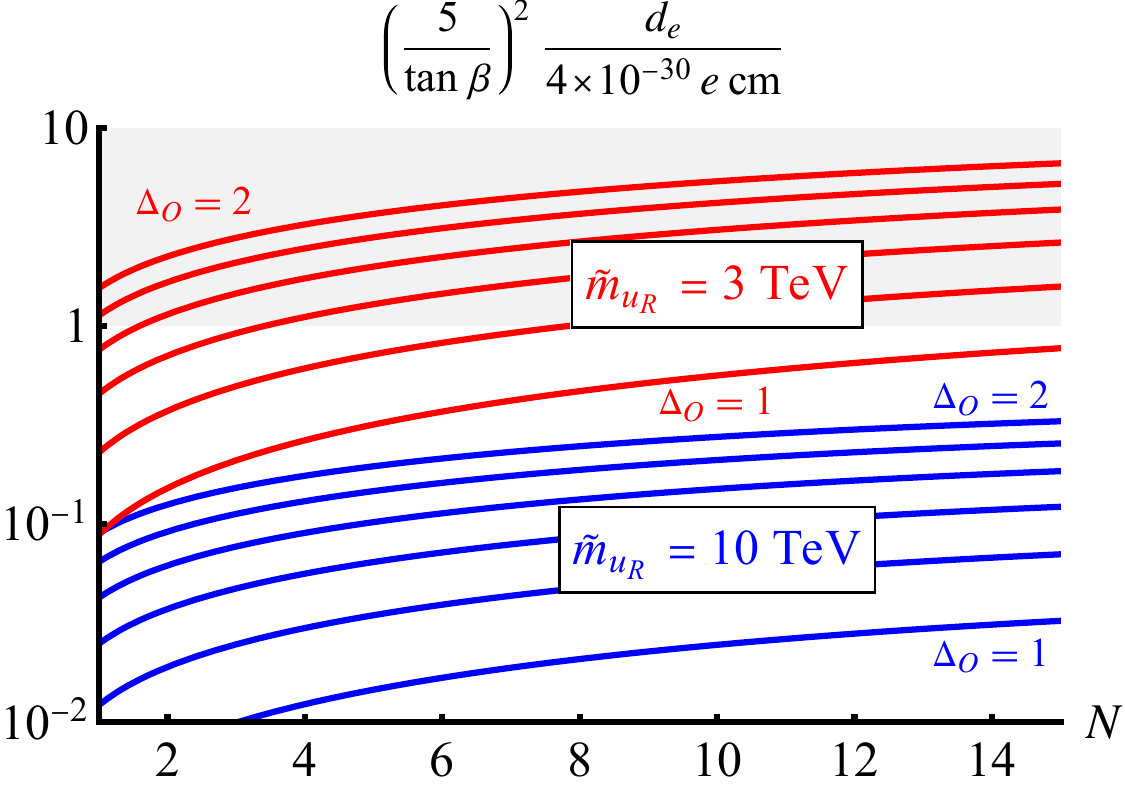}
\end{center}
\caption{The contribution to the electron EDM coming from Beyond Soft wrong-Higgs Yukawa interactions, as given by \eq{EDM3} and expressed in terms of the squark mass ${\tilde m}_q$, the index $N$ and the scaling dimension $\Delta_\O$ of the chiral operator $\O$. The EDM is normalized in units of $(\tan\beta /5)^2 \, 4\times 10^{-30} ~e\,{\rm cm}$ and the shaded region is experimentally excluded for $\tan\beta =5$.
Gauge mediation mass relations and radiative EW breaking are assumed. Blue (red) lines refer to ${\tilde m}_q=10$ TeV (${\tilde m}_q=3$ TeV) and each set corresponds to $\Delta_\O=$ 1, 1.2, 1.4, 1.6, 1.8, 2 (from bottom to top). We estimate that the theoretical uncertainties of these results could be as large as one order of magnitude.}
\label{fig:EDMBeyondSoft}
\end{figure}

To have a more transparent interpretation of the result, we can use the analysis presented in \sec{sec:masss} and recast \eq{EDM3} in terms of our fundamental inputs ${\tilde m}_q$, $N$ and $\Delta_\O$. This is done in \fig{fig:EDMBeyondSoft}, which shows the prediction for the electron EDM coming from Beyond Soft wrong-Higgs Yukawa interactions, expressed in units of $(\tan\beta /5)^2\, 4\times 10^{-30} ~e\,{\rm cm}$. The shaded region is experimentally excluded for $\tan\beta =5$. Blue lines refer to a squark mass of 10 TeV, while red lines refer to a squark mass of 3 TeV. In each set of lines, $\Delta_\O$ is varied in the range 1--2 (from bottom to top).

We stress that the lines shown in \fig{fig:EDMBeyondSoft} should not be interpreted as precise predictions, but only as order-of-magnitude estimates. Indeed, our calculation suffers from many sources of uncertainties, since we assume: {\it (i)} maximal allowed CP violating phase in the leptonic sector; {\it (ii)} full strength coupling of the non-holomorphic trilinear K\"ahler interaction in \eq{Klep}; {\it (iii)} exact gauge mediation mass relations; {\it (iv)} the $\mu$ parameter (and consequently $\mst$), which has been derived from the approximate expression of EW breaking in \eq{EEWW}. Summing up all these approximations, it is reasonable to expect that the theoretical uncertainties of the results shown in \fig{fig:EDMBeyondSoft} may reach one order of magnitude. 

The estimate presented in \fig{fig:EDMBeyondSoft} shows a parametric uncertainty of about two orders of magnitude, to which one must add a theoretical uncertainty from strongly-coupled dynamics of about another order of magnitude (upwards and downwards). Clearly, it is difficult to make definitive conclusions. However, in spite of these limitations, \fig{fig:EDMBeyondSoft} contains several useful messages.

The first message is that, for squarks close to their LHC mass limit,  the electron EDM may impose an upper limit on $\Delta_\O$. There is nothing surprising in this result. The solution of the flavor problem in the Super-Composite Higgs is based on a separation of roles between $\mS$, which controls naturalness, and $\mst$, which controls flavor. A basic hypothesis of the Super-Composite Higgs is the existence of a certain hierarchy in the ratio $\mst/\mS$. As shown in \eq{uffamu}, this hierarchy has an exponential dependence on $\Delta_\O$ and is depleted by large values of $\Delta_\O$. Therefore, a sufficient hierarchy in $\mst/\mS$ always requires an upper bound on $\Delta_\O$. The message from \fig{fig:EDMBeyondSoft} is that this upper bound may be more stringent than 2, in the extreme case of relatively light squarks with large $\tan\beta$ and $N$. However, the important result is that the theory is compatible with the electron EDM constraint for {\it any} squark mass above the LHC limit, and for $1<\Delta_\O< \Delta_\O^{\textrm{max}}$, where $\Delta_\O^{\textrm{max}}$ depends on ${\tilde m}_q$, $N$, $\tan\beta$, and the unknown factors due to strong dynamics.

The proximity of our electron EDM estimates to the current experimental limit of $4.1 \times 10^{-30}~e\,$cm~\cite{Roussy:2022cmp} is the most exciting message from \fig{fig:EDMBeyondSoft}, especially in view of the significant improvements expected from new experimental techniques using molecules, instead of the more traditional ultracold atoms. Upgradings of existing experiments, such as ACME or JILA, will soon allow for electron EDM searches below $10^{-30} ~e\,{\rm cm}$ \cite{Wu:2022yqp,Ang:2022ikd}. Moreover, advancements in molecular cooling, trapping and control techniques offer new routes to improve sensitivity up to $10^{-33} ~e\,{\rm cm}$, within about a decade~\cite{DeMille:2024djy}. The Super-Composite Higgs provides a valuable target for these ambitious searches. The unitarity upper bound on $\mst$, discussed in \sec{sec:masss}, guarantees a lower bound on the electron EDM well above the SM value and within future detectability.

\subsubsection*{Flavor transitions from Higgs exchange with wrong-Higgs Yukawa}

The wrong-Higgs Yukawa interaction in \eq{wronghiggs} leads to flavor transitions mediated by the heavy Higgs. Let us consider the interaction for down-type quarks
\beq
{\bar d}_{R_j} \, q_{L_i} \left( y_{d_{ij}}H_d+ \frac{\mu}{\mst}\,  \xi_{Q_{i}}\xi_{{\bar D}_{j}}\, 
H_u^\dagger \right) \, ,
\label{wronghiggsd}
\eeq
where, for simplicity, we have omitted unknown order-one and generation-dependent coefficients in the second term of \eq{wronghiggsd}.
In the limit of large $\tan\beta$, the two mass eigenstates of the neutral scalar Higgses are $h=H_d^0 \cos\beta + H_u^{0*}\sin \beta$, which corresponds to the SM Higgs, and $H=-H_d^0 \sin\beta + H_u^{0*}\cos \beta$, which has zero vev. In terms of these fields, the neutral component of the Yukawa interaction in \eq{wronghiggsd} becomes
\beq
{\bar d}_{R_j} \, d_{L_i} \left[ y_{d_{ij}}^{({\rm SM})} \, h +\Big( -\tan\beta \, y_{d_{ij}}^{({\rm SM})}+\frac{\mu}{\mst \cos\beta}\,  \xi_{Q_{i}}\xi_{{\bar D}_{j}}\Big)\, H \right] \, ,
\label{wronghiggsh}
\eeq
\beq
y_{d_{ij}}^{({\rm SM})} =\cos\beta \, y_{d_{ij}}+ \frac{\mu \sin\beta}{\mst}\,  \xi_{Q_{i}}\xi_{{\bar D}_{j}} \, ,
\eeq
where $y_{d_{ij}}^{({\rm SM})}$ is the SM Yukawa coupling.

Equation (\ref{wronghiggsh}) shows that the heavy Higgs $H$ develops flavor-violating couplings. Tree-level $H$ exchange leads to a 4-fermion interaction
\beq
\frac{\xi_{Q_{i}}^* \xi_{{\bar D}_{j}}^* \xi_{Q_{k}}\xi_{{\bar D}_{\ell}}}{\cos^2\beta \, \mst^2}\,
{\bar d}_{L_i} \, d_{R_j} \, {\bar d}_{R_\ell} \, d_{L_k} \, ,
\label{4fwrong}
\eeq
where we have taken the heavy Higgs mass to be $m_H^2 \approx \mu^2$. The interactions in \eq{4fwrong} have the same structure of the K\"ahler dim-6 contact interactions that we discussed at the beginning of this section. However, an important difference is that the Higgs-mediated 4-fermion interactions are enhanced by a $\tan^2\beta$ factor.

The limits on the compositeness scale can be read from \tab{tab:4ferm}, after accounting for the additional $\tan\beta$ effect. The strongest bound comes from $\varepsilon_K$ and is in the range
\beq
\mst \gsim \left(\frac{\nicefrac{1}{5}}{\cos\beta}\right)^2 \, 700~{\rm TeV}\, .
\label{limgrtan}
\eeq
This constraint is weaker than the one from the electron EDM in \eq{limEDMw}.

Besides $\Delta F=2$ transitions, the tree-level $H$ exchange can contribute to $\Delta F=1$ processes too. A good example is the decay $B_s \to \mu \mu$, which receives the following contribution 
\bea
{\rm BR} (B_s \to \mu \mu)&\approx &\frac{M_{B_s}^3 f_{B_s}^2 \tan^2\beta}{32 \pi \Gamma_{B_s}\, \mu^2 \mst^2 \cos^2\beta} \left( \xi_{Q_3}^2 \xi_{{\bar D}_2}^2 + \xi_{Q_2}^2 \xi_{{\bar D}_3}^2\right) 
{y_{\mu}^{({\rm SM})}}^2\, ,
\nonumber \\ &\approx &
\left( \frac{\nicefrac{1}{5}}{\cos\beta} \right)^4
\left( \frac{\tan\beta}{5} \right)^2
\left( \frac{\rm TeV}{\mu}\right)^2
\left( \frac{100~{\rm TeV}}{\mst}\right)^2
4 \times 10^{-12} \, ,
\label{bsmumu}
\eea
which is below current experimental sensitivity.

\boldmath
\subsubsection*{Flavor transitions and CP violation from soft terms}
\unboldmath

The last source of flavor and CP violation that we will consider is flavor misalignment in soft terms. This is the familiar effect present in all supersymmetric theories with general soft terms, usually parametrized in terms the mass insertions $\delta_{ij}$, defined as the ratio between the $ij$ element of the squark/slepton squared mass matrix and the typical mass eigenvalue. 

The mass-insertion method is particularly appropriate for the Super-Composite Higgs, as the squark/slepton squared mass matrix is the sum of a gauge-mediation-like term, which gives the leading contribution and is flavor diagonal, and a flavorful contribution, whose structure is governed by partial compositeness, as well as a supersymmetric term equal to the corresponding squark/slepton masses. Decomposing the scalar squared mass matrix into left-left ($LL$), right-right ($RR$), left-right ($LR$) and right-left ($RL$) $3\times 3$ submatrices, for up squarks we can write 
\beq
{\cal M}^2_{\tilde u} =
\begin{pmatrix}
{\cal M}^2_{{\tilde u}_L{\tilde u}_L} & 
{\cal M}^{2\, *}_{{\tilde u}_L{\tilde u}_R} \\
{\cal M}^{2\, \mathsf{T}}_{{\tilde u}_R{\tilde u}_L} & 
{\cal M}^{2\, *}_{{\tilde u}_R{\tilde u}_R} \\
\end{pmatrix}
\, ,
\eeq
\beq
{\cal M}^2_{{\tilde u}_L^i {\tilde u}_L^j} =
{\tilde m}_{q_L}^2 \delta_{ij} + (m_u^\dagger m_u)_{ij} + \xi_{Q_i}^* \xi_{Q_j} \, \frac{\sqrt{N} \, B_\mu}{4\pi} \, ,
\eeq
\beq
{\cal M}^2_{{\tilde u}_R^i {\tilde u}_R^j} =
{\tilde m}_{u_R}^2 \delta_{ij} + (m_u m_u^\dagger)_{ij} + \xi_{{\bar U}_i}^* \xi_{{\bar U}_j} \, \frac{\sqrt{N} \, B_\mu}{4\pi} \, ,
\eeq
where ${\tilde m}_{q_L,u_R}^2$ are the gauge-mediation-like soft masses and $m_u$ is the up-quark mass matrix. Analogous expressions can be written for down squarks and charged sleptons.

The $LR$ and $RL$ blocks are proportional to the corresponding Yukawa matrices. Therefore, they are diagonal are real, in the mass eigenstate basis for quarks and leptons. This has the important consequence that flavor transitions from soft terms occur only in the $LL$ and $RR$ sectors, and that CP violation occurs only through $LL$ and $RR$ generation mixings. 

Flavor non-diagonal left-right mixings are absent at leading order because of our assumption that the superpotential interactions involving $\O_3$ in \eq{pot1} are not generated at the scale $\mst$. Beyond Soft non-diagonal $A$-terms are induced at $\O(\msoft^2 /\mst)$ and their contribution, leading to the electron EDM given in \eq{EDM2}, has already been discussed.

In summary, in the mass eigenbasis for quarks and leptons, the relevant mass-insertion parameters are 
\beq
\delta^{(u,d)LL}_{ij}\approx \frac{\xi_{Q_i}^* \xi_{Q_j} \sqrt{N} B_\mu}{4\pi \, {\tilde m}_{q_L}^2} \, , ~~~~
\delta^{(u)RR}_{ij}\approx \frac{\xi_{{\bar U}_i} \xi_{{\bar U}_j}^* \sqrt{N} B_\mu}{4\pi \, {\tilde m}_{u_R}^2} \, , ~~~~
\delta^{(d)RR}_{ij}\approx \frac{\xi_{{\bar D}_i} \xi_{{\bar D}_j}^* \sqrt{N} B_\mu}{4\pi \, {\tilde m}_{d_R}^2} \, , 
\eeq
\beq
\delta^{(e )LL}_{ij}\approx \frac{\xi_{L_i}^* \xi_{L_j} \sqrt{N} B_\mu}{4\pi \, {\tilde m}_{\ell_L}^2} \, , ~~~~
\delta^{(e)RR}_{ij}\approx \frac{\xi_{{\bar E}_i} \xi_{{\bar E}_j}^* \sqrt{N} B_\mu}{4\pi \, {\tilde m}_{e_R}^2} \, , 
\eeq
where the masses in the denominators refer to the gauge-mediation expressions, up to order-one unknown factors.

The mass insertion parameters in the Super-Composite Higgs are sufficiently small to pass all experimental tests from flavor physics. The strongest bound comes from the mixed LL/RR contribution to $\varepsilon_K$~\cite{Ciuchini:1998ix} which imposes the following insignificant constraint (calculated for equal gluino and squark masses)
\beq
\sqrt{ \text{Im}\, \delta^{LL}_{ds} \delta^{RR}_{ds}}< \left( \frac{{\tilde m}_{q}}{\text{TeV}}\right)  \, 2.6 \times 10^{-4} ~~~\Rightarrow ~~~
{\tilde m}_{q} >  
\sqrt{\frac{N}{10}}  
\, \frac{B_\mu}{{\tilde m}_{q}^2 }
\left( \frac{\nicefrac{1}{5}}{\cos \beta} \right)  \, 300~ \text{GeV} \, .
\label{limfs}
\eeq
All other flavor processes give much weaker bounds.

\boldmath
\subsection{Summary of the constraints from flavor and CP violation}
\unboldmath
\label{flavorsummary}

As one of the aims of the Super-Composite Higgs is to provide a dynamical explanation for the structure of the Yukawa matrices, it is not surprising that the theory is rich in new sources of flavor violations, which can potentially lead to experimentally relevant effects. To summarize the results presented in this section, it is useful to classify the effects in three classes. 

\subsubsection*{I. Flavor/CP violation from compositeness}

The first class of effects comes from compositeness in the limit of exact supersymmetry. Higgs compositeness implies the existence of dim-6 K\"ahler operators, suppressed by $1/\mst^2$ and with a flavor structure governed by the pattern of partial compositeness. These operators lead to constraints on $\mst$ which are inescapable because their effects originate from the K\"ahler potential as a pure consequence of compositeness. They are a basic feature of the Super-Composite Higgs that cannot be removed using model-dependent assumptions. The strongest constraint from this class comes from $\varepsilon_K$ and is $\mst \gsim(\nicefrac{1}{5}/\cos\beta )\,130$ TeV (see \tab{tab:4ferm}). 

This class of effects reflects the composite nature of the Higgs and leads to results analogous to those found in non-supersymmetric composite models. There is an important difference, though. In non-supersymmetric composite models, the leading constraint comes from a dim-6 dipole operator, with coefficient $1/\mst^2$, which contributes to the electron EDM and gives the bound $\mst \gsim 2.2$~PeV. In the Super-Composite Higgs, as far as this class of effects is concerned, there are no dipole operators from pure compositeness, as they are forbidden by exact supersymmetry.

\subsubsection*{II. Flavor/CP violation from soft terms}

The second class of effects comes from the supersymmetric soft breaking terms. These contributions are analogous to those of generic supersymmetric models. However, this class of flavor violations in the Super-Composite Higgs are tamed by the pattern of partial compositeness, much in the fashion of flavorful supersymmetry. As a result, this class of effects does not provide any constraint relevant for the Super-Composite Higgs, see \eq{limfs}.

\subsubsection*{III. Flavor/CP violation from Beyond Soft terms}

The effects in class I and II reproduce the familiar behavior encountered in the two main ingredients of the Super-Composite Higgs: compositeness and supersymmetry. However, the combination of compositeness and supersymmetry leads to novel effects, which are not present in either theory in isolation. We have called this class of effects Beyond Soft. Beyond Soft terms are a distinctive feature of the Super-Composite Higgs, as they emerge from a close interplay between compositeness and supersymmetry breaking.

Beyond Soft terms come from supersymmetry breaking operators originating from compositeness, and therefore suppressed by powers of $\mst$. The novelty is that they introduce flavor-violations in the Lagrangian of the supersymmetric theory that cannot be described merely by soft terms. Examples are non-holomorphic $A$-terms or wrong-Higgs Yukawa couplings. Note that even the holomorphic trilinear terms originating from $Z_{y}$ and $Z_{\hat y}$ in \eq{kal1} should be classified as Beyond Soft, since they correspond to operators suppressed by $1/\mst$. Moreover, they are also accompanied by wrong-Higgs $A$-terms, which is another sign of being Beyond Soft.

Among Beyond Soft effects, the contributions to the electron EDM provide the strongest constraints. Direct dipole contact interactions yield the bound in \eq{EDMel}, while dipoles from Beyond Soft $A$-terms yield the bound in \eq{limEDMtr}. However, the most severe limits come from Beyond Soft wrong-Higgs Yukawa couplings. Heavy Higgs exchange contributes to $\varepsilon_K$, leading to the bound $\mst \gsim (\nicefrac{1}{5}/\cos\beta)^2$ 700 TeV, see \eq{limgrtan}. Especially important is the wrong-Higgs Yukawa contribution to the electron EDM, yielding the bound $\mst \gsim (\tan\beta/5)^2(\textrm{TeV}/M_{\tilde B})$ 4 PeV, see \eq{limEDMw}. The effectiveness of these constraints is due to the extra power of $\tan\beta$.

\subsubsection*{The electron EDM as a probe of the Super-Composite Higgs}
As summarized above, the leading constraint on the Super-Composite Higgs comes from the contribution of Beyond Soft wrong-Higgs Yukawa couplings to the electron EDM. The exciting result is that this contribution is typically close to the experimental limit on the EDM, as illustrated in \fig{fig:EDMBeyondSoft}. This is particularly interesting in view of the expected experimental improvements that should soon allow for an electron EDM exploration below the $10^{-30}~e\,{\rm cm}$ level, with a longer-term target of $10^{-33}~e\,{\rm cm}$ (see {\it e.g.} ref.~\cite{DeMille:2024djy}). 

Although theoretical uncertainties prevent us from making precise predictions, our estimates are in the right ballpark for future discovery. This is not the result of a special choice or an adjustment of parameters. The Super-Composite Higgs predicts a rather rigid hierarchical structure of scales, with a strict upper bound on $\mst$ based on unitarity. The intriguing coincidence is that the unitarity upper bound places the estimate for the electron EDM in the right range for future observability. 

\subsubsection*{Dipoles and Beyond Soft terms}
We would like to elaborate  on the presence of dipoles in a supersymmetric theory. 
These effects can only exist after supersymmetry breaking, but they arise from more than one
channel. Overall, we can identify {\it {short}} and {\it {long}} distance effects. The first class, exemplified by \eq{EDMel}, arises directly at
the shortest possible scale that permits dipoles, $m_S$, and also necessitates  matter and Higgs compositeness. The Wilson coefficient of the resulting dim-6 effective operators is $O(\mu/\mst^3)$, see \eq{EDMel}, which makes it
 subdominant. 
 
 The leading effects are instead the long-distance ones, determined by quantum fluctuations (one-loop diagrams) at the sparticle mass scale $\msoft$. The microscopic seeds of these effects above $\msoft$ are both soft and Beyond Soft, with the latter giving the dominant contribution. The Wilson coefficient of the resulting  dim-6 dipole operator is parametrically of order
\beq
\frac{g^2}{16 \pi^2\msoft \mst} \sim \frac{1}{N \mS \mst} \sim \frac{R_\mu^{1/\Delta_\O}}{N\,\mst^2}\,.
\eeq
According to eq.~\eqref{stima2}, for $\Delta_\O\sim 2$ this is comparable to the $1/\mst^2$ Wilson coefficient
of the same operator in ordinary composite Higgs, while for $\Delta_\O\sim 1$ is slightly larger. Even if exact supersymmetry forbids dipole operators, the actual size of dipole Wilson coefficients in the Super-Composite Higgs turn out to be comparable to those in ordinary compositeness. It must also be noted that, numerically, in the supersymmetric case the loop functions introduce a numerical reduction by about an order of magnitude to the above estimate, but  that is essentially neutralized by a $\tan^2\beta$ enhancement. 

Therefore, in both scenarios, the electron EDM places the scale of compositeness in the PeV range.
 There is an important difference, though. In ordinary composite Higgs, the $O$(PeV) limits derived from EDM are lethal, making the theory highly unnatural and requiring {\it ad hoc} adjustments. On the contrary, in the Super-Composite Higgs, values of $\mst$ in the PeV range not only are acceptable, as Higgs naturalness is governed by supersymmetry (and not by compositeness), but are a necessary feature, since a scale separation $\msoft < \mS < \mst$ is a basic ingredient of our setup.

\boldmath
\subsubsection*{The role of $\tan\beta$}
\unboldmath
Finally, we would like to comment on the role of $\tan\beta$ in flavor observables. Relatively large values of $\tan\beta$ are usually advocated to obtain the correct value of the Higgs mass without exceedingly heavy stops. Large $\tan\beta$ enhances Beyond Soft effects as well. The most striking effects occur in the case of the wrong-Higgs Yukawa interaction: the contribution to the EDM scales as $\tan^2\beta /\mst$, see \eq{EDM3}, and the contribution to $\varepsilon_K$ scales as $(\tan^2\beta /\mst)^2$, see \eq{limgrtan}. Even more pronounced is the $\tan\beta$ enhancement in $B_s \to \mu \mu$, whose branching ratio scales as $(\tan^3\beta /\mst)^2$, see \eq{bsmumu}. For this reason, the bounds on $\mst$ from flavor processes become particularly strong at large $\tan\beta$. It should be kept in mind that the relation between $\tan\beta$ and the Higgs mass is specific to the minimal Higgs structure and could be altered by variations of the minimal model.

\section{Beyond minimality}
\label{sec:extrasym}

The minimal setup described so far offers a spectacularly successful realization of the Super-Composite Higgs framework.
The theory is based on the existence of an operator dimension gap ensuring that the supersymmetry breaking effects of compositeness are well described by the lowest lying chiral operator $\O$ and by the associated vector $\R\sim \O\O^\dagger$. The resulting scenario is essentially governed by only 3 free parameters that describe the entire theoretical edifice. The theory makes qualitative but robust predictions about the hierarchy of scales and, using the criterion of unitarity, dictates an upper bound on the compositeness scale $\mst$ of at most 100 PeV. The Super-Composite Higgs satisfies all constraints from flavor and CP violation, and makes the exciting prediction that the electron EDM is within experimental reach.

In spite of these unquestionable successes of the minimal setup, here we would like to explore possible variations of the Super-Composite Higgs framework.

\subsection{Additional symmetries}
\label{sec:addit}
Our formulation of the theory is based on a SILSH Lagrangian in which interactions are weighted by appropriate powers of couplings and scales, but no additional (approximate) symmetries are imposed at the scale $\mst$. Introducing new symmetries can forbid certain interactions and affect the results of the minimal setup.

In our construction, we have granted ourselves the freedom to omit operators from the superpotential, on the basis of holomorphy.
This is justified by the non-renormalization theorem and technically acceptable. However, there is no theoretical control over the reasons why these operators do not exist. Having a symmetry reason would be more satisfactory. 

Appealing to the non-genericity of the superpotential, the terms that were excluded from our analysis are (in a schematic notation)
\beq
\begin{array}{c}
{\cal W} ~~= \\[-0.3cm]
\textrm{\scriptsize \phantom{uffa}}
\\[-0.5cm]
\textrm{\scriptsize \phantom{uffa}}
\end{array}
\begin{array}{c}
H_uH_d \\[-0.3cm]
\textrm{\scriptsize dangerous}\\[-0.5cm]
\textrm{\scriptsize $\mu$-term}
\end{array}
\begin{array}{c}
+ \\[-0.3cm]
\textrm{\scriptsize \phantom{uffa}}
\\[-0.5cm]
\textrm{\scriptsize \phantom{uffa}}
\end{array}
\begin{array}{c}
H_{u,d}\, \Phi_L \Phi_{\bar R}\, \O \\[-0.3cm]
\textrm{\scriptsize dangerous}\\[-0.5cm]
\textrm{\scriptsize $A$-terms}
\end{array}
\label{danger}
\eeq
where $\O$ is the supersymmetry breaking chiral operator.
The goal here is to identify global, $R$ or discrete symmetries that forbid the terms in \eq{danger}. At the same time, we have to make sure that the symmetry does not exclude beneficial operators in the K\"ahler, of the kind
\bea
\begin{array}{c}
{\cal K}  \\[-0.3cm]
\textrm{\scriptsize \phantom{uffa}}
\end{array}
&
\begin{array}{c}
= \\[-0.3cm]
\textrm{\scriptsize \phantom{uffa}}
\end{array}
&
\begin{array}{c}
(\Phi^\dagger \Phi+ H^\dagger H)\,{\cal R}
 \\[-0.3cm]
\textrm{\scriptsize matter+Higgs soft masses}
\end{array}
\begin{array}{c}
+ \\[-0.3cm]
\textrm{\scriptsize \phantom{uffa}}
\end{array}
\begin{array}{c}
H_uH_d \,\O^\dagger \\[-0.3cm]
\textrm{\scriptsize $\mu$-term}
\end{array}
\begin{array}{c}
+ \\[-0.3cm]
\textrm{\scriptsize \phantom{uffa}}
\end{array}
\begin{array}{c}
H^\dagger H\,{\cal O}
 \\[-0.3cm]
\textrm{\scriptsize $A$-terms}
\end{array}
\nonumber \\
&
\begin{array}{c}
+ \\[-0.3cm]
\textrm{\scriptsize \phantom{uffa}}
\end{array}
&
\begin{array}{c}
H_{u,d}\, \Phi_L \Phi_{\bar R}\, \O^\dagger \\[-0.3cm]
\textrm{\scriptsize non-holomorphic $A$-terms}\\[-0.45cm]
\textrm{\scriptsize EDM effect}
\end{array} 
\begin{array}{c}
+ \\[-0.3cm]
\textrm{\scriptsize \phantom{uffa}}\\[-0.45cm]
\textrm{\scriptsize \phantom{uffa}}
\end{array}
\begin{array}{c}
H_{d,u}^\dagger\, \Phi_L \Phi_{\bar R}\, \O^\dagger \\[-0.3cm]
\textrm{\scriptsize wrong-Higgs Yukawa}\\[-0.45cm]
\textrm{\scriptsize EDM effect}
\end{array} 
\begin{array}{c}
+~\hc \\[-0.3cm]
\textrm{\scriptsize \phantom{uffa}}\\[-0.45cm]
\textrm{\scriptsize \phantom{uffa}}
\end{array}
\label{tengo}
\eea
where $\cal R$ denotes the composite operator $\O^\dagger \O$. The last term in the first line of \eq{tengo} leads to flavor-universal trilinear $A$-terms that are useful to obtain the right Higgs quartic without relying on stop masses above 10--14 TeV. The two terms in the second line of \eq{tengo} 
describe the operators responsible for the contributions to the electron EDM in eqs.~(\ref{EDM2}) and  (\ref{EDM3}). The wrong-Higgs Yukawa interaction is responsible for the prediction that the electron EDM is within experimental reach. We wish to preserve the corresponding operator since it provides one the most appealing features of the Super-Composite Higgs, phenomenologically speaking. 
However, if the operator is missing, the non-holomorphic $A$-term provides a contribution to the electron EDM that is about $4\times \tan\beta$ times smaller than the one from wrong-Higgs Yukawa interactions. Given the large parametric and theoretical uncertainties in our estimate, this could also lead to a measurable effect.

We can now explore some examples of symmetries that address the program of forbidding \eq{danger} while retaining \eq{tengo}. 

\subsubsection*{PQ symmetry}

The simplest example of an additional symmetry is a global Peccei-Quinn $U(1)_{\textrm{PQ}}$. The Higgs superfields $H_{u,d}$ have both charge $1$ while all matter superfields have charge $-1/2$, compatibly with the existence of the Yukawa interactions. The idea is that 
 $U(1)_{\textrm{PQ}}$, or possibly one of its ${\mathbb Z}_N$ subgroups, emerges as an accidental symmetry below the far UV scale $\LUV$ and is broken spontaneously at $m_S$. Its validity at $m_*$ then provides selection rules on the  effective action, in particular on the coupling between matter fields and the ${\CFTs}$ composites. The spectrum of dimensions and PQ charges of these operators crucially matters. The  generation of  a $\mu$ term
 of the proper size requires the mediation of supersymmetry breaking to be dominated by a ${\CFTs}$ chiral operator with  PQ charge 2. Let us indicate this operator by ${\tilde \O_2}$ and its real vector associate ${\tilde \O_2}^\dagger {\tilde \O_2}$ by $\tilde \R_0$. Moreover, let us assume that there exists a dimension gap sufficient  to neglect the effects of operators of higher dimension. One can then easily check that, out of all the structures listed in eqs.~(\ref{kalH}), (\ref{kal1})--(\ref{pot1}), those that survive in the K\"ahler potential are
 \beq
 {\cal K}=(\Phi^\dagger \Phi+ H^\dagger H + H_{u,d}\, \Phi_L \Phi_{\bar R})\,  \tilde \R_0 + H_uH_d\, \tilde\O_2^\dagger + H_{d,u}^\dagger\, \Phi_L \Phi_{\bar R} \, \tilde \O_2 +{\mathrm{h.c.}}
 \label{eq:PQkahler1}
 \eeq
Both the $\mu$ term and the dangerous $A$ terms in \eq{danger} are now barred from the superpotential  for symmetry reasons, regardless of holomorphy. However, the disappointing aspect is that the PQ symmetry forbids also the last three terms in \eq{tengo}. This makes it problematic to obtain the right Higgs mass and completely kills any observable effect in flavor/CP physics. The result is unsatisfactory.

A possible way out is to assume yet another chiral operator $\tilde\O_0$ of similar dimension, $\Delta_{\tilde \O_0}\simeq \Delta_{\tilde \O_2}$. Adding it to the set of composites, along with the vectors $\tilde \R_2=\tilde \O_0^\dagger \tilde \O_2$ and $\tilde \R_{-2}=\tilde \O_0 \tilde \O_2^\dagger$, we obtain the additional K\"ahler structures
  \beq
  {\cal K}= H^\dagger H (\tilde\O_0+\tilde\O_0^\dagger)+H_uH_d\, \tilde\R_{-2}+ H_{d,u}^\dagger \Phi_L \Phi_{\bar R}\, \tilde \R_2 +{\mathrm{h.c.}}
  \eeq
This allows for sizable $A$-terms, beneficial for the Higgs quartic but still forbidding the phenomenologically interesting contribution to the electron EDM from wrong-Higgs Yukawa interactions. On a positive note, non-holomorphic $A$-terms are allowed, although their contribution to the electron EDM is somehow suppressed, see \eq{EDM2}. Another problem with this solution is that the dangerous $A$-terms in the \eq{danger} are now allowed by symmetry and we must appeal to the non-genericity of the superpotential to eliminate them. 

Overall, the gain has been modest. A global PQ symmetry cannot accomplish the task of forbidding \eq{danger} while, at the same time, allowing \eq{tengo}. The best it can do is to explain the absence of $\mu$, but the absence of the dangerous $A$-terms must come from the properties of the superpotential. Moreover, the price for this limited gain is that the leading contribution to the electron EDM is lost.

 \subsubsection*{PQ-axion}  

An interesting phenomenological consequence of the global PQ symmetry is the presence of a light axion-like scalar $a$ that emerges after spontaneous symmetry breaking. The PQ-axion interacts with ordinary quarks, leptons, gluons and photons and is subjected to the various constraints from ALPs searches, including beam-dump experiments, $B$ and $K$ rare decays, and other processes~\cite{Bauer:2021mvw,Antel:2023hkf}. The resulting limits on the decay constant $f_a$ are fairly model dependent, especially because the PQ-axion decay modes are affected by the mixings with the various SM pseudoscalar excitations, which introduce large unknowns~\cite{Ovchynnikov:2025gpx}, and because the flavor-changing effects due to PQ-axions are determined by the RG evolution from the symmetry-breaking scale~\cite{Bauer:2021mvw}. Nevertheless, the limits on $f_a$ can be significant, even reaching the PeV. This would practically rule out our scenario, as our power counting  implies $f_a\sim \mS/g_* \lsim 100$ TeV. 

However, most of the experimental limits on $f_a$ evaporate for PQ-axions heavier than the $B$ meson. In our scheme, $U(1)_{\textrm{PQ}}$ is accidental
below $\LUV$ and broken by irrelevant operators. Assuming that the leading effect comes  at dimension  $\Delta_{a}$, simple power counting implies 
\be
m_a^2 \sim \frac{\mS^{\Delta_a-2}}{\LUV^{\Delta_a-4}}\, .
\ee
For $\LUV> M_{GUT}$, the PQ-axion mass can be above the $B$ meson mass, as long as
$\Delta_A< 7$. The forthcoming explorations on the intensity frontier will extend existing searches and the PQ-axion of Super-Composite Higgs models could provide an interesting experimental target.

An alternative option is to settle for a discrete subgroup of $U(1)_{PG}$, which squarely eliminates the axion. The cosmological problem from domain walls can be easily avoided by taking into account the explicit symmetry breaking associated with small irrelevant effects.

\boldmath
 \subsubsection*{$R$-symmetry} 
 \unboldmath
 
We can also contemplate the possibility that the theory at $m_*$ be endowed with an $R$-symmetry spontaneously broken at $m_S$. Here things are trickier because the CFT above 
$m_*$ possesses a $U(1)_R$ which is part of the conformal group. In particular the chiral composites $\O_{{\bar \Phi}_a}\equiv (\O_{{\bar Q}_a},\O_{{U}_a},\O_{{D}_a}, \O_{{\bar L}_a}, \O_{{E}_a})$ have $R$ charges dictated by their scaling dimensions: $R^\Phi_a =\nicefrac{2}{3}\, \Delta^\Phi_a $. Any $R$-symmetry surviving at $m_*$ will have to be a subgroup of the product of this $U(1)_R$ times any other global symmetry of the CFT. The breaking of $U(1)_R$  at $m_*$ can arise from the same relevant deformation that gaps the CFT. In other words, the breaking to a discrete subgroup can be explict, with no associated light axion.
The most minimal option is that there are no other global symmetries and just a discrete subgroup of $U(1)_R$ survives. 

Consider, for instance, a ${\mathbb Z}_4$ subgroup under which any superfield transforms as
\be
\Psi(\theta,\bar \theta,x) \to i^q \Psi(-i\theta,i\bar \theta,x)
\ee
with integer $q$. The choice $q=2$ for all matter fields and Higgses, allows the Yukawa interactions and forbids the $\mu$ term.
By this choice, invariance of the elementary-composite mixings at the origin of partial compositeness implies the composites $\O_{{\bar \Phi}_a}$ must all have $q=0$ under the residual ${\mathbb Z}_4$. Under $U(1)_R$ they transform as
\be
\O_{{\bar \Phi}_a}(\theta,\bar\theta,x)\to e^{iR^\Phi_a  \alpha}\O_{{\bar \Phi}_a}(e^{-i\alpha}\theta,e^{-i\alpha}\bar\theta,x)
\ee
and the embedding of ${\mathbb Z}_4$ into $U(1)_R$ gives a non-trivial constraint on the R-charges. Indeed the generator of ${\mathbb Z}_4$ corresponds to a $U(1)_R$ element with 
$\alpha= \pi(n+1/2)$ for $n$ integer.  The request that all $\O_{{\bar \Phi}_a}$ have $q=0$ under ${\mathbb Z}_4$ then becomes
\be
R^\Phi_a  =\frac{4 n^\Phi_a }{2n+1} \qquad \Rightarrow \qquad \Delta^\Phi_a =\frac{6 n^\Phi_a }{2n+1}
\label{rational}
\ee
with $n^\Phi_a $ integers. This result must be matched with the spectrum of dimensions that is 
implied by the observed fermion masses through partial compositeness. In practice, for reasonable values of the UV scale
where the RG flow starts, one needs the dimensions $\Delta^\Phi_a$ to be distributed between 2 and 3 with separations of roughly 0.2--0.4. This requires an inverse unit of charge $2n+1$ of at least 15, and 29 is a good choice. This seems undoubtedly a bit baroque, though not inconsistent.

Before discussing the consequences of ${\mathbb Z}_4$, one last thing to worry about is its relation with the $U(1)_R$ of ${\CFTs}$. In the absence of other global $U(1)$ symmetries, it will have to be a subgroup of this $U(1)_R$. The breaking of the latter at $m_S$ could well be explict, and associated with the relevant deformation that gaps ${\CFTs}$. Still, the breaking of ${\mathbb Z}_4$ could be purely spontaneous, in which case its selection rules are controlled only by the vevs of the ${\CFTs}$ operators. We shall work under this hypothesis. Again, we must assume that some irrelevant UV breaking of 
${\mathbb Z}_4$ rids us of the cosmological domain wall problem, but that is easily accomplished.

Finally, consider the phenomenological consequences of ${\mathbb Z}_4$. Notice that this symmetry forbids a straight $\mu$ term at the scale $m_*$, so it is also a PQ symmetry. The $\mu$ term is however generated from the mechanism of ref.~\cite{Giudice:1988yz}
if the leading composite $\O$  mediating soft terms carries ${\mathbb Z}_4$ charge $q=0$. With this assumption,
 ${\mathbb Z}_4$, all observable flavor effects are lost. However, as the K\"ahler terms $H^\dagger H(\O+\O^\dagger)$ are allowed, large universal $A$-terms are generated, which help with the Higgs quartic. 
 
While  the  contributions to the electron EDM associated with $\O$ vanish, we generically  expect   effects  mediated by composites with higher dimension and the suitable ${\mathbb Z}_4$ quantum number. The size of these effects is model dependent, but the prospects for improvent in its  experimental sensitivity still singles out $d_e$ as the leading indirect probe of the scenario.

\bigskip

In conclusion, the exploration of additional symmetries presented in this section has shown that there is no simple solution to accomplish the goal of finding a symmetry justification for the absence in the superpotential of the terms in \eq{danger} while retaining all K\"ahler interactions in \eq{tengo}. Partial solutions can be found but some of them suppress the operators that give the leading effects to the electron EDM. This study motivates us to investigate how robust the prediction on the electron EDM is. This question will be addressed in the next section.

\subsection{How robust is the prediction on the electron EDM?}
\label{sec:robust}
 
The observability of the electron EDM is a central prediction of the Super-Composite Higgs of great phenomenological and experimental significance. Thus, it is important to study the conditions under which this prediction may not hold. We discuss here four possible deformations of the theory in which the contributions to the electron EDM may be reduced or eliminated.

\subsubsection*{Extended Higgs sectors}

The leading effects of the wrong-Higgs trilinear are enhanced by more than one power of $\tan\beta$ and so do the resulting bounds on $m_*$. 
In our estimates we favored relatively large $\tan\beta$ in order to be able to reproduce the observed value of the Higgs mass with stops well within the reach of FCC-$hh$. One direction to explore is to change the game for the Higgs quartic by considering variants of the model where, along with the two Higgs doublets, other composite chiral superfields, such as $SU(2)_W$ singlets or triplets, can arise.

In this case, the existence of superpotential trilinears controlled by $g_*$ would drastically change the dynamics underlying the Higgs quartic, possibly leading to plausible scenarios with $\tan\beta\sim 1$, light stops and smaller but still interesting $d_e$. An important open question is to see how an extended Higgs sector modifies the upper bound on $\mst$ from unitarity and how this affects the prediction on the electron EDM.

The study of extended Higgs sectors is a project on its own, which we do not pursue in this paper. We leave this study for a future publication.

\subsubsection*{Additional symmetries} 

Unlike the K\"ahler interaction proportional to $c_{\Phi \Phi'}$ in \eq{eccok}, whose presence is robust as it cannot be forbidden by any obvious symmetry, the wrong-Higgs trilinear can be controlled by symmetries. We discussed some examples in \sec{sec:addit}. The simplest option is a PQ symmetry that eliminates the second operator in the second line of \eq{tengo}, but not the first one. As discussed above, this still has a chance of providing an observable electron EDM.

A more radical option is given by the discrete $R$-symmetry discussed in \sec{sec:addit}. This eliminates all contributions to $d_e$
that arise from leading, lowest dimension, Beyond Soft effects. We still expect $d_e$ to be generated by higher-dimensional composites, but the prediction for $d_e$ will depend on further assumptions.

\boldmath
\subsubsection*{$1/N$ suppression}
\unboldmath

In warped compactifications, $m_*$ corresponds to the curvature scale or, equivalently, to the mass of the Kaluza-Klein modes. In such constructions, one can consistently work
with ``minimal 5D Lagrangians'', where higher-derivative terms and some non-linear terms in the fields are controlled by a scale $M\gg m_*$, and thus suppressed. The Wilson coefficients of these operators are nonetheless generated from loop corrections when matching the 5D theory to the 4D theory at $m_*$, but they appear with a suppression $g_*^2/16\pi^2\sim 1/N$. Dipole operators in composite Higgs are known to be in this class (see for instance appendix
A of \cite{Liu:2016idz}). It seems consistent to assume that the wrong-Higgs trilinear may have this fate and come with a $1/N$ suppressed Wilson coefficient. If this were the case, the contribution to the electron EDM from wrong-Higgs Yukawa interactions would be reduced.

\subsubsection*{Messenger threshold}

Our study was restricted to the scenario with the least number of necessary mass scales.
A small variant would feature,  within the $\CFTs$ sector, an additional threshold $m_M$
below which only SM neutral states survive. Notice that, above and below $m_M$, the $\CFTs$ sector will be described by different fixed points. By making some minimal assumptions, we can work out the way eqs.~\eqref{softm}, \eqref{eqmu} and \eqref{eqbmu} would be modified in such scenario.
Let us assume that the chiral operator
of lowest dimension $\O$ dominates the scene. This operator will have different scaling dimensions $\Delta_{\O}$ and $\Delta_{\O}^\prime$ , respectively below and above $m_M$. The operator matching below and above the messenger threshold is
$\O_{below}=(m_M)^{\Delta_{\O}-\Delta_{\O}^\prime}\,\O_{above} $. Then, by repeating our EFT construction, we find that the gauge-mediated gaugino masses and $\mu$ term are
\beq
M_{\lambda}\sim \frac{g^2\, m_S^{\Delta_{\O}+1}}{g_*^2\, m_M^{\Delta_{\O}}}\,,\qquad\qquad \mu \sim \left(\frac{m_M}{\mst}\right )^{\Delta_{\O}^\prime}
\frac{m_S^{\Delta_{\O}+1}}{m_M^{\Delta_{\O}}} \, ,
\label{messag}
\eeq
with the scalar masses and $B_\mu$ following suit. Again the relation  $\mu \sim \msoft $ can be obtained by the accidental
relation $g^2/g_*^2\sim (m_M/m_*)^{\Delta_{\O}^\prime}$.

Note that, by taking the limit $m_M\to \mS$ and $\Delta_{\O}^\prime \to \Delta_{\O}$ in \eq{messag}, we recover the results valid in the minimal formulation of the Super-Composite Higgs for the gaugino mass, see \eq{softm}, and $\mu$, see \eq{eqmu}. Moreover, by taking the limit $\Delta_{\O}\to 1$, we recover the familiar expression for gaugino masses in gauge mediation.

It is interesting to
compare this result to \eq{okada}. From a 5D perspective, the separation between the brane at which the Higgs is localized and the brane at which supersymmetry is broken is what makes the difference. Now, with the additional parameter $m_M$, we are on the grounds of standard gauge mediation. 

From \eq{messag}, we see that the hierarchies between the scales $\mst > m_M>\mS$ are given by
\beq
\frac{\mst}{m_M}\sim \left( \frac{16\pi^2}{g^2N}\right)^{1/\Delta_{\O}^\prime}\, ,~~~~
\frac{m_M}{\mS}\sim \left( \frac{g^2N\, \mS}{16\pi^2\, \msoft}\right)^{1/\Delta_{\O}}\, .
\label{rapmes}
\eeq
With the introduction of the additional scale $m_M$, the upper bound on $\mst$ is lost. Now, $\mst$, $m_M$ and $\mS$ can slide to arbitrarily large values, while keeping $\msoft$ fixed, as long as the ratios in \eq{rapmes} are maintained. In this limit, all Beyond Soft effects disappear, together with their phenomenological consequences. 

\boldmath
\section{Ultra-light gravitino}
\unboldmath
\label{sec:cosmo}

The only particle belonging to the supersymmetry-breaking sector that survives below the scale $\mS$ is the massless spin-1/2 Goldstino. Once the theory is coupled to gravity by promoting supersymmetry to a local symmetry, the Goldstino degrees of freedom provide the longitudinal modes of the spin-3/2 gravitino, generating a mass
\beq
\mgra = \frac{\Fs}{\sqrt{3} \, M_P} = 
 \left( \frac{\sqrt{\Fs}}{100~{\rm TeV}}\right)^2 2.4~{\rm eV}\, .
\eeq
Here $M_P$ is the reduced Planck mass and $\Fs$ is the Goldstino decay constant. Besides generating a gravitino mass, gravity is irrelevant for our considerations and the gravitino can be effectively treated as a spin-1/2 particle with interactions determined by the non-linear realization of supersymmetry. This is why in this section we will use the words `gravitino' and `Goldstino' interchangeably.

$\Fs$ is the $F$-term of the canonically normalized Goldstino superfield $X_c$, which is related to the mass normalized $X$ by $X_c=X/g_*$ (see \sec{sec:allthat}). Hence we can write
\beq
\Fs = \frac{\mS^2}{\Ks\, g_*}=\frac{ \sqrt{N}\, \mS^2}{4\pi\,\Ks} \, ,
\label{defkappa}
\eeq
where $\Ks$ is an order-one parameter that takes into account uncertainties from strong dynamics. In principle, we could use \eq{defkappa} as the definition of $\mS$ and set $\Ks =1$. However, in order to keep track explicitly of the different sources of theoretical uncertainties, we prefer not to rely on a specific choice for $\mS$ and thus we retain the (redundant) parameter $\Ks$.

Using \eq{defkappa}, the gravitino mass becomes
\beq
\mgra = \frac{ \sqrt{N}\, \mS^2}{4\sqrt{3}\pi\, \Ks\,M_P} =
\sqrt{\frac{N}{10}} \left( \frac{\mS}{100~{\rm TeV}}\right)^2 \frac{0.6~{\rm eV}}{\Ks}\, .
\label{grame}
\eeq

It is interesting to analyze the parametric structure of the expression for the gravitino mass. We can define $g_P(E)$ as the gravitational coupling at the energy scale $E$, and $\ell_P(E)$ as the dimensionless gravitational loop factor
\beq
g_P(E)=\frac{E}{M_P} \, , ~~~~~\ell_P(E)= \frac{N\, g_P^2(E)}{16\pi^2} \, .
\eeq
The gravitino mass, in units of $\mS$, can then be expressed as the square root of the gravitational loop expansion parameter evaluated at the scale $\mS$
\beq
\frac{\mgra}{\mS} \approx \sqrt{\ell_P(\mS)} \, .
\eeq
This is to be compared with the gaugino mass $M_\lambda$ which, in units of $\mS$, is given by a SM gauge loop factor
\beq
\frac{M_\lambda}{\mS} \approx \ell_g \, , ~~~~~\ell_g = \frac{N\, g_{\rm SM}^2}{16\pi^2}\, .
\eeq 
The single power of $g_P\sqrt N$ for the gravitino as opposed to the double power of $g_{\rm SM}\sqrt N$ for the gaugino is the hallmark of the super-Higgs mechanism in supergravity: $\mgra$ originates from the mixing between two  massless fields, one elementary, the gravitino $\psi_{\mu \alpha}$, and one composite,  the Goldstino $\chi_\alpha$. Instead, $M_\lambda$ arises  from the mixing between the elementary gaugino $\lambda_\alpha$ and gapped composite states (see \fig{fig:feynCFTs}).  

In the Super-Composite Higgs, the gravitino is always the lightest supersymmetric particle (LSP) . Hence, $R$-parity implies that the gravitino is stable and can be present today in the universe as a cosmological relic.

\subsection{Gravitino cosmology}
\label{sec:gravcos}

The present cosmic abundance of gravitinos depends critically on whether the reheating temperature at the end of inflation $\TR$ is larger or smaller than the typical soft mass $\msoft$. Let us consider the two cases separately.

\boldmath
\subsubsection*{\textit{(i)} Case $\TR > \msoft$}
\unboldmath

At high temperature ($T\gg \msoft \gg \mgra$), the gravitino production rate is \cite{Pradler:2006qh,Rychkov:2007uq}
\beq
\Gamma_{\tilde G} \approx 3\,{T^3}\left(\frac{1}{M_P^2} + \frac{\msoft^2}{\Fs^2}\right)\, .
\label{rateG}
\eeq
The first term in parenthesis is the production of the spin-3/2 components, whose interaction is purely gravitational. The second term is the production of the spin-1/2 components, whose interaction can be easily derived from the Goldstino effective Lagrangian and are controlled by the scale $\Fs$. For the range of $\mgra$ we are interested in, we can safety identify the gravitino with the spin-1/2 Goldstino and drop the spin-3/2 components.

The production rate in units of Hubble is
\beq
\frac{\Gamma_{\tilde G}}{H} \approx \sqrt{\frac{\gSM}{g_*}}\frac{T\,\msoft^2}{M_P\,\mgra^2}\, ,
\label{eqGH}
\eeq
where $g_*$ is the weighted number of degrees of freedom in the thermal bath which, for the SM, is $\gSM=106.75$. Of course, this standard definition of $g_*$ should not be confused with the coupling of the strongly-coupled sector that has been used in the rest of the paper.
The ratio in \eq{eqGH} decreases as the universe expands, but it is still much larger than one at $T\approx \msoft$. Therefore, until then, gravitinos remain in thermal equilibrium.

For $T< \msoft$, \eq{rateG} is not valid since sparticles are no longer present in the thermal bath. Goldstinos can only be pair-produced by contact interactions obtained by integrating out the supersymmetric partners of SM particles. The supersymmetry Ward identity dictates the Goldstino to decouple at zero momentum with the leading derivative interaction fully controlled by $\Fs$ and  insensitive to the value of $\msoft$ (this result arises from an at-first-sight remarkable diagrammatic cancellation upon integrating out sparticles~\cite{Brignole:1997pe}). Using the Goldstino effective Lagrangian, we thus find that the pair-production rate is
\beq
\Gamma_{{\tilde G}{\tilde G}}  \approx \frac{T^9}{4\pi \mgra^4 M_P^4}
\, , ~~~~
\frac{\Gamma_{{\tilde G}{\tilde G}} }{H} \approx 20
\left( \frac{\gSM}{g_*}\right)^{1/2}  \left( \frac{T}{\rm TeV}\right)^7 \left( \frac{\rm eV}{\mgra} \right)^4\, .
\label{ratt}
\eeq
This shows that, at $T=\msoft$, the ratio $\Gamma_{{\tilde G}{\tilde G}} /H$  is order-one but, because of the steep temperature dependence, gravitino pair-annihilation goes rapidly out of equilibrium immediately below the sparticle threshold. Therefore, gravitinos freeze out while being relativistic.

For $T<\msoft$, gravitinos populate a decoupled thermal distribution, whose present temperature is determined by the entropy injection from the time they decouple to today
\beq
T_{3/2} = T_\gamma \left( \frac{43}{11\, g_*} \right)^{1/3}  = \left( \frac{\gSM}{g_*}\right)^{1/3}0.9~{\rm K}\, ,
\eeq
where $g_*$ is the number of degrees of freedom at gravitino decoupling.

As long as gravitinos remain relativistic, it is customary to express their energy density in terms of the contribution to the effective number of neutrinos $N_{\textrm{eff}}$
\beq
\Delta N_{\textrm{eff}} = \frac{T_{3/2}^4}{T_\nu^4}=\left( \frac{43}{4\,g_*}\right)^{4/3}= \left( \frac{\gSM}{g_*}\right)^{4/3} 0.047\, .
\label{deltn}
\eeq

Eventually, gravitinos become non relativistic and their present energy density is given by
 \beq
 \Omega_{3/2} h^2 = 1.0 \times 10^{-3} \left( \frac{\gSM}{g_*} \right) \left( \frac{\mgra}{\rm eV} \right)\, ,
 \label{omega}
 \eeq
 where $h$ is the dimensionless Hubble constant.
 
Relativistic gravitinos act as hot dark matter, impeding clustering and the growth of matter perturbations beyond the free-streaming scale
\beq
k_{FS} \approx  \left( \frac{0.9~{\rm K}}{T_{3/2}}\right) \left( \frac{\mgra}{\rm eV}\right)  \frac{ 2\, h}{\sqrt{1+z}}\, {\rm Mpc}^{-1}\, ,
\eeq
where $z$ is the corresponding redshift. A comprehensive analysis of CMB, weak lensing and galaxy clustering data has led to an upper bound $\mgra < 2.3$~eV~\cite{Xu:2021rwg}. This limit was derived with simulations that extend only up to about 10~eV because, beyond this value of $\mgra$, the distance $k^{-1}_{FS}$ becomes too small to be easily probed by cosmological data. For larger masses, data from the Lyman-$\alpha$ forest provide the constraint $\mgra < 16$~eV~\cite{Viel:2005qj}. Finally, for $\mgra > 100$~ eV, gravitinos saturate the observed cold dark matter energy density, as shown by \eq{omega}. We are not aware of a study covering the window 10~eV$<\mgra < 16$~eV. However, considering the significant improvement of cosmological data since the publication of~\cite{Viel:2005qj}, it is possible that this mass window is now closed.

In summary, in the case  $\TR > \msoft$ and $\mgra \lsim $~10--16 eV, astronomical observations impose the upper limit
\beq
\mgra < 2.3~{\rm eV} ~~\Rightarrow~~ \mS<\sqrt{\Ks}\left( \frac{10}{N}\right)^{\nicefrac{1}{4}}
200~\tev \, .
\label{limk1}
\eeq
As we will discuss in \sec{sec:collider}, this current limit on $\mS$
has formidable consequences for collider physics.

The limits on relic gravitinos from observational cosmology are expected to improve at a rapid and impressive pace. The current limit on new relativistic species is $\Delta N_{\textrm{eff}}  <$ 0.2--0.3 (depending on assumptions)~\cite{ParticleDataGroup:2024cfk} and is not yet sensitive enough to probe a nearly-massless gravitino, whose prediction is given in \eq{deltn}. However, the target of future experiments is to explore $\Delta N_{\textrm{eff}}$ up to 0.03--0.05.
Combining large-scale structure surveys (such as DESI, Vera Rubin, Euclid) and CMB experiments (such as Simons Observatory, CMB-S4, LiteBIRD), the full mass range of ultra-light gravitinos will be explored in the near future~\cite{Xu:2021rwg}. The Super-Composite Higgs offers an interesting target for these experiments, since it naturally predicts a gravitino in the still unexplored mass region.

\boldmath
\subsubsection*{\textit{(ii)} Case $\TR < \msoft$}
\unboldmath

When $\TR < \msoft$, gravitinos can only be produced in pairs with the rate given in \eq{ratt}. Production of relativistic gravitinos is dominated by the highest possible temperature, and therefore their present cosmic energy density is
\beq
 \Omega_{3/2} h^2 =\frac{\mgra}{3.6~{\rm eV}}\, 
 \left. \frac{\Gamma_{{\tilde G}{\tilde G}}}{H} \right|_{T=\TR} \approx
 \left( \frac{\gSM}{g_*} \right)^{1/2} \left( \frac{\rm eV}{\mgra} \right)^3 \left( \frac{\TR}{300~{\rm GeV}}\right)^7 \, 10^{-3}\, .
 \eeq
By choosing $\TR$ appropriately, the gravitino density can always be made sufficiently small to evade the cosmological limits previously discussed.

The condition $\TR < \msoft$ removes cosmic constraints on gravitinos, but requires a rather unusual cosmological history, with a reheat temperature after inflation well below the TeV. This requires an inflationary dynamics at particularly low energy scales. Moreover, the condition $\TR < \msoft$ prohibits  mechanisms for the cosmic baryon asymmetry based on high-energy processes, such as conventional leptogenesis and GUT baryogenesis. However, since $\TR$ is allowed to be larger than the temperature of the EW phase transition (which is about 100 GeV), mechanisms of EW baryogenesis are allowed, and so are those based on a baryon asymmetry created directly by the late decay of the inflaton or metastable dark-matter particles~\cite{Cui:2012jh,Cui:2015eba}. Another cosmological option is to retain high $\TR$ and high-scale baryogengesis, but to have a late entropy dump into the thermal bath by a heavy metastable particle after gravitino decoupling, thereby relatively depleting the gravitino abundance.

\subsection{Searches at high-energy colliders}
\label{sec:collider}

 While the gravitino plays the role of the LSP, collider phenomenology is governed by the nature of the next-to-lightest supersymmetric particle (NLSP). 
The two most plausible candidates for NLSP are the Bino and the right-handed stau. The ratio between the gauge-mediation expressions of their masses is shown in \fig{fig:NLSP}.
Since our preference is for relatively large $N$ ($N \gsim 3$), the stau is the most likely NLSP.
However, given the uncertainties from strong dynamics, we cannot exclude the case of a Bino NLSP.

\begin{figure}[t]
\begin{minipage}{0.5\textwidth}
\begin{center}
\includegraphics[width=0.95\columnwidth]{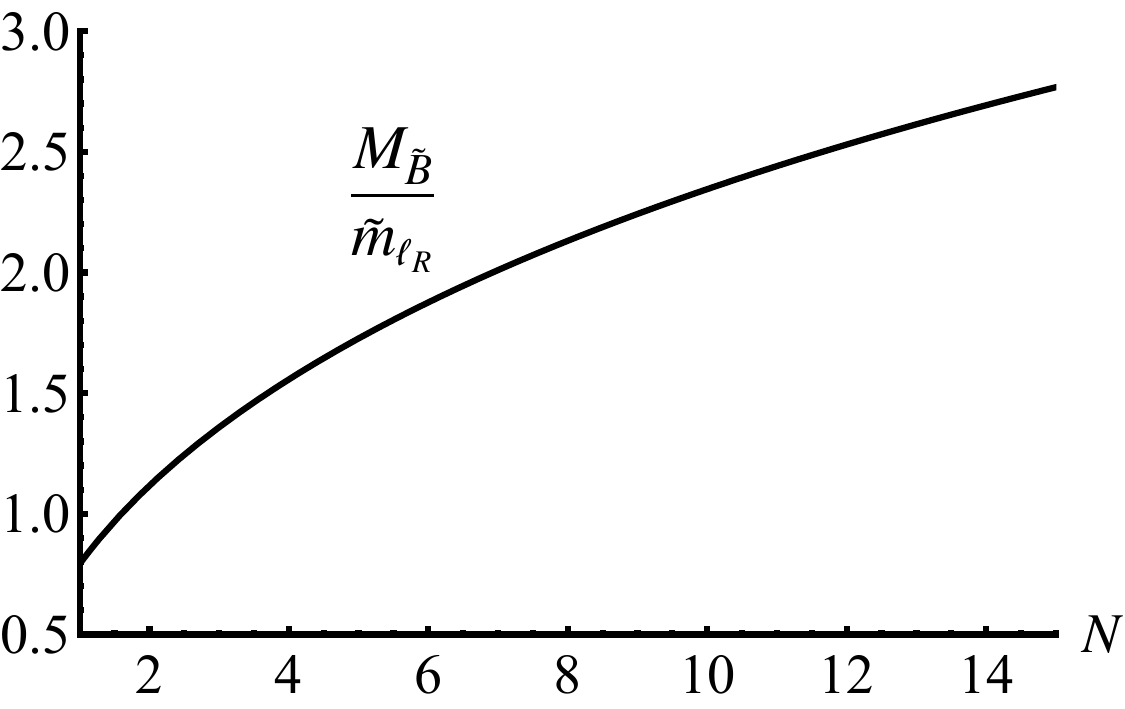}
\end{center}
\end{minipage}
\begin{minipage}{0.5\textwidth}
\begin{center}
\includegraphics[width=0.95\columnwidth]{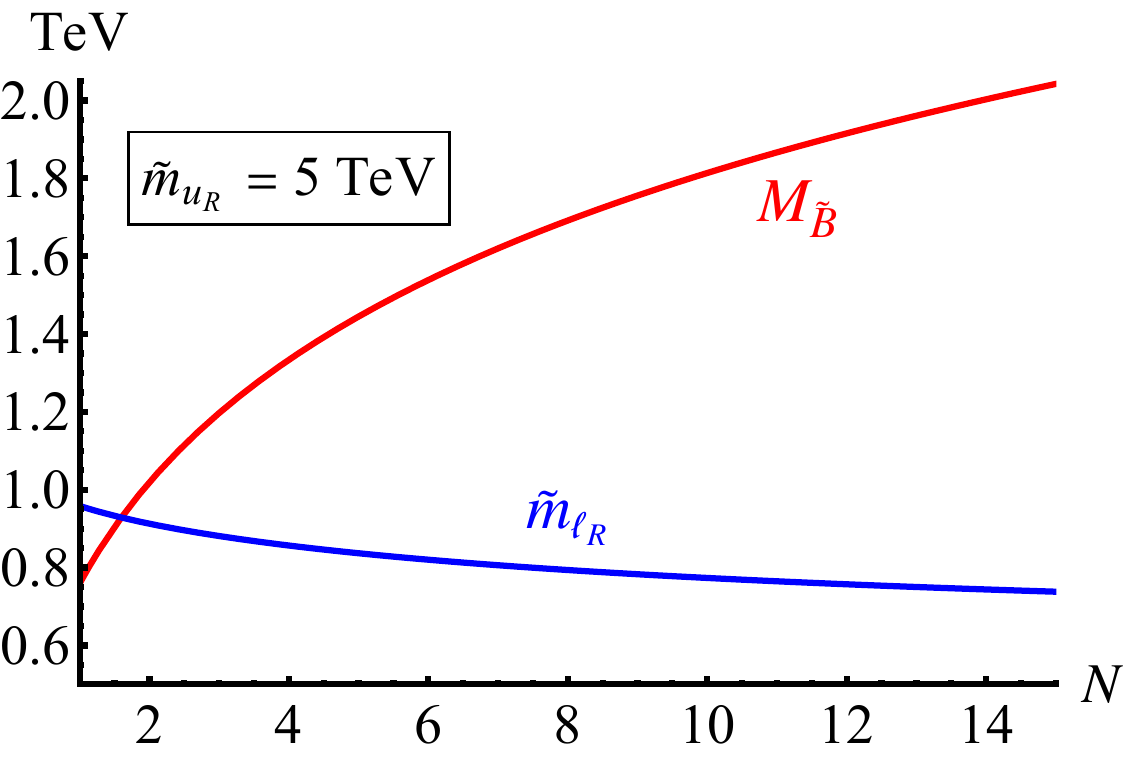}
\end{center}
\end{minipage}
\caption{The ratio between the Bino and right-handed slepton masses as a function of $N$, using the gauge-mediation relations at leading-order accuracy (left panel), and their individual values corresponding to a right up-squark of 5 TeV (right panel).}
\label{fig:NLSP}
\end{figure}

In these two cases, the decay width of the NLSP into gravitino is
\beq
\Gamma ({\tilde \tau}_R \to \tau {\tilde G}) =\frac{{\tilde m}_{\tau_R}^5}{16\pi \,\Fs^2}\, ,
\label{decstau}
\eeq
\beq
\Gamma ({\tilde B} \to \gamma {\tilde G}) =\frac{M_{{\tilde B}}^5 \cos^2\theta_W}{16\pi \, \Fs^2}\, , ~~~~ 
\Gamma ({\tilde B} \to Z {\tilde G}) =\frac{M_{{\tilde B}}^5 \sin^2\theta_W}{16\pi \, \Fs^2}\left(1- \frac{M_Z^2}{M_{{\tilde B}^2}}\right)^4\, .
\eeq
Hence, the decay length of an NLSP with energy $E$ is
\beq
L_{\rm NLSP} = \left( \frac{\sqrt{\Fs}}{100~{\rm TeV}}\right)^4 \left( \frac{\rm TeV}{{\tilde m}_{\tau_R}}\right)^5 \sqrt{\frac{E^2}{{\tilde m}_{\tau_R}^2}-1} ~ {\rm nm}\, .
\eeq
This shows that, in the range of scales relevant to the Super-Composite Higgs, the NLSP always decays promptly. Let us focus now on the most likely case of stau NLSP.

Because of flavor near-universality, we expect the right-handed selectron and smuon (here collectively denoted as ${\tilde \ell}_R$) to be very close in mass to the stau NLSP. Their decay modes into the NLSP have widths
\beq
\Gamma ({\tilde \ell}_R^- \to \ell^- \tau^+ {\tilde \tau}_R^-)=
\Gamma ({\tilde \ell}_R^- \to \ell^- \tau^- {\tilde \tau}_R^+)= 
\frac{g^{\prime 4} \, ({\tilde m}_{\ell_R}^2 - {\tilde m}_{\tau_R}^2)^5}{15360 \,\pi^3 M_{\tilde B}^2 {\tilde m}_{\ell_R}^7}\, .
\label{dec3}
\eeq
Here we have taken $M_{\tilde B}\gg {\tilde m}_{\ell_R}$ and $({\tilde m}_{\ell_R}^2 - {\tilde m}_{\tau_R}^2)\ll {\tilde m}_{\ell_R}^2$, which is well justified as we expect
\beq
({\tilde m}_{\ell_R}^2 - {\tilde m}_{\tau_R}^2) \approx m_\tau^2 \tan^2\beta\, .
\eeq
Equation (\ref{dec3}) shows that the decay modes of ${\tilde \ell}_R$ into the NLSP are so much suppressed by phase space that, for the values of $\Fs$ relevant to our study, they are negligible with respect to the direct decay into gravitinos, whose width is analogous to \eq{decstau}.

\begin{figure}[t]
\begin{center}
\includegraphics[width=0.6\columnwidth]{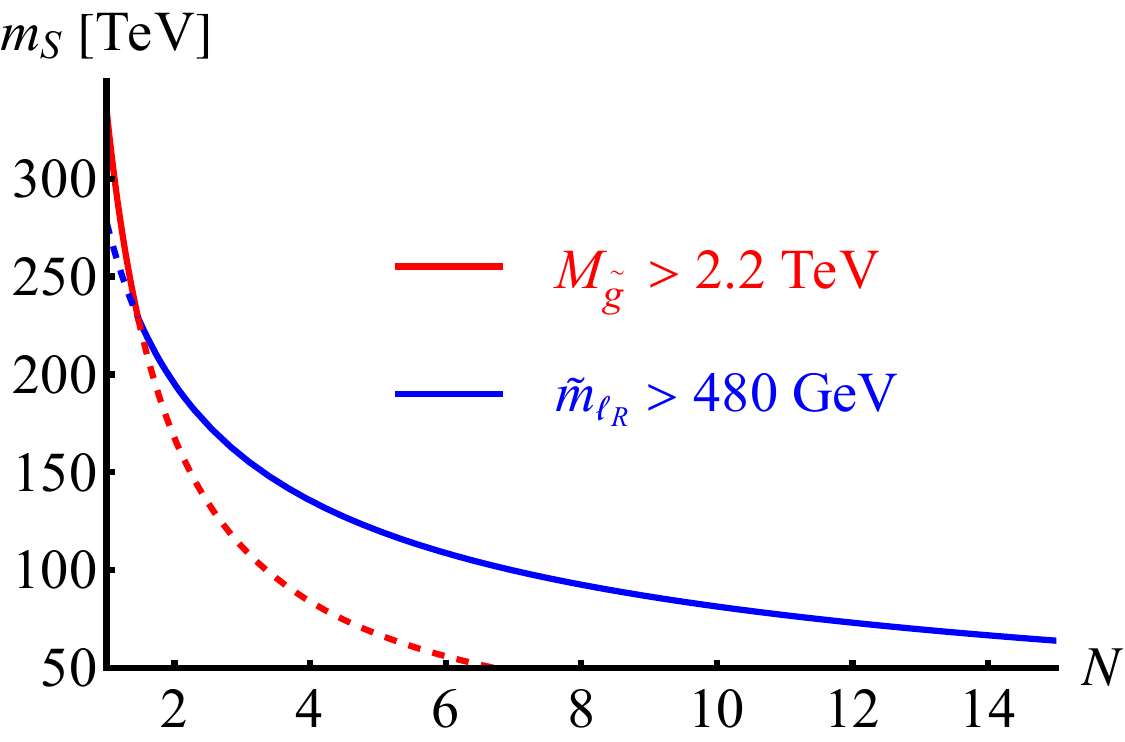}
\end{center}
\caption{The lower limit on the supersymmetry-breaking scale $\mS$ from the LHC bounds on the masses of the gluino (red line) and the right-handed slepton (blue line). Solid lines indicate the most stringent limit, while dashed lines refer to subdominant constraints.}
\label{fig:Limitms}
\end{figure}

This result has important consequences for collider searches. Despite ${\tilde \tau}_R$ being the NLSP, supersymmetric particle production at colliders is {\it not} characterized by a dominant number of $\tau$'s in final states. On the contrary, the decay chains of supersymmetric particles will end with sleptons of all three generations, each promptly producing energetic leptons of their own family, as well as missing energy. The three right-handed sleptons act as co-NLSP. Lepton universality is maintained, also at the level of the decay chains, with electrons, muons and taus predicted to be equally abundant in the final states of high-energy proton collisions. Energetic leptons and missing energy is a promising, and relatively clean, collider signature that will allow a thorough exploration of the Super-Composite Higgs at future facilities.

It is useful to present the lower limit on the supersymmetry-breaking mass $\mS$ from current LHC data. Using the gauge-mediation mass relations, we find that the strongest collider bounds on $\mS$ come from the gluino at small $N$ and from the right sleptons at large $N$. The current LHC limit of the gluino mass is 2.2~TeV~\cite{ParticleDataGroup:2024cfk}. LHC experiments have also obtained a limit of 700~GeV on the mass of degenerate left and right sleptons. This limit does not readily apply to our case because, according to gauge mediation, left sleptons are about twice as heavy as right sleptons. Estimating the ratio between the R-only and L+R channels from the analyses in refs.~\cite{CMS:2018eqb,ATLAS:2019lff}, we derive a limit on right sleptons of about 480~GeV. In summary, from LHC data and gauge-mediation mass relations, we obtain the lower bounds shown in \fig{fig:Limitms}.

\begin{figure}[t]
\begin{center}
\includegraphics[width=0.7\columnwidth]{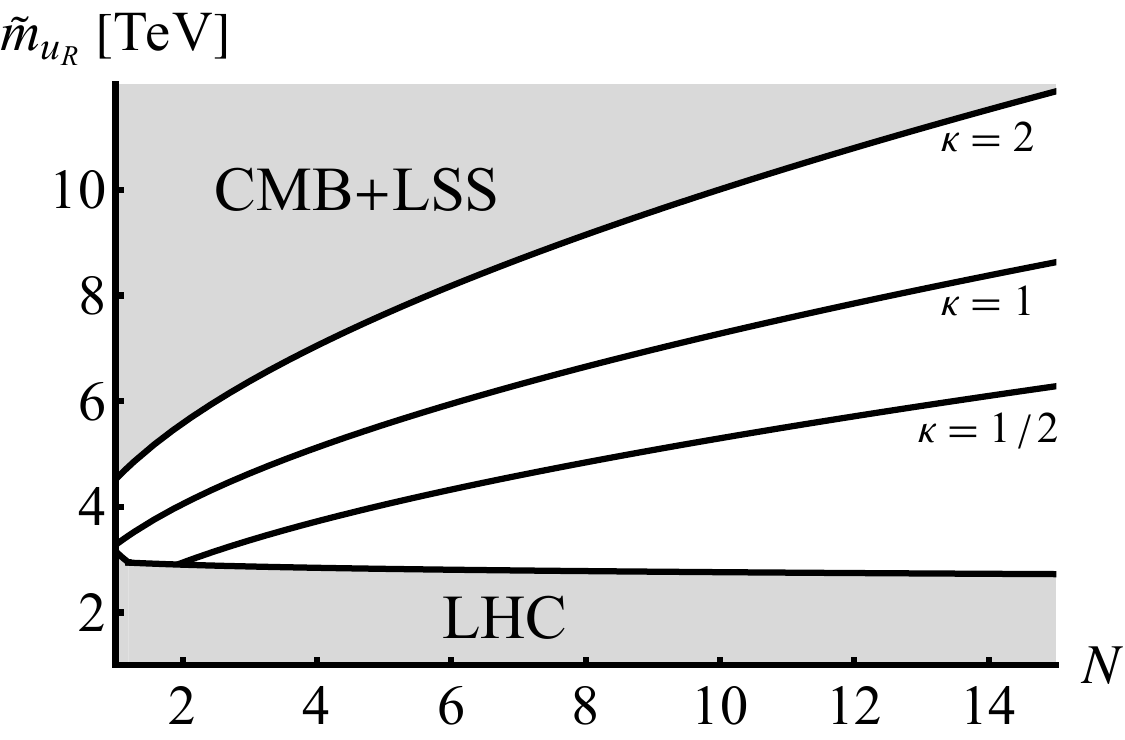}
\end{center}
\caption{The shaded regions are excluded by {\it (i)} LHC searches on squarks and gluinos, assuming exact gauge-mediation mass relations; {\it (ii)} a combination of data from CMB, weak lensing and galaxy clustering in large scale structures for three values of $\kappa$, the coefficient that combines the uncertainties due to strong dynamics as defined in \eq{kappafin}. The white region defines the squark mass range allowed in Super-Composite Higgs.}
\label{fig:cosmo}
\end{figure}

Dismissing the unusual case of cosmological histories in which the reheating temperature is below $\msoft$, we can compare the LHC lower limit on $\mS$  with the upper limit on $\mS$ in \eq{limk1}. This is shown in \fig{fig:cosmo}, where the bounds are expressed in terms of the squark mass, which is a more transparent parameter, phenomenologically speaking. The translation between $\mS$ and ${\tilde m}_q$ is based on the gauge-mediation mass relation, modulated by an overall factor $\kappa_{\tilde q}$ that accounts for uncertainties from strong dynamics,
\beq
{\tilde m}_q^2= \kappa_{\tilde q}\,
{\tilde m}_{ q_{\rm (GM)}}^{2} \, ,
\label{eqq2bis}
\eeq
where ${\tilde m}_{ q_{\rm (GM)}}^{2} $ is given in \eq{eqq2}.

The shaded LHC region in \fig{fig:cosmo} corresponds to the exclusion from searches for gluinos and squarks. For gluinos, we have used the limit $M_{\tilde g}>2.2$ TeV, which becomes the leading constraint only in a thin slice close to $N=1$. For squarks, we have used the ATLAS combined limit~\cite{ATLAS:2020syg}, which takes into account the effect of virtual gluinos in squark production. This explains the mild decrease of the limit as $N$ grows, since gluinos get heavier. For $N>3$ right-handed sleptons give a slightly stronger limit but we prefer not to use it, as it is vulnerable to modifications of the gauge-mediation mass relations. The shaded CMB+LSS region corresponds to the limit in \eq{limk1}, discussed in \sec{sec:gravcos}. The three lines correspond to $\kappa=\nicefrac{1}{2},1,2$, where $\kappa$ combines the strong-scale effects on the relation between $\mS$ and $\Fs$, see \eq{defkappa}, together with the effects on the gauge-mediation relation between $\mS$ and the squark mass, see \eq{eqq2bis}, 
\beq
\kappa \equiv \Ks \, \kappa_{\tilde q} \, .
\label{kappafin}
\eeq
It is not unreasonable to expect that, in our theory, $\kappa$ can deviate from the NDA prediction $\kappa =1$ by at least a factor of 2, upwards or downwards. While in a perturbative setup $\kappa$ could be made arbitrarily small, the strongly-interacting regime we are interested in will entail an upper bound on $\kappa$. To determine the absolute upper bound on squark masses, we will adopt a conservative choice for the theoretical uncertainties and require $\kappa <2$.

\begin{figure}[t]
\begin{minipage}{0.5\textwidth}
\begin{center}
\includegraphics[width=0.97\columnwidth]{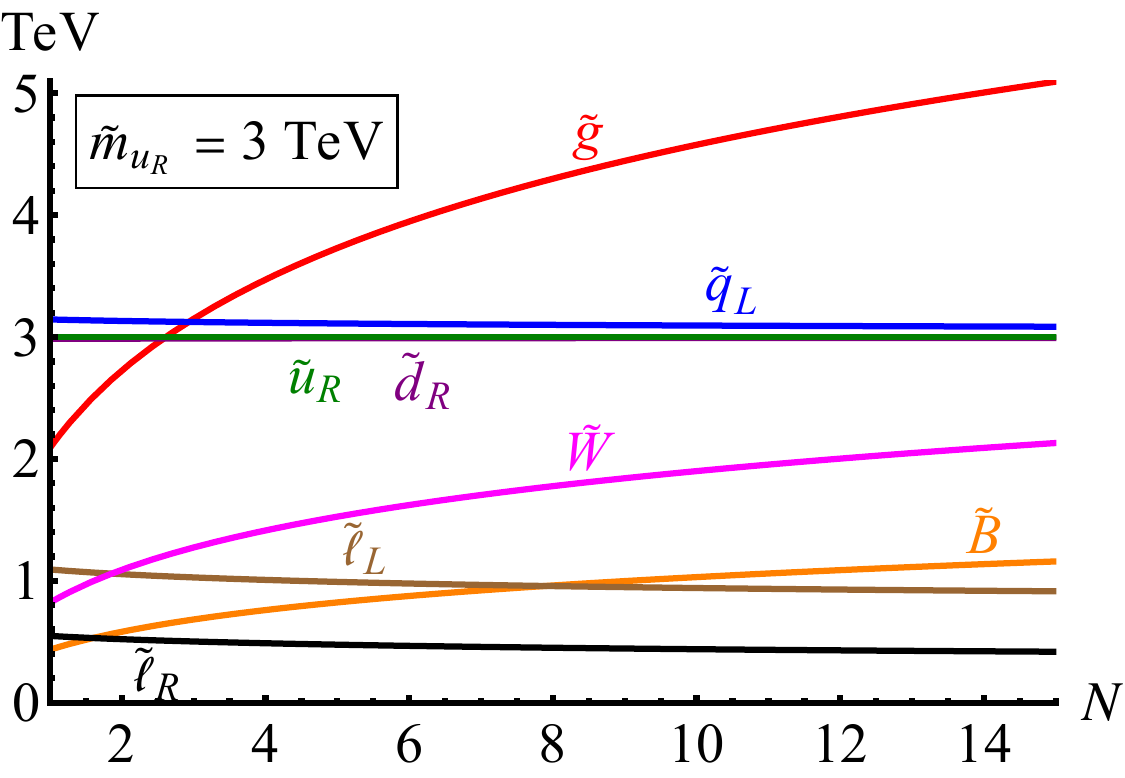}
\end{center}
\end{minipage}
\begin{minipage}{0.5\textwidth}
\begin{center}
\includegraphics[width=0.97\columnwidth]{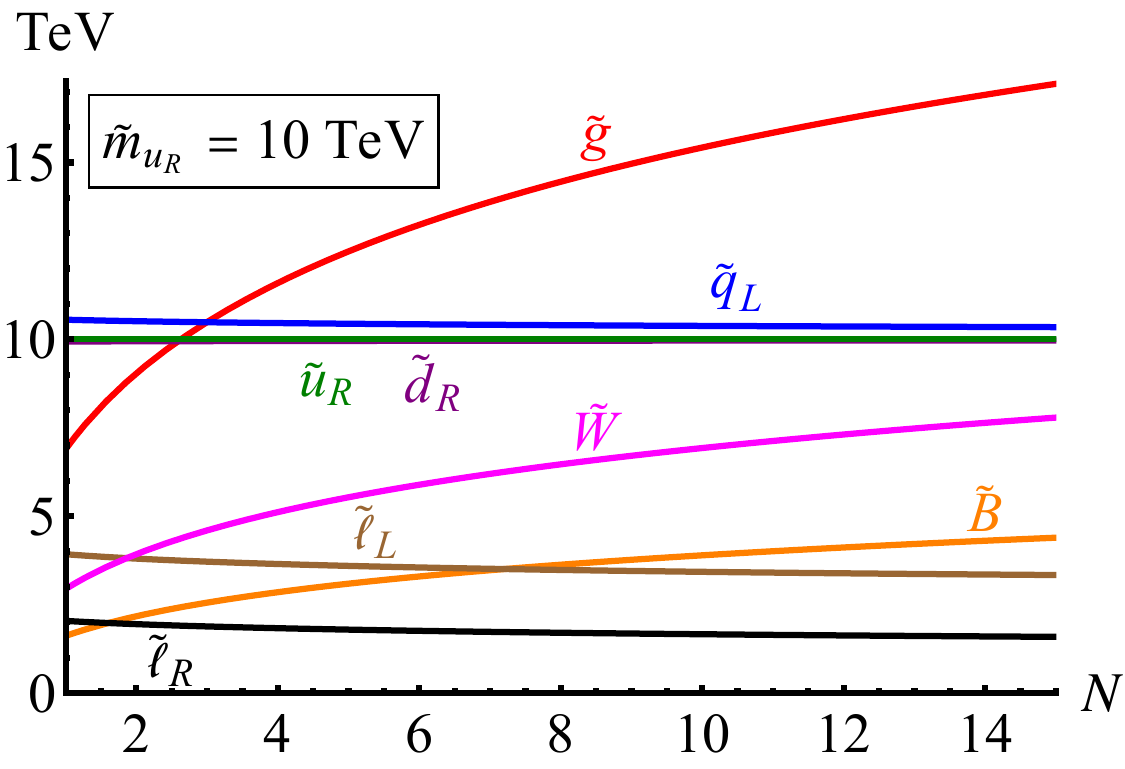}
\end{center}
\end{minipage}
\caption{The mass spectrum of the supersymmetric particles, using gauge-mediation relations at leading order. The spectrum is shown as a function of $N$ and normalized such that ${\tilde m}_{u_R} = 3$~TeV (left panel) or  ${\tilde m}_{u_R} = 10$~TeV (right panel), which we regard as the smallest and largest values of the preferred mass range in the Super-Composite Higgs. The spectra scale linearly with the overall mass normalization, up to logarithmic effects.}
\label{fig:GM}
\end{figure}

The result in \fig{fig:cosmo} is striking. The Super-Composite Higgs predicts that the spectrum  of supersymmetric particles must lie within a well-defined window, which becomes broader in the theoretically preferred region of large $N$. The theory gives an upper bound on squark masses, which is between 6 TeV (for $N=3$) and 12 TeV (for $N=15$). This upper bound is not determined by naturalness criteria or special parameter choices. It is determined purely by current experimental data applied to the broad Super-Composite Higgs framework.

Deriving upper bounds on the gravitino mass (and consequently on squark masses, in our theoretical framework) may seem counterintuitive from a particle-physics point of view, where experiments usually set lower bounds. However, the cosmic contribution of gravitinos decreases for smaller $\mgra$, as shown in \eq{omega}, and therefore their observational effects diminishes as they get lighter.

In the context of the Super-Composite Higgs, the current bounds on the gravitino or possible future measurements of its cosmological effects imply an encouraging message for future high-energy colliders, pointing to the 10 TeV mass region as a crucial target for experimental exploration. This energy region is indeed within reach of some of the most ambitious future experimental projects. For example, the FCC-$hh$ with 85~TeV of energy and 30~ab$^{-1}$ of integrated luminosity can cover this mass range, as its typical mass reach is 16 TeV for gluinos, 13 TeV for squarks, and 8 TeV for Wino. A 10 TeV Muon Collider can explore EW sparticles up to about 5 TeV.

\begin{figure}[t]
\begin{center}
\includegraphics[width=0.85\columnwidth]{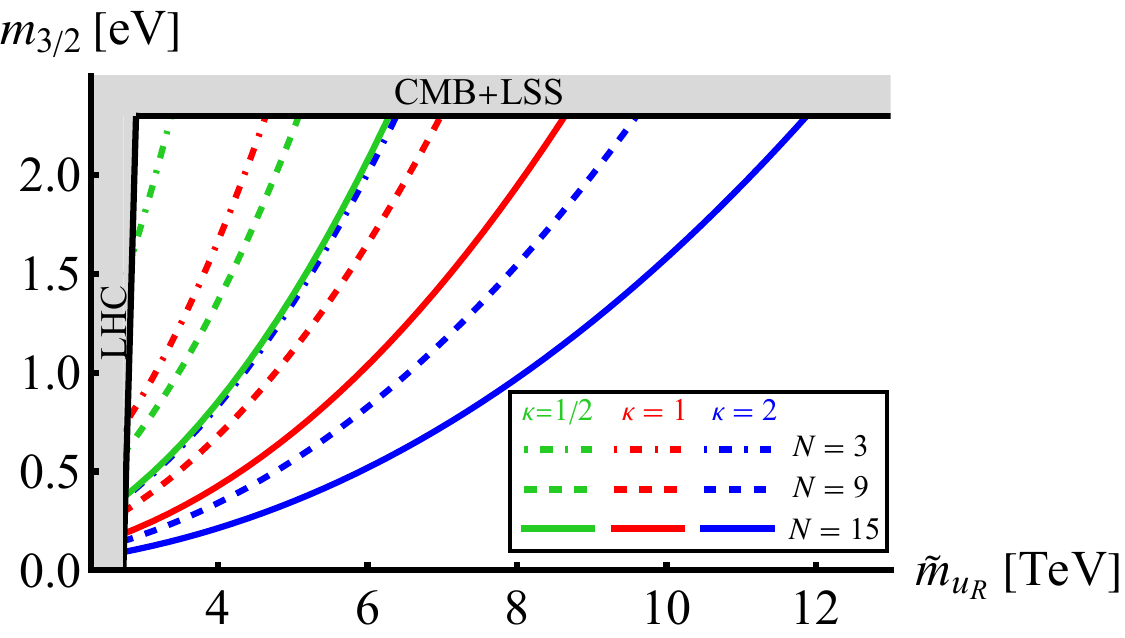}
\end{center}
\caption{The relation between the gravitino and squark masses for various choices of the parameters $N$ and $\kappa$. The upper bound on the gravitino mass from astronomical observations implies an upper bound on squark masses.}
\label{fig:grasqu}
\end{figure}

In \fig{fig:GM}, we show the prediction of the supersymmetric mass spectrum, based on gauge-mediation relations. We are showing the results for ${\tilde m}_{u_R} = 3$~TeV and 10~TeV, which are roughly the two extremes of the preferred mass range for the Super-Composite Higgs, as determined by data. Future collider can perform a thorough exploration of the full spectrum of supersymmetric particles predicted by the Super-Composite Higgs.

To emphasize the correlation between high-energy experiments and astronomical observation, we show in \fig{fig:grasqu} the parameter dependence of the relation between gravitino and squark masses 
\beq
\mgra = \frac{{\tilde m}_q^2}{\kappa\,\mathscr{M}}\, , ~~~~~\mathscr{M}\equiv \frac{4\sqrt{3}\pi \,M_P}{\sqrt{N}}\left( \frac{{\tilde m}_{ q_{\rm (GM)}}^{2}}{\mS^2}\right) 
\, .
\eeq

It is interesting that collider searches and astronomical observations give complementary bounds, which single out the 1--10~TeV region as special for supersymmetric particles. This is the window of opportunity where the Super-Composite Higgs is within reach of both cosmic surveys and high-energy colliders.

\subsection{Dark matter}

The gravitino LSP in the Super-Composite Higgs is not a viable dark matter candidate. Its relic abundance can be adjusted by dialing the value of $\TR$, but its low mass can only lead to hot dark matter. Nevertheless, as far as dark matter is concerned, the Super-Composite Higgs holds a surprising coincidence.

Suppose that the $\CFTs$ sector carries an accidental global or discrete symmetry, which determines the stability of the lightest composite particle charged under that symmetry. We can imagine something similar to baryon number in QCD, which guarantees proton stability, although the dynamics of our theory is quite different. Let us indicate the stable $\CFTs$ composite by $S$ and assume it is neutral under the full SM gauge group, or else it would be ruled out by data on dark-matter direct detection. The cosmic density of $S$ particles is determined by the standard thermal relic calculation.

The $\CFTs$ sector is automatically endowed with a nearly massless state, the  Goldstino/gravitino $\tilde G$, into which $S$ can annihilate in the early universe through strongly-coupled dynamics. Other annihilation channels into relatively light $\CFTs$ states may be possible, but their existence is model-dependent. On the other hand, the annihilation channel $SS\to \tilde G \tilde G$ is theoretically robust and has the largest phase space.

In the absence of long-range interactions associated with light mediators and in the regime of low velocity $v$, which dominates the relic abundance, the amplitude for $SS\to \tilde G \tilde G$ is  described by a local interaction. Using our power-counting rules, we can estimate the annihilation cross section as
\beq
\sigma(SS\to \tilde G\tilde G )~ v\approx \frac{g_*^4}{4\pi \, m_S^2}=\frac{(4\pi)^3}{N^2\, m_S^2} ~~~~~~({\rm for}~N\gsim  4\pi)
\, .
\label{annihilationNDA}
\eeq
This  result is  reliable at small enough $g_*$, but one should keep in mind that scattering phases in the strong sector scale roughly like  $\delta \sim g_*^2/4\pi$. This implies a saturation in the growth  of \eq{annihilationNDA} with $g_*$ at $g_*^2\sim 4\pi$ or, equivalently, $N\sim 4\pi$, slightly before
full-fledged strong coupling, which arises at $N\sim 1$.
Therefore, for $N\lsim 4\pi $ the annihilation cross section saturates at the value
\beq
\sigma(SS\to \tilde G\tilde G )~ v\approx \frac{4\pi }{m_S^2}~~~~~~({\rm for}~N\lsim  4\pi)
\, .
\label{annihilationMAX}
\eeq
From this, we derive that today's cosmic energy density in $S$ particles is
\beq
\Omega_S h^2 \approx 0.1 \left( \frac{\mS}{100~\tev}\right)^2\times \left\{
\begin{array}{cl}
1 & ~~~{\rm for}~N\lsim  4\pi \\
( \frac{N}{4\pi})^2& ~~~ {\rm for}~N\gsim  4\pi
\end{array} 
\right.
\eeq

It is remarkable that the correct dark-matter density is obtained for values of $\mS$ and $N$ precisely within the range preferred by the Super-Composite Higgs from independent considerations. 

The composite dark-matter particle $S$ interact with SM matter through higher-dimensional operators suppressed by powers of $\mS$. Let us consider, for simplicity, the case in which $S$ has spin zero. Its coupling to gluons will be of the form dictated by power-counting rules
\beq
\frac{g_3^2}{\mS^2}\, G^a_{\mu\nu}G^{a\, \mu\nu}\, S^\dagger S \, ,
\label{DMglu}
\eeq
where $g_3$ is the QCD coupling constant.

Direct dark-matter detection is governed by the interaction between $S$ and nucleons, which can be derived from \eq{DMglu} by recalling that the matrix element of the trace of the energy-momentum tensor is related to the nucleon mass $m_N$ by
\beq
\langle N | T^\mu_\mu | N\rangle = m_N\, {\bar \psi}_N \psi_N \, ,
\eeq
\beq
T^\mu_\mu = \frac{b_3 \alpha_s}{8\pi} \, G^a_{\mu\nu}G^{a\, \mu\nu} +\sum_q m_q \, {\bar q} q \, ,
\eeq
where $b_3$ is the QCD $\beta$-function coefficient as defined in \eq{betafun} and $b_3=-7$ in the SM. Then, from \eq{DMglu}, we obtain the coupling between $S$ and nucleons
\beq
 \frac{k_g\, (4\pi)^2 m_N}{\mS^2}\, {\bar \psi}_N \psi_N \, S^\dagger S \, , ~~~~ k_g =\frac{2}{b_3}=-0.3 \, .
 \label{coupg}
\eeq

Another important interaction occurs between $S$ and the SM Higgs doublet $H$. This interaction arises at the Higgs compositeness scale $\mst$ and involves the composite operators $\O$ and $\cal R$. After replacing their background values $\O/\mst^{\Delta_\O}\to \mu \,\theta^2$ and ${\cal R}/\mst^{\Delta_{\cal R}}\to \mu^2 \,\theta^4$, where $\mu$ is the higgsino mass parameter, we obtain
\beq
\frac{g_*^2\, \mu^2}{\mS^2}\, H^\dagger H \, S^\dagger S \, .
\label{DMHig}
\eeq
Equation~(\ref{DMHig}) can be converted into a coupling between $S$ and nucleons, after replacing one Higgs with its vev $v$ and connecting the other Higgs to a line of light quarks. Using the low-energy coupling of Higgs to nucleons~\cite{Shifman:1978zn}, we obtain the effective $S$--nucleon interaction
\beq
\frac{k_h\, (4\pi)^2 m_N}{ \mS^2}\, {\bar \psi}_N \psi_N \, S^\dagger S \, , ~~~~ k_h =\frac{2\,  \mu^2}{9 N\, m_h^2}=1.4\left(\frac{ 10}{N}\right) \left(\frac{\mu}{\tev}\right)^2\, .
 \label{couph}
\eeq

Needless to say, both eqs.~(\ref{coupg}) and (\ref{couph}) are affected by uncertainties of order a few, due to strong dynamics. Therefore, we cannot definitively determine whether gluons or Higgs give the largest effect.

The spin-independent $S$ scattering cross section off a nucleon is
\beq
\sigma_{\rm SI} \approx \frac{k_{g,h}^2\, (4\pi)^2 m_N^2}{ \mS^4}=k_{g,h}^2\, \left( \frac{100~\tev}{\mS}\right)^4 0.6 \times 10^{-45}~{\rm cm}^2 \, .
\label{DMrate}
\eeq
Since the current experimental sensitivity in direct detection is $10^{-45}~{\rm cm}^2$ for a particle mass of 100~TeV~\cite{LZ:2024zvo}, \eq{DMrate} shows that composite $S$ dark matter is at the edge of the discovery reach, albeit with large uncertainties.

We would also like to point out that a peculiarity of $S$ dark matter is a strong suppression of indirect signals, since the main annihilation channel is into gravitinos, which are practically invisible and inert particles. Annihilations into SM particles (and their supersymmetric partners) occur, but at a reduced rate. For instance, using the normalization in \eq{DMglu}, the annihilation cross sections into gluons, relative to gravitinos, is
\beq
\frac{\sigma(SS\to gg)}{\sigma(SS\to \tilde G\tilde G )}\approx 
4\, \alpha_s^2(\mS) 
\times \left\{
\begin{array}{cl}
1 & ~{\rm for}~~~N\lsim  4\pi \\
( \frac{N}{4\pi})^2& ~ {\rm for}~~~N\gsim  4\pi
\end{array} 
\right.
\eeq
The annihilation rate into Higgs/higgsinos is much more suppressed, since it doesn't benefit from the $(\mu/ m_h)^2$ enhancement that occurs in the elastic scattering at low momentum transfer. 

In conclusion, in the Super-Composite Higgs, a stable and neutral composite state of the $\CFTs$ sector is a viable dark-matter candidate with the following properties: {\it (i)} its thermal relic abundance is in the correct range; {\it (ii)} its scattering rate with nuclei is at the edge of current limits from direct searches and within the sensitivity of future experiments; {\it (iii)} its annihilation occurs mostly in an invisible channel, thus suppressing indirect detection signals. These three predictions, while affected by significant uncertainties stemming from strongly-coupled dynamics, are not the result of any specific adjustments to the underlying parameters. Instead, they arise purely from the structure of the theory, in which the input parameters are determined from particle physics considerations unrelated to dark matter.

\boldmath
\section{Summary}
\unboldmath
\label{sec:conc}

Given the length of our paper, we provide a comprehensive summary of our main results for the reader's convenience.

\subsubsection*{The structure of the Super-Composite Higgs}
The Super-Composite Higgs is a new theoretical framework that simultaneously addresses the naturalness and flavor issues. It combines the properties of supersymmetry and compositeness, allowing each theory to strengthen the other while mitigating their individual shortcomings. Supersymmetry is responsible for resolving the big hierarchy problem (while the little-hierarchy puzzle remains unexplained). By relieving Higgs compositeness of the burden of addressing naturalness, the theory can successfully generate the pattern of SM Yukawa couplings without conflicting with experimental data on flavor processes. Moreover, compositeness assists supersymmetry by providing a novel solution to the $\mu$--$B_\mu$ problem; and supersymmetry assists compositeness by offering a justification for the lightness of the Higgs, based on the non-renormalization theorem, without relying on spontaneously broken approximate global symmetries.

The AdS/CFT duality allows us to look at the Super-Composite Higgs from two different and complementary perspectives. From the viewpoint of 4D theories (see \sec{sec:frameCFT} and \fig{fig:Overall}), the theory is constructed out of a CFT-like strongly coupled sector where there is no notion of flavor symmetry, which is indeed maximally violated, although baryon and lepton numbers and CP are assumed to be accidental symmetries. The CFT sector communicates with the Higgsless SM sector through small mixing parameters, which act as seeds of proto-Yukawa couplings, violate CP and implement the mechanism of partial compositeness~\cite{Kaplan:1991dc}. The strongly-coupled sector undergoes a dynamical tumbling process in two successive steps (for a toy model of a strongly-coupled sector that generates multiple dynamical scales, see \sec{sec:tumbling}). 

At the first stage of confinement, at an energy scale $\mst$, the Higgs fields emerge as light composite particles and a hierarchical pattern of Yukawa couplings is generated by the RG evolution of CFT operators from the UV cutoff scale $\LUV$ to $\mst$ (see \sec{sec:allthat} for technical details). At a second stage of confinement, at an energy scale $\mS$, supersymmetry is broken and soft terms start to be generated. Finally, after the supersymmetric particles are integrated out at a scale $\msoft$, only the SM particles remain, together with a light gravitino, which contains the degrees of freedom of the Goldstino, the remnant of the dynamical spontaneous breaking of supersymmetry. As the various stages are interconnected and separated only by small energy differences, the entire structure functions as a single dynamical mechanism, capable of explaining naturalness and flavor.

The 5D perspective (see \sec{sec:frame5D} and \fig{fig:susy}) offers a geometrical interpretation of the theory, with three branes (called UV, IR$_*$ and IR$_S$) located at different points on a slice of the fifth warped dimension. The positions of the IR$_*$ and IR$_S$ branes are dual to respectively $\mst$ and $\mS$.
Gauge fields propagate in the entire space, while  the SM matter fields are restricted to the space between the UV and IR$_*$ branes. Furthermore, the Higgs fields and the supersymmetry breaking dynamics (along with the Goldstino) are respectively localized at IR$_*$ and IR$_S$. The bulk matter field profiles determine the hierarchical pattern in the Yukawa matrices, while the localizations of the Higgs and Goldstino on different branes provide the minimal plausible scenario for the origin of $\mu$ and $B_\mu$, as discussed below.

The two perspectives are dual to each other and the usual vocabulary of the AdS/CFT correspondence can be used to translate between the two languages. In particular the locations of the UV, IR$_*$ and IR$_S$ branes along the warped dimension correspond to the energy scales $\LUV$, $\mst$ and $\mS$ in the 4D picture.

The non-perturbative character of the theory prevents us from attempting a fully quantitative analysis. Worse yet, we cannot even be sure that a CFT with the desired properties actually exists, as discussed in \sec{sec:intro}. 
However, this doesn't deter us from speculating about this scenario, with the goal of establishing the structural foundations of the Super-Composite Higgs. Our guide is the 5D picture, which provides the only concrete realization of the scheme we have in mind.

\subsubsection*{The SILSH approach}

To describe the dynamics of the Super-Composite Higgs we have 
extended the SILH effective Lagrangian~\cite{Giudice:2007fh} to the case of supersymmetry, thereby calling it SILSH (see \sec{sec:SILSH}). This procedure allows us to make computations and obtain results which, although only qualitative, are based on systematic expansions and well-defined hypotheses. The SILH is not merely an agnostic EFT approach in which all interactions with the same dimensionalities are treated equally. Instead, the SILH relies on power counting of both mass scales and couplings, and makes precise assumptions about the combinations of couplings, mass scales and fields that are allowed to appear in the Lagrangian. These assumptions reflect our basic knowledge about the structure of the fundamental theory. 

To facilitate the analysis, we assume a gap in scaling dimensions among operators in the $\CFTs$ sector such that supersymmetry breaking can be ascribed to a single lowest-dimension chiral operator $\O$ and a vector superfield $\cal R$, which corresponds to the composite operator $\O^\dagger \O$. Under this simplifying hypothesis, the SILSH methodology enables us to classify operators, estimate their magnitudes, and extract the pattern of the flavor structure and soft terms. Our formulation of the SILSH has a value in itself, since our analysis can be applied to a larger class of strongly-interacting supersymmetric theories than the Super-Composite Higgs alone.

\subsubsection*{Supersymmetry breaking: I. Gauge mediation}

The soft terms have a specific structure that differs from any other known class of supersymmetric theories, but shares common elements with many familiar cases (see \sec{sec:SILSH} and table~\ref{tab:soft} for a schematic summary). 

The first contribution to sparticle masses has a pattern that resembles gauge mediation, although the effect comes from direct interaction with the strongly-coupled supersymmetry-breaking sector. As a result, gaugino masses are generated at tree-level, through an effective mixing with composite states of the $\CFTs$ sector. This can be interpreted as a partial compositeness of gauginos. Scalar masses are produced at one-loop level through SM gauge interactions. Expressing the strong coupling as $g_*=4\pi/\sqrt{N}$, one recovers the parametric structure of the familiar gauge-mediation soft terms, with gaugino masses generated at one loop and scalar masses at two loops. As a result, soft masses have the same form as gauge mediation, although deformed by unknown (but flavor universal) order-one coefficients, which parametrize the effect of strong dynamics.

This result has a clear 5D interpretation. Since gauge fields live everywhere in the extra dimension, they have a full overlap with the IR$_S$ brane where supersymmetry breaking resides. Scalar masses are generated at one loop, in a fashion analogous to gaugino mediation. However, due to warping, the KK excitations are strongly localized near the IR$_S$ brane, enhancing the one-loop gauge effect. Thus, the soft terms acquire the same parametric dependence of familiar gauge mediation, in perfect agreement with the 4D result.

\subsubsection*{Supersymmetry breaking: II. Higgs compositeness}

Higgs compositeness gives an additional contribution to the soft terms in the Higgs sector, which provides a solution to the $\mu$--$B_\mu$ problem. Appealing to holomorphy, we exclude from the superpotential a $\mu$ term at the fundamental level. This corresponds to our hypothesis that, at the confinement scale $\mst$, the CFT generates two massless and weakly-coupled composite states to be interpreted as the Higgs doublets. Then, the $\mu$ parameter comes purely from supersymmetry breaking and can be considered a soft term. 

Applying the SILSH rules, we find $\mu \sim (\mS / \mst )^{\Delta_{\O}} \mS$, where $\Delta_{\O}$ is the scaling dimension of the primary chiral operator responsible for supersymmetry breaking at the scale $\mS$. For $\mu$ to be of the same order as the gauge-mediated masses, we must impose that the suppression factor $(\mS / \mst )^{\Delta_{\O}}$ is roughly of the same size as a SM gauge loop factor $g^2/16\pi^2$. This requires an accidental coincidence. However, as we know that the tumbling mechanism can generate a certain hierarchy between $\mst$ and $\mS$, it is not implausible to obtain a value in the right ballpark.

The interesting surprise comes from the calculation of $B_\mu$, which leads to  $B_\mu /\mu^2\sim (\mS / \mst )^{\Delta_{\cal R}-2\Delta_{\O}}$, where $\Delta_{\cal R}$ is the scaling dimension of the vector superfield that corresponds to the composite operator $\O^\dagger \O$. Since $\Delta_{\cal R}=2\Delta_{\O} +O(1/N)$, the $\mu$--$B_\mu$ problem is solved at large $N$.

To better explain the physical meaning of this result, we have shown that, had the Higgs been only partially composite, $\mu^2/B_\mu$ would be suppressed by a loop-factor, exactly as in simple variants of gauge mediation. The $\mu$--$B_\mu$ problem is solved only because of the fully composite nature of the Higgs.

A geometric interpretation of our $\mu$--$B_\mu$ solution is offered by the 5D picture. Since the Higgs sector is localized at the IR$_*$ brane, while supersymmetry breaking occurs at the IR$_S$ brane, the Higgs feels supersymmetry breaking only through the tail of the profile of some chiral bulk field, see \fig{fig:susy}. The fact that $\mu$ depends on the bulk field and $B_\mu$ on the bulk field squared provides the proper scaling $B_\mu /\mu^2\sim 1$.

Besides $\mu$ and $B_\mu$, compositeness also generates soft masses for the Higgs and, more importantly, trilinear $A$-terms, which are aligned to Yukawa couplings. This is a novelty with respect to gauge mediation, which predicts vanishing $A$ coefficients at the mediation scale. This novelty is a welcome twist because a large top $A$-term helps obtaining the Higgs mass without resorting to excessively heavy stops.

\subsubsection*{Supersymmetry breaking: III. Partial compositeness}

Partial compositeness of matter fields leads to another class of contributions to soft masses which, although smaller in size than the effects from gauge mediation, is important because it is not flavor universal. The corresponding soft masses and $A$-terms come with coefficients that reflect the pattern of Yukawa couplings but are not aligned with them. This gives rise to a non-trivial structure of flavor violations, which is analogous to the one encountered in theories with flavorful supersymmetry~\cite{Nomura:2008pt}.

An important novelty introduced by the Super-Composite Higgs comes from a subtle interplay between compositeness and supersymmetry. The theory contains supersymmetry-breaking effective interactions originating from compositeness and therefore suppressed by powers of $\mst$. As such, these interactions manifest themselves differently from the usual soft terms and cannot be described by the ordinary classification of low-energy soft terms. For this reason, we call them Beyond Soft. Examples are non-holomorphic $A$-terms and wrong-Higgs Yukawa couplings. Beyond Soft terms give the largest effects in flavor transitions, also because they can be enhanced by powers of $\tan\beta$.

\subsubsection*{Parameters and mass scales in the Super-Composite Higgs}

As discussed in \sec{sec:Minimal}, the fundamental parameters of the theory are the two mass scales $\mst$, $\mS$ and the two quantities $N$ and $\Delta_\O$, pertaining to the $\CFTs$ sector. Moreover, one of the mass scales can be eliminated by imposing the condition of radiative EW breaking. In order to relate our results to a physical observable, we have chosen to express all mass scales in terms of the squark mass ${\tilde m}_q$. This allows us to take ${\tilde m}_q$, $N$ and $\Delta_\O$ as our 3 input parameters. In principle, all other quantities in the theory can be calculated in terms of these 3 parameters. In practice, however, our results should be regarded only as estimates because they are affected by order-one uncertainties due to incalculable contributions from the strongly-coupled sector.

Following this procedure, we obtain the predictions for the different mass scales shown in figs.~\ref{fig:mu} and \ref{fig:Spectrum}. It is remarkable how much we can learn about the pattern of mass scales in the Super-Composite Higgs, in spite of its non-perturbative nature. However, our results should not be over-interpreted, as they are affected by uncertainties due the strong dynamics. In particular, the radiative breaking condition has been calculated ignoring the contributions from Higgs compositeness, which depend on additional unknown coefficients and could significantly change our results.

Nevertheless, a pattern emerges, with supersymmetric partners in the range 1--10 TeV, the supersymmetry-breaking scale $\mS$ in the range 50--500 TeV, and the compositeness scale $\mst$ in the range 1--100 PeV.

The Super-Composite Higgs predicts an intricate pattern of intertwined dynamical mechanisms that explain the naturalness and flavor issues, and that populate the energy domain from the TeV to multi-PeV. This is not dissimilar to the SM, which describes different but related mechanisms that explain the properties of elementary particles and that populate the energy domain from chiral symmetry breaking at a few hundreds of MeV to spontaneous EW breaking at a hundred GeV.

With our minimal set of IR scales in the CFT sector needed to solve the $\mu$--$B_{\mu}$ problem ($\mst$ and $\mS$), the theory predicts the upper bound $\mst < \mS^2/\mu$ based on solid unitarity considerations. Numerically, the result depends on ${\tilde m}_q$ and $N$. For instance, for ${\tilde m}_q = 10$~TeV, the upper bound on $\mst$ ranges from 20 to 100 PeV, depending on $N$.

\subsubsection*{Flavor constraints}

Since flavor has a dynamical origin in the Super-Composite Higgs, it is not surprising that the theory predicts a variety of new effects in processes with flavor transitions and/or CP violation, see \sec{sec:flavor}.

In the Super-Composite Higgs, there are three classes of contributions to flavor processes. The first class is a purely supersymmetric effect of compositeness associated to dim-6 effective operators suppressed by $1/\mst^2$. The second class consists of standard soft terms but with a flavorful structure arising from partial compositeness. Both of these effects are unavoidable, as they are essential components of the theoretical edifice. They have counterparts in ordinary theories with partial compositeness and supersymmetry, respectively. The novel aspect introduced by the Super-Composite Higgs is that exact supersymmetry prevents compositeness from generating dipole operators. When these two classes of effects are compared with experimental data, they pass all tests on flavor processes by a substantial margin, leading us to conclude that the theory is safe, regardless of uncertainties arising from strong dynamics.

The third class consists of Beyond Soft terms, which are a peculiarity of supersymmetry breaking and compositeness coming together at relatively low scales. 
They are responsible for the largest contributions to flavor processes in the theory, possibly within the range of detectability. 

The essential feature of the Super-Composite Higgs in addressing the flavor problem is the separation of scales, with $\mS$ responsible for naturalness and $\mst$ governing flavor. A certain hierarchy in the ratio $\mst/\mS$ provides an acceptable degree of naturalness, while also adequately suppressing contributions to flavor and CP violating processes. However, flavor effects can never fully decouple in the Super-Composite Higgs, as unitarity imposes a strict upper bound on the ratio $\mst/\mS$ or, equivalently, a lower bound on new contributions to flavor and CP violating processes.

\subsubsection*{The electron EDM}

Among all flavor/CP constraints, the most significant concerns the electron EDM.
Beyond Soft terms contribute to the electron EDM through non-holomorphic $A$-terms and wrong-Higgs Yukawa interactions, with the latter giving the largest effect. 

While theoretical uncertainties due to strong dynamics inhibit definitive conclusions, it appears that the theory not only aligns with current experimental constraints but also suggests that the electron EDM is within reach of future searches. The situation is especially encouraging, as upcoming experimental upgrades are expected to improve the experimental sensitivity by several orders of magnitudes. Our findings highlight the electron EDM as the most promising flavor probe for the Super-Composite Higgs.

Beyond Soft contributions to the electron EDM are a distinctive feature of the Super-Composite Higgs. However, these contributions could potentially be eliminated or suppressed, though this would require  elaborate modifications to the theory, see \sec{sec:extrasym}.

\subsubsection*{EW naturalness}

The Super-Composite Higgs cures the big hierarchy problem posed by the Higgs mass, but suffers from a remaining tuning given by $(3~\tev / {\tilde m}_t)^2 \, 0.3\%$, using the parameter sensitivity criterion of ref.~\cite{Barbieri:1987fn}. Moreover, the condition that determines the ratio of Higgs vevs gives an additional tuning of order
 $1/\tan\beta $. Undoubtedly, the situation is unsatisfactory.

However, the merit of the Super-Composite Higgs is to provide a microscopic description that meets the expectations of the bottom-up approach discussed in \sec{sec:intro}. In the Super-Composite Higgs, the scale of supersymmetry breaking $\mS$ minimizes the logarithm that measures the Higgs sensitivity to heavy colored sparticles. From a bottom-up perspective, this is the least conceivable tuning in the context of low-energy supersymmetry. 

The comparison with non-supersymmetric Higgs compositeness is even more striking. In typical composite models, in which flavor origins are addressed  at the compositeness scale $m_*$, the amount of tuning in the EW breaking condition is roughly $8\pi^2 m_h^2/(3 y_t^3 m_*^2)\sim (0.5\, \tev/m_*)^2$. Since flavor constraints require $m_*> 2.2$ PeV, the tuning is at the level of one in ten millions.

The Super-Composite Higgs does not completely solve the little hierarchy problem, but it achieves the best one can hope for, given the current limits on supersymmetric particles.

\subsubsection*{Ultra-light gravitino and its cosmological signatures}

A characteristic feature of the Super-Composite Higgs is an ultra-light gravitino, with mass in the eV range or below (see \sec{sec:cosmo}). Being the LSP, the gravitino is a stable particle, which populates the universe as a cosmic relic. If the reheating temperature at the end of inflation is larger than the typical scale of soft masses, gravitinos get in thermal equilibrium in the early universe and decouple when they are still relativistic. Therefore, their current density is determined only by their mass and by the effective number of degrees of freedom present in the thermal bath at the time of decoupling. 

Ultra-light gravitinos give a cosmological imprint that can be tested by CMB experiments and galaxy surveys. A combination of current data leads to an upper bound $\mgra <2.3$~eV~\cite{Xu:2021rwg}, which lies in a very significant mass range for the Super-Composite Higgs. Future experiments in observational cosmology are expected to fully cover the mass window of ultra-light gravitinos, providing a crucial test of our theory. Super-Composite Higgs predicts an effect that can be measured by upcoming CMB and large-scale structure experiments.

\subsubsection*{High-energy collider prospects}

Nearly-degenerate right-handed sleptons are the most likely form of NLSP, although the case of a Bino NLSP is still possible, especially if $N$ is not too large. Therefore, the most common situation leads to  distinctive experimental signatures, with collider final states characterized by missing energy and charged leptons of all three generations.  

The limit on the gravitino mass from data on the CMB and large-scale structures can be translated into an upper bound on the spectrum of supersymmetric particles. In particular, the corresponding upper limit on the squark mass is shown in \fig{fig:cosmo}. These upper bounds are based solely on experimental data and do not depend on the naturalness criterion or similar arguments. However, they assume that the universe has attained temperatures of at least a few TeV at some point in its cosmological history.  

The result is striking. Astronomical observations indirectly impose a stringent upper bound on the mass spectrum of supersymmetric particles, predicting that squark masses must lie within a window, approximately between 3 and 10 TeV.  This is within reach of future colliders exploring the 10 TeV mass range. Combining this result with the upper bound on the compositeness scale $\mst$ from unitarity, we conclude that all mass scales of the Super-Composite Higgs cannot be arbitrarily large. 

\subsubsection*{Dark matter}

In the Super-Composite Higgs, the LSP is not a viable dark matter candidate.
However, an intriguing coincidence occurs if the supersymmetry-breaking sector contains a stable and gauge-neutral composite state. This composite particle possesses the correct properties to qualify as a dark matter candidate, with the appropriate thermal relic abundance achieved in the preferred parameter range of the Super-Composite Higgs. Its spin-independent scattering cross section off nuclei is at the edge of currents limits from direct dark-matter detection experiments and within the sensitivity of future searches. Indirect detection signals are suppressed, as dark matter annihilates primarily through an invisible channel.

\subsubsection*{What is gained with the Super-Composite Higgs?}

The Super-Composite Higgs aims to tackle both naturalness and flavor within a single realistic framework. This goal is achieved by combining supersymmetry and compositeness: supersymmetry cures the big hierarchy problem, while partial compositeness explains the flavor pattern. The advantage of our approach, compared to ordinary Higgs compositeness, is that supersymmetry allows for a separation of scales between the dynamics that resolve naturalness and flavor issues. Nevertheless, this separation does not merely amount to a trivial decoupling of compositeness, as can be understood from the following two observations.

The first crucial observation is that, in the Super-Composite Higgs, the scale separation between the two sectors is not arbitrary: flavor cannot be decoupled from supersymmetry breaking. Specifically,
the compositeness scale $\mst$ cannot be arbitrarily larger than the supersymmetry breaking scale $\mS$ because unitarity, through the structure of the theory, imposes an upper bound on the ratio $\mst /\mS$. In other words, the solutions of naturalness and flavor are inextricably connected.

The second crucial observation is that supersymmetry is a space-time symmetry, and therefore connects particle physics to gravity. The low scale of supersymmetry breaking leads to the existence of an ultralight gravitino, which faces stringent cosmological constraints. Current observations imply an upper bound on $\mS$, the scale that governs naturalness. Consequently,  the scales of naturalness and flavor are not only intimately related, but they are also bounded from above, making the theory experimentally testable and falsifiable.

\subsubsection*{Conclusions}

The Super-Composite Higgs offers a new and attractive framework to simultaneously tackle the big hierarchy problem and the flavor puzzle. It satisfies all constraints from flavor/CP processes and opens up new theoretical perspectives in model building. It also motivates explorations of strongly-coupled QFTs with the ambitious goal of finding explicit constructions that feature the appropriate operator spectrum and tumbling dynamics.

The Super-Composite Higgs makes striking experimental predictions across three different areas of fundamental physics. At the observational cosmology frontier, it predicts detectable signals in future CMB and large-scale structure experiments, with the effects arising from relic ultra-light gravitinos. At the precision frontier, it predicts an electron EDM within reach of future experimental upgrades. At the high-energy frontier, it predicts discoveries at future colliders such as the FCC-$hh$, since the masses of supersymmetric particles must necessarily respect an upper bound determined by current cosmological data, which correspond to about 10 TeV for squarks and 15 TeV for gluinos. These predictions follow from relatively robust considerations, evaded only by introducing non-minimal modifications to the theory or the cosmological history.  

\subsection*{Acknowledgments}

KA is supported by NSF grants PHY-2210361 and PHY-2609893. RR is partially supported by the Swiss National Science Foundation under contract 200020-213104 and through the National Center of Competence in Research SwissMAP. RS is supported by NSF grants PHY-2210361 and PHY-2514660.  We acknowledge useful discussions with Tony Gherghetta, Jacqueline Lodman, Alex Pomarol, Lisa Randall, and Luca Vecchi.

\addcontentsline{toc}{section}{References}
\bibliographystyle{mine}
\bibliography{SuperC}

\end{document}